\documentclass[preprint,11pt]{JHEP3} 

\JHEPspecialurl{http://jhep.sissa.it/JOURNAL/JHEP3.tar.gz}

\usepackage{epsfig,multicol,amsmath,slashed}

\def\beq{\begin{equation}}
\def\eeq{\end{equation}}
\def\bea{\begin{eqnarray}}
\def\eea{\end{eqnarray}}

\def\eq#1{{Eq.~(\ref{#1})}}
\def\fig#1{{Fig.~\ref{#1}}}
\def\sec#1{{section~\ref{#1}}}
\def\secc#1{{Section~\ref{#1}}}

\def\Tab#1{{Table~\ref{#1}}}

\parskip=\itemsep               

\newcommand{\nn}{\nonumber}
\newcommand{\D}{\partial}

\newcommand{\h}{\frac{1}{2}}

\newcommand{\be}{\beta}

\newcommand{\ga}{\gamma}
\newcommand{\de}{\delta}
\newcommand{\De}{\Delta}

\newcommand{\la}{\lambda}

\newcommand{\Ga}{\Gamma}

\newcommand{\La}{\Lambda}

\newcommand{\f}{\frac}

\newcommand{\SPE}{<|S^2_{\mbox{\footnotesize{enh}}}|>}

\newcommand{\bas}{\bar{\alpha}_s}
\newcommand{\as}{\alpha_s}

\newcommand{\Lb}{\left(}
\newcommand{\Rb}{\right)}

\newcommand{\GV}{\mbox{GeV}/\mbox{c}^2}

\def\npb#1#2#3{    {\it Nucl. Phys. }{\bf B#1} (19#2) #3}

\def\prd#1#2#3{    {\it Phys. Rev. }{\bf D#1} (19#2) #3}
\def\prdc#1#2#3{    {\it Phys. Rev. }{\bf D#1} (20#2) #3}

\title{\LARGE \bf Summing loops in QCD:\\
Exclusive Higgs production and the survival probability}
\author{\large  J.~Miller\thanks{Email:
jeremymi@post.tau.ac.il; miller@physics.org}\,\, \\
CENTRA, Departamento de F$\acute{i}$sica, Instituto Superior T$\acute{e}$cnico (IST),\\  Av. Rovisco Pais,\\1049-001 Lisboa,\\Portugal}

\abstract{The Pomeron loop amplitude and the amplitude for the diagram with any general number of loops is derived in the QCD dipole approach. It was 
found that the major contribution to the amplitude of an arbitrary Pomeron enhanced diagram, is equivalent to the amplitude of the diagram
with non interacting Pomerons.
This provides the necessary tools for solving the long standing theoretical problem of summing over Pomeron loop diagrams. 
In this theoretical framework the contribution of the full set of enhanced diagrams to the survival probability is estimated.
This enables an accurate prediction for the exclusive cross section for diffractive Higgs production, which includes the suppression factor needed to
screen out the full set of hard re-scattering corrections, in QCD.}

\keywords{ BFKL Pomeron, Triple Pomeron vertex, Higgs boson, summing Pomeron loops, QCD}
\preprint{ \today}

\begin{document}

\section{Introduction}
\label{sintro}\!

\DOUBLEFIGURE[h]{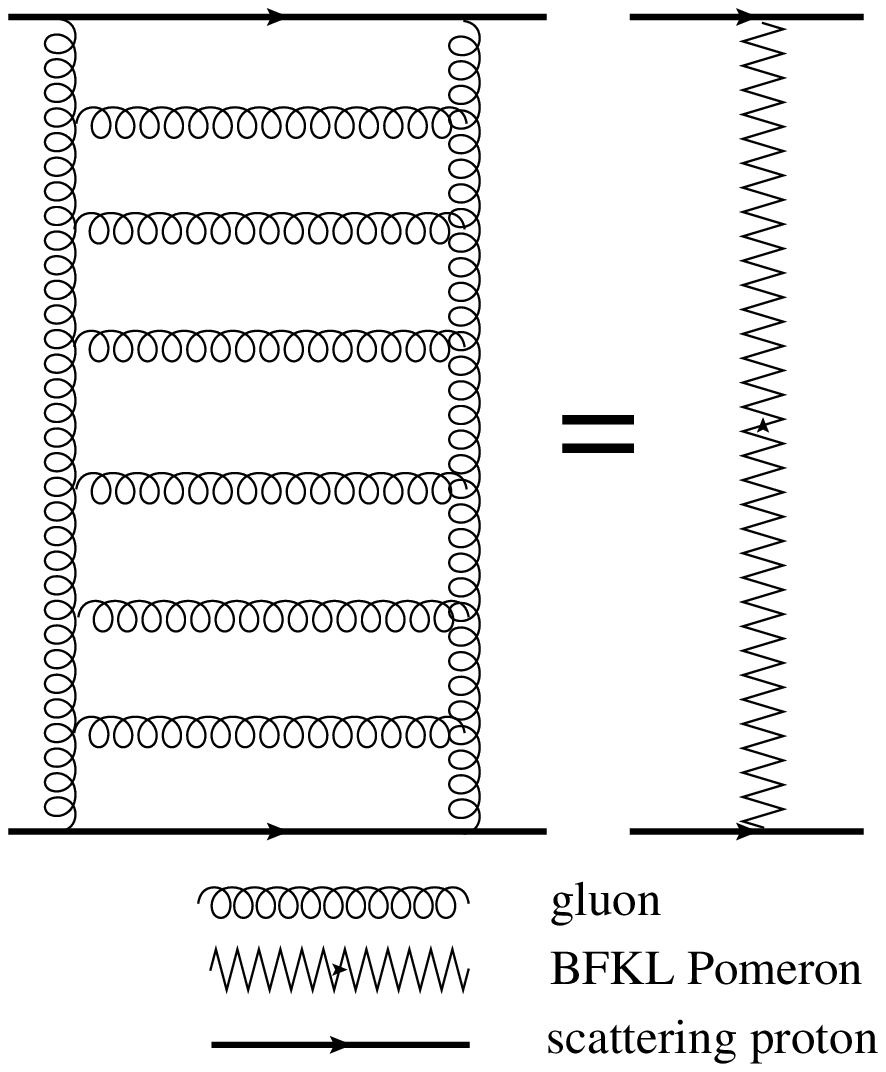,width=50mm,height=45mm}{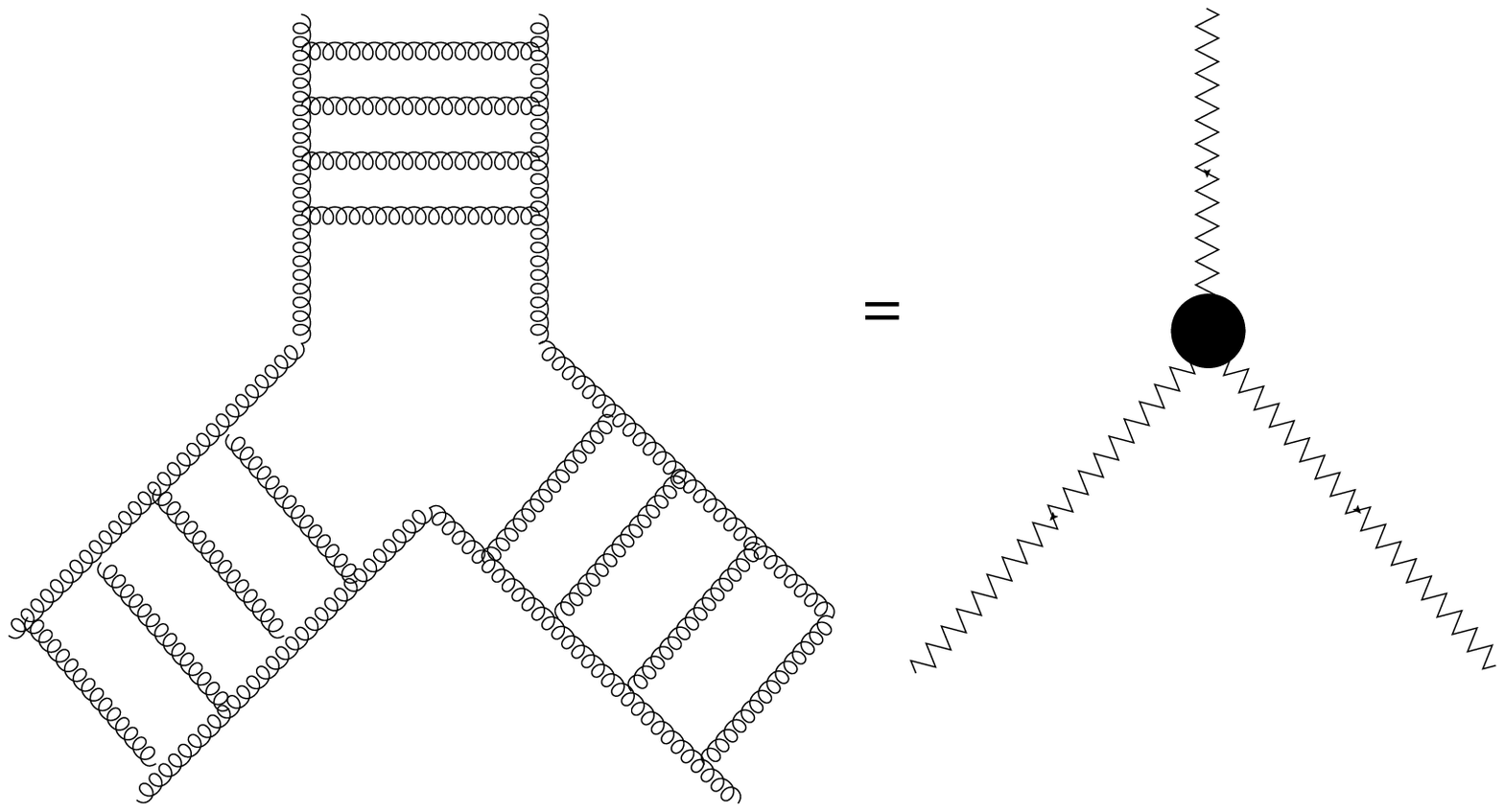,width=80mm,height=45mm}
{The BFKL Pomeron structure.
\label{fBFKLPomeron}}{The triple Pomeron vertex.\label{ftriplePomeronvertex} }

The main goal of this paper is to solve the unsolved problem of the summation over Pomeron loops, in QCD. 
In proton proton collisions the main t-channel exchange is the BFKL Pomeron, which is a double t channel gluon exchange with ``ladder''
gluons in between, as shown in \fig{fBFKLPomeron}. The BFKL Pomeron can split into two daughter Pomerons and re-merge through the triple Pomeron vertex shown in \fig{ftriplePomeronvertex}, forming Pomeron
loops. Due to the large size of the triple Pomeron vertex (see for example refs. \cite{Korchemsky:1997fy, Navelet:1997fy}) Pomeron loop diagrams give a significant contribution
to the high energy scattering amplitude in proton proton collisions, which is comparable to the amplitude of the basic diagram of \fig{fBFKLPomeron}. As such, the scattering amplitude of hadronic
reactions requires an accurate estimate for the summation over Pomeron loop diagrams to be taken into account (for an example of such diagrams, see \fig{fPomeronloop2}, \fig{fembeddedpict} and \fig{fnpict}).
Hence the correct algorithm for the summation over Pomeron loops is a result which is in high demand, not just from a theoretical perspective, but also from an experimental point of view.
The main practical application of this result is in exclusive diffractive Higgs production, at the LHC. \\

\FIGURE[h]{ \centerline {\epsfig{file=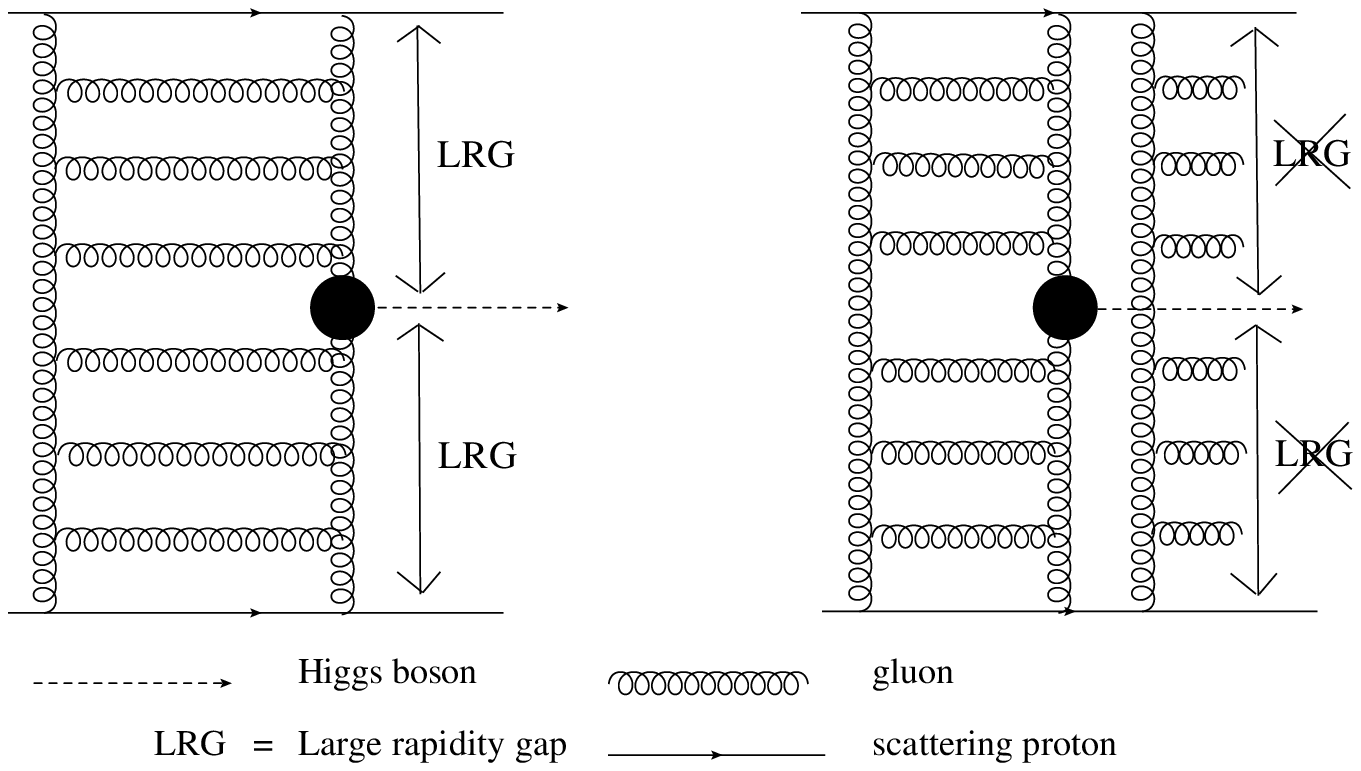,width=100mm}}\caption{ Diffractive production of the Higgs boson through double t-channel gluon exchange.
On the left is exclusive Higgs production with large rapidity gaps (LRG) between the Higgs and the emerging protons. On the right is the production of Higgs with extra production which spoils the LRGs, arising from
additional inelastic scattering.} \label{flrg} }

The keenly awaited result of the discovery of the Higgs boson at the LHC, is expected to emerge from smashing together two protons,
in so-called diffractive Higgs production.  Unfortunately, proton-proton scattering results in the production of many other unwanted particles.
This makes the process of detecting the Higgs very problematic. The desired result is the production of the Higgs with two large rapidity gaps (LRG) between
 the Higgs and the emerging protons as shown in the left hand diagram of \fig{flrg}, which ensures that there is no additional production.
The survival of these two large rapidity gaps is quantified by the survival probability.
The difficulty in isolating the Higgs signal is characterized by the small value of the survival probability,
which has so far been estimated to be small, and could be even less than 1 \% (see detailed estimates in refs. \cite{Miller:2006bi,Gotsman:2008tr}).

Due to the size of the strong coupling, the dominant mechanism for diffractive Higgs production is through the exchange of
a t channel gluon between the scattering protons. To ensure that there is no further production, a second t - channel gluon exchanged between the scattering protons
is needed to cancel the color flow, as shown in \fig{flrg}. This colorless 
double gluon exchange cancels the possibility of the production of additional particles.\\

The double gluon exchange evolves to the ``ladder'' structure, as gluons are exchanged between the
two t channel gluons, forming the rungs of the ladder (see \fig{flrg}). This structure is the so called BFKL Pomeron. The energy levels of the
BFKL Pomeron are labeled by the BFKL eigenfunction $\omega\Lb n,\nu\Rb$, where $n$ is an integer and $\nu$ is a conformal variable
which one integrates over, when calculating Feynman diagrams.\\

Unfortunately, extra parton showers between the scattering protons is inevitable, leading to
the production of additional particles that fill up the large rapidity gaps (see \fig{flrg} right).
Hard re-scattering corrections, namely Pomeron branching which forms loop diagrams and fan diagrams of the type shown in \fig{flfs} also contribute to the problem of additional unwanted production.
Pomeron loop diagrams and fan diagrams of this kind are called ``enhanced'' diagrams. 
The suppression factor needed, to remove the effect of enhanced diagrams on the Higgs cross section, is the ``enhanced survival probability''  $<|S^2_{\mbox{\footnotesize{enh}}}|>$.\\

Pomeron loop diagrams are extremely useful because they include the contribution of  ``fan diagrams", Pomeron
loops and the contribution of non interacting Pomerons as shown in \fig{flfs}.
As such, summing over the complete set of Pomeron loop diagrams will provide the suppression factor needed for screening out all
unwanted particle production, which stems from Pomeron enhanced diagrams. In other words, the summation of Pomeron loops provides the
full set of hard re-scattering corrections to the enhanced survival probability $<|S^2_{\mbox{\footnotesize{enh}}}|>$.\\

\FIGURE[h]{ \centerline {\epsfig{file=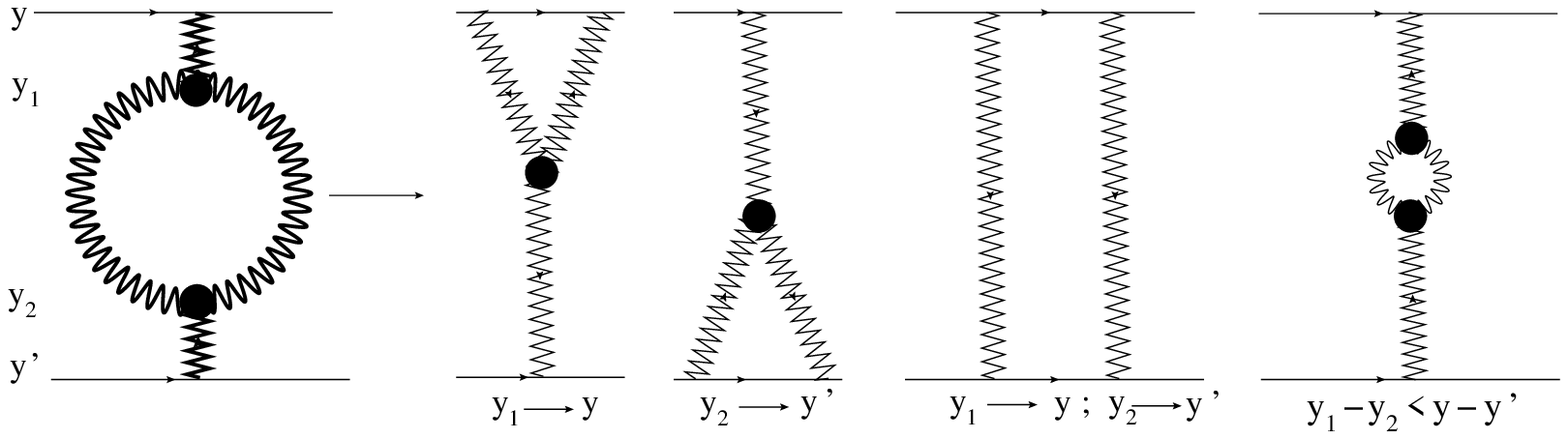,width=180mm}}\caption{ The first Pomeron loop diagram gives the simultaneous contribution of the fan diagram, the loop diagram and the contribution from non interacting
Pomerons.} \label{flfs} }

 In ref. \cite{Miller:2006bi} the one loop Pomeron diagram was calculated in QCD.
The result was used to fix the parameters of the mean field approximation (MFA) of Mueller, Patel, Salam and Iancu (MPSI) (see ref. \cite{toy}), and using this the summation over Pomeron loops
 was estimated using the toy model of Kovchegov \cite{20}.
In this way the survival probability was found to be potentially as low as $0.4 \%$. This implied that Pomeron loop diagrams give a significant contribution to the survival probability.
In ref. \cite{Levin:2007wc} we derived a useful expression for the triple Pomeron vertex in the momentum representation, which is the crucial ingredient
for calculating any diagram with Pomeron branching and Pomeron loops. Using this we showed that a diagram with an arbitrary number of Pomeron loops reduces to the
diagram of non interacting Pomerons. This important step forward, outlined the first method for the summation over Pomeron loop diagrams completely
in the QCD approach, and not in the mean field approximation.\\

In a recent publication \cite{Braun:2009fy} by M. Braun, the Pomeron loop diagram was calculated using the original exact form of the triple Pomeron
vertex, which was first derived by Korchemsky \cite{Korchemsky:1997fy} and at the same time by Bialas, Navelet and Peschanski (BNP) in ref.
\cite{Navelet:1997fy}. It was suggested in ref. \cite{Braun:2009fy} that use of the Korchemsky expression for the triple Pomeron vertex,
leads to a more accurate result for the Pomeron loop diagram. Therefore the approach suggested in ref. \cite{Braun:2009fy} has been adopted in this paper.\\

The approach used to find the Pomeron loop amplitude, 
can be extended to the calculation of more complicated diagrams  (see for example \fig{figN}), with an arbitrary number of Pomeron loops in the QCD approach instead of the MFA formalism. 
Using the new algorithm developed in this paper, the amplitude for multiple loop diagrams derives from an iterative expansion of the simple Pomeron loop amplitude. In this framework,
a general formula can be found for the amplitude of the diagram with an arbitrary number of Pomeron loops
in QCD, which is a function
of the number of loops in the diagram, and the rapidity gap between the scattering protons. If the production of the Higgs boson is included in the
diagram, the formula is also a function of the mass of the Higgs boson.  This provides the
mechanism for summing over Pomeron loop diagrams in the precise QCD formalism, instead of the toy model approach of ref. \cite{toy,20}. In this 
approach it was found that the main contribution to the amplitude of the general enhanced diagram, is equivalent to the
amplitude of the diagram with non interacting Pomerons, with renormalized Pomeron vertices. \\

This is a considerable step forward theoretically, since it makes possible the summation over Pomeron loop diagrams in QCD to any order.
Experimentally this means that an accurate prediction for the exclusive cross section for diffractive Higgs production is possible that 
includes the suppression factor needed to screen out the full set of re-scattering corrections, namely the enhanced survival probability.
In this new formalism it was found that the contribution of enhanced diagrams to the survival probability is substantial,
confirming the results found in refs. \cite{Miller:2006bi, Gotsman:2008tr}.
The results show also that the survival probability is very sensitive to the choice of the strong coupling,
 and as such it decreases as $\as$ increases, in agreement with ref. \cite{Miller:2006bi}.

This paper is organized in the following way. In \sec{s1} the notation and conventions used throughout this paper are listed. The amplitude
for the basic diagram of \fig{f1p}, which shows diffractive Higgs production through single t-channel
Pomeron exchange is derived. \secc{s2} is devoted to the derivation of the amplitude of the Pomeron loop, shown in \fig{fPomeronloop2}.
A pedagogical approach is taken to explain the style of integration over the conformal variables in the loop. This forms the
basis of the techniques used for more complicated diagrams in later sections. In \sec{smi} the main idea used to derive the formula for the general multiple Pomeron loop diagram 
is outlined, which is an iterative technique. 
 Using this iterative approach, in \sec{s3}, more complicated multiple Pomeron loop diagrams, such as \fig{fembeddedpict}, \fig{fnpict} and \fig{figN} are calculated.
Finally in \sec{s4}  the general expression for the amplitude
of the diagram with an arbitrary number of Pomeron loops is derived. This provides the tools necessary to sum over the complete set of
Pomeron loop diagrams up to any order. The summation over enhanced diagrams forms the basis for \sec{ssurvivalprobability}, 
 to estimate the enhanced survival probability $<|S^2_{\mbox{\footnotesize{enh}}}|>$. These results are discussed in the conclusion in \sec{sc}. In the appendix section a brief overview of the derivation
of the triple Pomeron vertex is given, mostly following the strategy of Korchemsky  in ref. \cite{Korchemsky:1997fy}, and the amplitude of the first fan diagram of \fig{ffan} is calculated.

\section{The Pomeron propagator}
\label{s1}

The following conventions and notations will be used, some of which are based on the paper of ref. \cite{Braun:2009fy} by M. Braun.
The Pomeron propagator shown in \fig{fPomeronpropagator1}, which is a t channel exchange between two color dipoles in QCD, with
rapidity values $y$ and $y^{\,\prime}$, is denoted by the expression;

\beq
g_{y-y^{\,\prime}}\,=\,g_{y-y^{\,\prime}}\Lb R\,,\, r_1,r_2\,|\,R^{\,\prime}\,,\,r_1^{\,\prime}\,,\,r_2^{\,\prime}\Rb\,
\label{Pomeronpropagator}\eeq

\FIGURE[h]{ \centerline {\epsfig{file=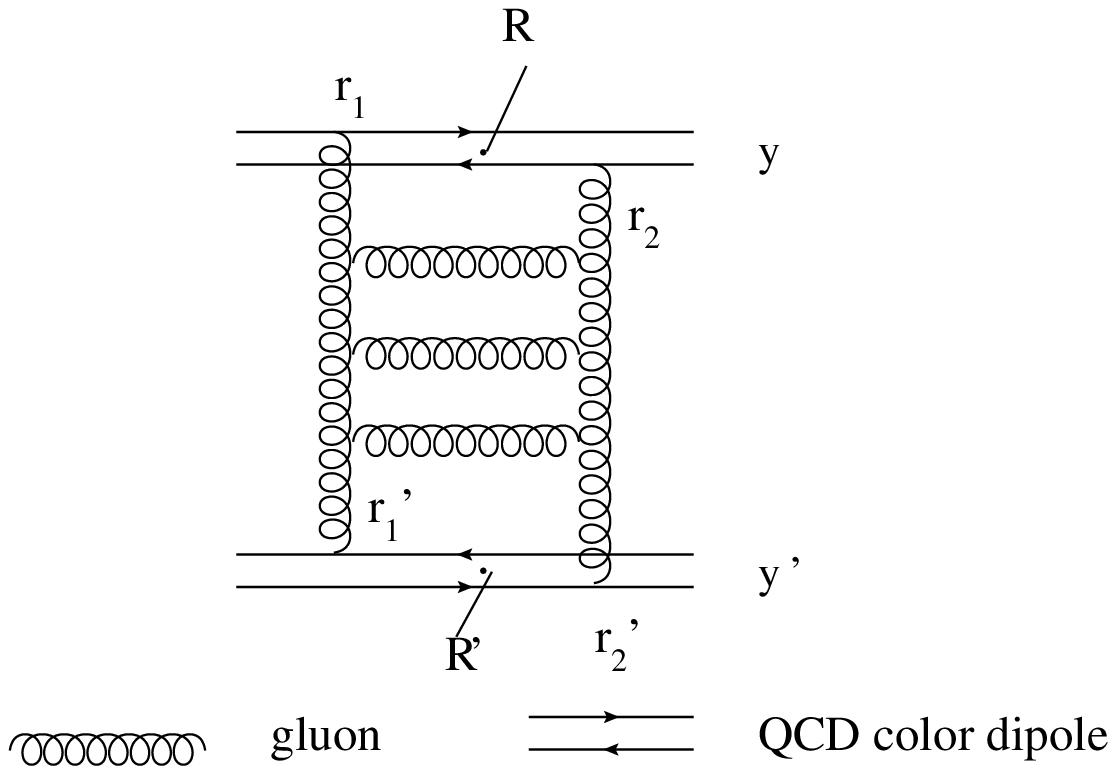,width=70mm}}\caption{ Coordinates for the two reggeized gluons in the BFKL Pomeron propagator, in color dipole scattering.} \label{fPomeronpropagator1} }

where as shown in \fig{fPomeronpropagator1}, $r_1$ and $r_2$ denote the initial and final coordinates of the two reggeized gluons, and $R$ is the center of mass coordinate of the two reggeized gluons. The conformally invariant expression, for the
coupling of the BFKL Pomeron to the QCD color dipole is;

\bea
E_\ga\,=\,\Lb\f{r_{12}}{r_{10}r_{20}}\Rb^\ga\Lb\f{r^\ast_{12}}{r_{10}^\ast r_{20}^\ast}\Rb^{\tilde{\ga}}\label{coupling}\eea

where $r_{ij}\,=\,r_i-r_j$ and $r_{i0}\,=\,r_i-R$. The conformal weights are given by the expression

\bea
\ga\,&&=\,\f{1+n}{2}+i\nu; \hspace{1.5cm} \tilde{\ga}\,=\,1-\ga^{\ast};\hspace{3cm}n\,\in\,\mathbb{Z};\,\,\,\nu\,\in\mathbb{R} \label{conformalweights}
\eea

Two important properties of the conformal weights are

\bea
\ga\,-\tilde{\ga}\,=\,n\,\hspace{1.5cm}\,\ga\,+\,\tilde{\ga}\,=\,1+2i\nu\label{properties}
\eea

where $n$ is the conformal spin, and $\nu$ is the scaling dimension of the state. Wherever $\ga$ appears, it is intended to be a shorthand for the set of two numbers $\left\{n\,,\,\nu\,\right\}$. The above expression for the Pomeron propagator in
\eq{Pomeronpropagator} can also be written in the representation of complex angular momentum $j\,=\,1+\omega$ instead of rapidity $y$ by the following Mellin transformation;

\bea
g_{y-y^{\,\prime}}\,=\,\int^{a+i\infty}_{a-i\infty}\f{d\omega}{2\pi i}\,e^{\,\omega \Lb y-y^{\,\prime}\Rb}\,g_\omega\label{omegarepresentation}\eea

where in $\omega$ representation, the Pomeron propagator is given by the following expression;

\bea
g_\omega\,=\,\int\!d^2R\int\!d^2R^{\,\prime}\,\int\!\mathcal{D}\ga\,E_\ga \,E^{\,\prime}_{\,-\,\tilde{\ga}}\, g_\omega\Lb\ga\Rb\label{omrep}\eea

where $\mathcal{D}\ga$ is a shorthand notation for

\bea
\int\mathcal{D}\ga\,&&=\,\oint_{C_\ga}\,d\ga\label{oint}\,h\Lb\ga\Rb;\hspace{1.5cm}\mbox{where}\hspace{1.5cm}h\Lb\,\ga\,\Rb\,=\,\f{2}{\pi^4}\,\begin{vmatrix}\ga\,-\,\f{1}{2}\,\end{vmatrix}^2\label{hgam}\eea

\FIGURE[h]{\begin{minipage}{70mm}
\centerline{\epsfig{file=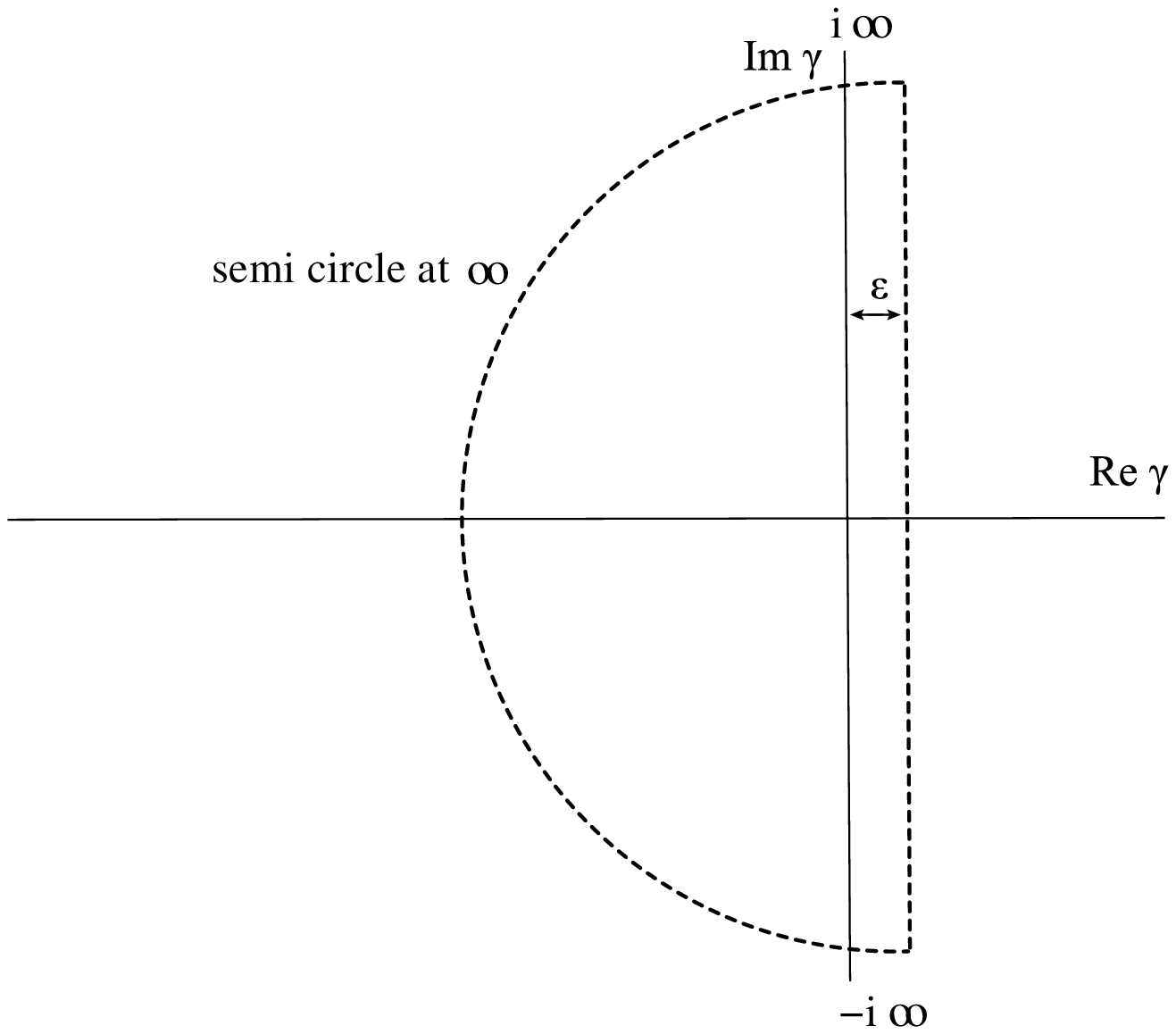,width=70mm}}
\end{minipage}
\caption{Contour enclosing singularities for the integration over the
conformal variable $\gamma$.  } \label{fgraph} } The contour $C_\ga$ shown in \fig{fgraph}
consists of the imaginary $\gamma$ axis from $\pm\,i\infty$, and the
semi circle at infinity, to the left of the imaginary $\gamma$ axis. $C_\ga$ encloses all singularities in the
integrand of \eq{omrep}. The
integrand vanishes on the semi circle at infinity, such that
it is sufficient just to replace

 \beq
\oint_{C_\ga}\,d\gamma\,\rightarrow\,\int^{\epsilon\,+i\infty}_{\epsilon
- i\,\infty}d\gamma\,\label{ig} \eeq

It is more economical to calculate diagrams with BFKL Pomeron states in terms of $\nu$, instead  $\gamma$.  For the
integration limits
$\epsilon+i\infty\leq\gamma\leq\epsilon+i\,\infty$ (as
$\epsilon\to\,0$), \eq{conformalweights} gives the corresponding limits
of integration for the variable $\nu$, as $-\infty\leq\nu\leq\infty$,
and one should sum over all real positive integers $n$. In this way in \eq{omrep}, one can replace the
integration over $\gamma$ with the integration over $\nu$ using
the notation

\beq \sum^{\infty}_{n=-\infty}\,\int^{\epsilon\,+i\infty}_{\epsilon-
i\,\infty}d\gamma\,\,=\,\sum^{\infty}_{n=-\infty}\,\int^\infty_{-\infty}d\nu \label{gamnu}
\eeq

The notation $\sum^{\infty}_{n=-\infty}\int^\infty_{-\infty}d\nu$, corresponds to the
integration over the quantum numbers associated with the continuous
unitary variable irreducible representations of
$SL\,\Lb\,2\,,\,\mathbb{C}\,\Rb\,$ , defined in \eq{conformalweights} (see refs. \cite{Navelet:2002zz,Navelet:1998yv,Bialas:1997xp} for a detailed explanation). Thus in terms of $\nu$;

\bea
\int\mathcal{D}\ga\,&&=\,\sum^\infty_{n\,=\,-\infty}\int^\infty_{-\infty}d\nu\,h\Lb n\,,\,\nu\Rb\label{dnu}
\eea

\vspace{1cm}
\bea
\mbox{where}\hspace{1.5cm}h\Lb n\,,\,\nu\Rb\,&&=\,\f{2}{\pi^4}\begin{vmatrix}\ga-\f{1}{2}\end{vmatrix}^2\,=\,\f{2}{\pi^4}\Lb \nu^2+\f{n^2}{4}\Rb\label{h}\eea

The conformal propagator $g_\omega\Lb\ga\Rb$ takes the form;

\bea
g_\omega\Lb\ga\Rb\,&&=\,\f{1}{\omega-\omega\Lb n\,,\nu\Rb}\,\,\,\la\Lb n\,,\,\nu\Rb\label{conformalpropagator}\\
\nn\\
\mbox{where}\hspace{1.5cm}\la\Lb n\,,\,\nu\Rb\,&&=\,\f{1}{16}\,\times\,\f{1}{h\Lb n+1\,,\,\nu\Rb\,h\Lb n-1\,,\,\nu\Rb}\nn\\
&&=\,\f{1}{16}\,\f{1}{\left\{\nu^2+\Lb n+1\Rb^2/4\right\}\,\left\{\nu^2+\Lb n-1\Rb^2/4\right\}}\label{lmbda}\eea

$\omega\Lb n\,,\,\nu\Rb$ are the eigenfunctions of the BFKL equation, which represent the energy levels of the BFKL Pomeron, and these are given by the expression

\bea
\omega\Lb n\,,\,\nu\Rb\,&&=\,\bas\left\{\psi\Lb 1\Rb-\Re e\,\psi\Lb\ga\Rb\right\}\,=\,\bas\left\{2\psi\Lb 1\Rb-\psi\Lb\ga\Rb-\psi\Lb 1-\ga\Rb\right\}\label{BFKL}\eea

where the function

\bea
\psi\Lb x\Rb\,&&=\,\f{d\,\ln\Ga\Lb x\Rb}{dx}\,=\,\f{1}{\Ga\Lb x\Rb}\,\f{d\Ga\Lb x\Rb}{dx}\label{psi}\eea

is the Di-gamma function, and where $\Ga\Lb x\Rb\,=\,\Lb x-1\Rb!\,\,$ is the Euler Gamma function. Throughout this paper, the notation

\bea
\bas\,=\,\f{\as\,N_c}{\pi};\label{bas}\eea

where $N_c$ is the number of colors in the $SU\Lb N_c\Rb$ color group, or abstractly speaking, $N_c$ is the number of generators for the standard representation of $SU\Lb N_c\Rb$. For QCD, $N_c\,=\,3$. Here, $\as$ is the strong coupling for QCD
interactions, and it depends inversely on energy. Wherever a value for $\as$ is used in a calculation, its value for a specific choice of energy scale will be stated explicitly.\\

\FIGURE[ht]{ \centerline {\epsfig{file=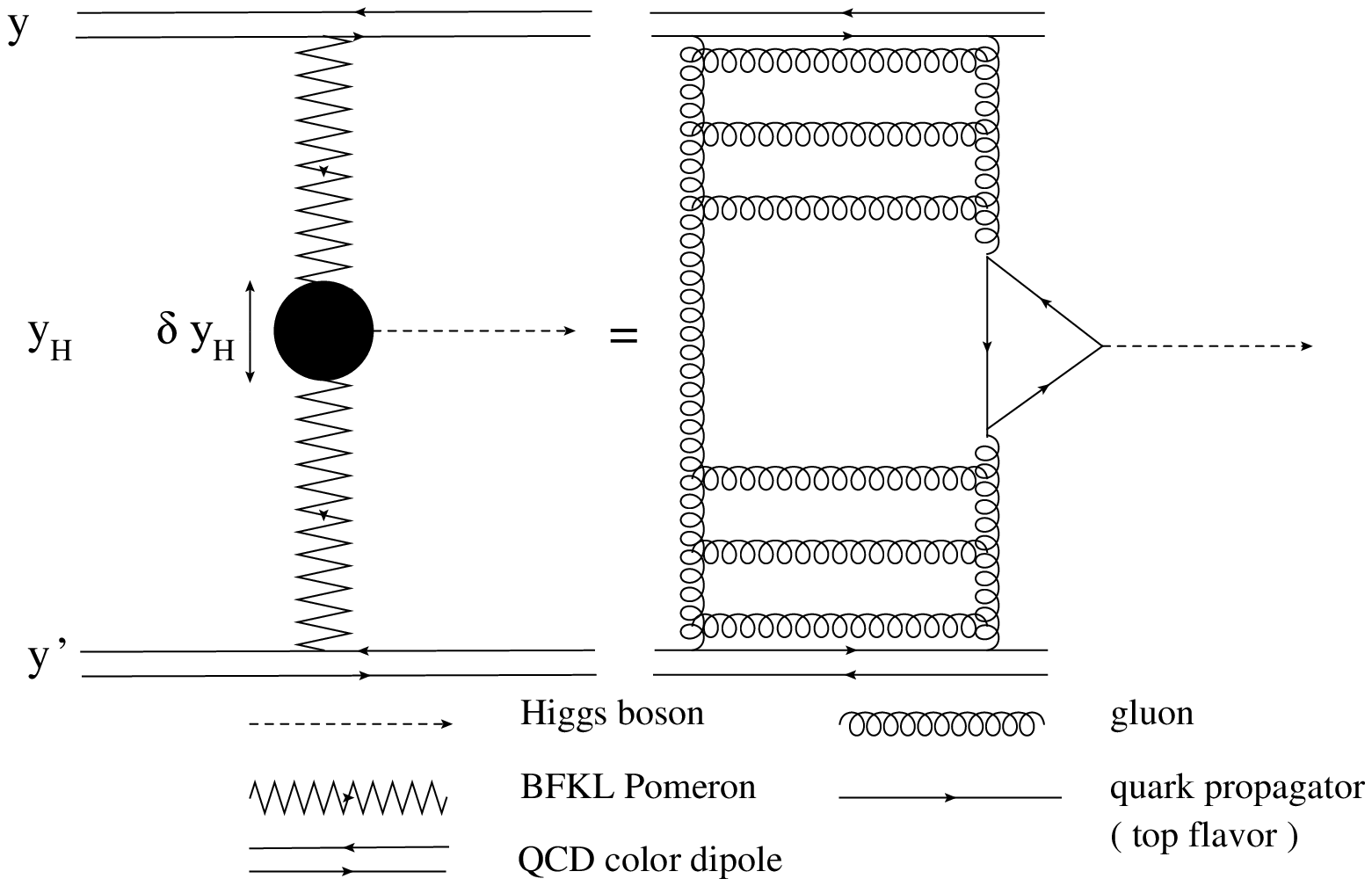,width=100mm}}\caption{ Diffractive production of the Higgs boson through t-channel Pomeron exchange.
The process Pomeron + Pomeron $\to$ Higgs proceeds mostly through the top quark triangle.} \label{f1p} }

Substituting for the conformal propagator \eq{conformalpropagator} in \eq{omegarepresentation}, and using the integration measure of \eq{dnu} to integrate over the conformal variable,
the amplitude for diffractive Higgs production through the t channel Pomeron exchange process shown in \fig{f1p} is;

\bea
A_{(0)}\Lb\De,\de y_H|\mbox{\fig{f1p}}\Rb\!&&\!=\f{\as^2}{4}\!\!\sum^\infty_{n=-\infty}\!\int^\infty_{-\infty}\! d\nu\ h\Lb n, \nu\Rb \la\Lb n,\nu\Rb
\exp\left\{\omega\Lb n,\nu\Rb \Lb \De-\de y_H\Rb\right\}
E_\ga E^{\,\prime}_{\,-\,\tilde{\ga}^{\,\prime}} A_H\label{1}\hspace{0.8cm}\eea

where $\De\,=\,y\,-\,y^{\,\prime}$ is the rapidity gap between the scattering dipoles shown in \fig{f1p}, and $\de y_H=\ln\Lb M_H^2/4s_0\Rb$ is the window of rapidity occupied by the heavy Higgs boson ($M_H$ is the mass of the Higgs 
boson and $s_0\,=\, 1\,\mbox{GeV}^2$). The notation $A_{(0)}$ represents the fact that
 there are $N\,=\,0$ generations of Pomeron branching,
since later on the notation $A_{(N)}$ is used to label the diagram with $N$ generations of Pomeron branching.  The numerical
factor of $\as^2/4$ in front of \eq{1} takes into account the couplings of the two reggeized gluons in the BFKL Pomeron structure in \fig{f1p}, to the color dipoles. Since there are two gluons which each
couple twice at both ends to the QCD color dipole, this leads to a factor which is proportional to $\as^2$. The factor of $1/4$ accounts for the 4 degenerate diagrams in \fig{f1p}, which arise due to the different ways of coupling the
reggeized gluons to the color dipoles. Hence one divides by a symmetry factor which ensures that identical diagrams are only counted once.  A more detailed explanation of this coefficient
 can be found in refs. \cite{Navelet:2002zz,Navelet:1998yv,Kozlov:2004sh}.  $A_H$ in \eq{1} represents the amplitude for the subprocess where Pomeron + Pomeron $\to$ Higgs,
which is given by the expression \cite{Higgs,Rizz,daws,22,2,Miller:2007pc}

\bea
A_H\,&&=\,\f{1}{3}\,\f{2^{1/4}\,G_F^{1/2}\,\as\Lb M_H\Rb}{\pi}\,\Lb\,N_c^2\,-\,1\,\Rb\hspace{1cm}\,\Lb\,G_F\,=\,1.166\times\,10^{-5}\,\hspace{0.35cm}\mbox{GeV}^{-2}\hspace{0.2cm}\Rb\nn\\
&&=\,6.89
\,\times\,10^{-4}\,\,\,\mbox{GeV}^{-2}\label{AH}\eea

where $G_F$ is the Fermi constant. Here a typical value for
$\alpha{}_{s}$, at the scale of $M_Z$, the mass of the $Z$ boson, is
used. It is expected that the Higgs will be produced with a mass of
approximately $100\,\GV$, which would give a value for the strong
coupling constant $\alpha{}_{s}\sim{}0.12{}$. This corresponds to a
$Z$ particle mass \cite{25}, of $M_{Z}=90.8\pm{}0.6
$ $\GV$.
The leading order contribution to this process is where one of the two reggeized gluons in the BFKL Pomeron states produces
 the Higgs boson through the quark triangle shown in \fig{f1p}, for which the top flavour dominates. This is on account of the quark + quark $\to $ Higgs vertex in \fig{f1p} which is proportional to the mass of the quark flavour,
so the top flavour which is the heaviest ($m_{\mbox{\footnotesize{top}}}\,=\,175\,\GV$)
contributes the most.

The BFKL eigenfunction specified in \eq{BFKL} decreases sharply as $n$ decreases. In fact,
the BFKL eigenfunction remains positive at high energy,
only  when $n=0$. For this reason, solutions with $n\neq 0$ are negligible, and from here onwards will be ignored.
Note that from the definition of \eq{BFKL} in the case when $n\,=\,0$;

 \bea\omega\Lb\,\nu\,\Rb\,=\,\bas\,\left\{\,2\psi\Lb\,1\Rb-\psi\Lb\,\h+i\nu\,\Rb-\psi\Lb\,\h-i\nu\,\Rb\,\right\}\label{BFKLnu(1)}\eea

which has a saddle point at $\nu\,=\,0$. Hence for the BFKL eigenfunction one can use the expansion around the point $\nu\,=\,0$

\bea
\omega\Lb\nu\Rb\,=\,\omega\Lb\,0\Rb\,-\h\,\nu^2\,\omega^{\,\prime\,\prime}\Lb 0\Rb\,+\mathcal{O}\Lb\nu^3\Rb\label{BFKLexpansion(1)}\eea

so that \eq{1} simplifies to

\bea
A_{(0)}\Lb\,\De,\de y_H|\mbox{\fig{f1p}}\Rb\,&&=\,\f{\as^2}{2\pi^4}\int^\infty_{-\infty}d\nu \f{\nu^2}{\Lb1+4\nu^2\Rb^2}\nn\\&&\times\,
\exp\left\{\omega\Lb 0\Rb\,\Lb \De-\de y_H\Rb-\,\h\nu^2\,\omega^{\,\prime\,\prime}\Lb\,0\Rb\,\Lb \De-\de y_H\Rb+i\nu\ln\Lb E\,E^{\,\prime}\Rb\right\}
\,A_H\hspace{1cm}\label{2}\\
\nn\\
\mbox{where}\hspace{0.4cm}E&&=\Lb\f{r_{12}}{r_{10}r_{20}}\Rb\Lb\f{r_{12}}{r_{10}r_{20}}\Rb^\ast; \hspace{0.5cm}
E^{\,\prime}=\Lb\f{r^{\,\prime}_{12}}{r^{\,\prime}_{10}r^{\,\prime}_{20}}\Rb\Lb\f{r^{\,\prime}_{12}}{r^{\,\prime}_{10}r^{\,\prime}_{20}}\Rb^\ast\label{EE}\eea

The integration over $\nu$ in \eq{2} can be solved by using the method of steepest descents. In this technique the exponential is expanded around
the saddle point $\nu_{sp}$ (in this case $\nu_{sp}\,=\,\ln\Lb E \,E^{\,\prime}\Rb/\omega^{\,\prime\,\prime}\Lb 0\Rb\,\Lb \De-\de y_H\Rb$)
and the remaining part of the integrand is fixed by setting the integration variable $\nu\,=\,\nu_{sp}$. 
In this way the integration reduces to a Gaussian type integration over $\nu$, and using the result that
$\int^\infty_{-\infty}dx \,x^2\, e^{-ax^2}\,=\,\pi^{1/2}/2\,\times\,a^{-3/2}$  ($a\,\neq\,0$), \eq{2} yields the result;

\bea
A_{(0)}\Lb\,\De,\de y_H|\mbox{\fig{f1p}}\Rb\,&&=\,\f{\as^2\pi^{1/2}}{4\pi^4}\Lb\f{2}{\omega^{\,\prime\,\prime}\Lb 0\Rb\Lb\De-\de y_H\Rb}\Rb^{3/2}
\exp\left\{\omega\Lb 0\Rb\,\Lb \De-\de y_H\Rb-\f{\ln^2\Lb E\, E^{\,\prime\,}\Rb}{2\omega^{\,\prime\,\prime}\Lb 0\Rb\,\Lb\De-\de y_H\Rb}\right\}\,\nn\\
&&\times\,\Lb 1+\mathcal{O}\Lb\f{\ln\Lb E\,E^{\,\prime\,}\Rb}{\omega^{\,\prime\,\prime}\Lb 0\Rb\,\Lb\De-\de y_H\Rb}\Rb\Rb
\,A_H\,\label{3}\eea

In the case where the $\nu$ saddle point $\nu_{sp}=\ln\Lb E\,E^{\,\prime}\Rb/
 (\omega^{\,\prime\,\prime}  (0) \Lb \De -\de y_H\Rb)$ is small, then one can recast \eq{3}
in the following asymptotic form, namely

\bea
A_{(0)}\Lb\,\De,\de y_H|\mbox{\fig{f1p}}\Rb\,&&=\,\f{\Lb2\pi\Rb^{1/2}\bas^2\,A_H}{2\pi^2N_c^2}\f{e^{\omega\Lb 0\Rb\,\Lb \De-\de y_H\Rb}}{\Lb\omega^{\,\prime\,\prime}\Lb 0\Rb\Lb\De-\de y_H\Rb\Rb^{3/2}}
\nn\\
\nn\\
&&=\,1.57
\,\times\,10^{-8}\,\,\mbox{GeV}^{-2}\hspace{1cm}(\as\,=\,0.12)\nn\\
&&\hspace{0.5cm}2.17
\,\times\,10^{-7}\,\,\mbox{GeV}^{-2}\hspace{1cm}(\as\,=\,0.2)\label{3(1)}\eea

\FIGURE[h]{\begin{minipage}{90mm}
\centerline{\epsfig{file=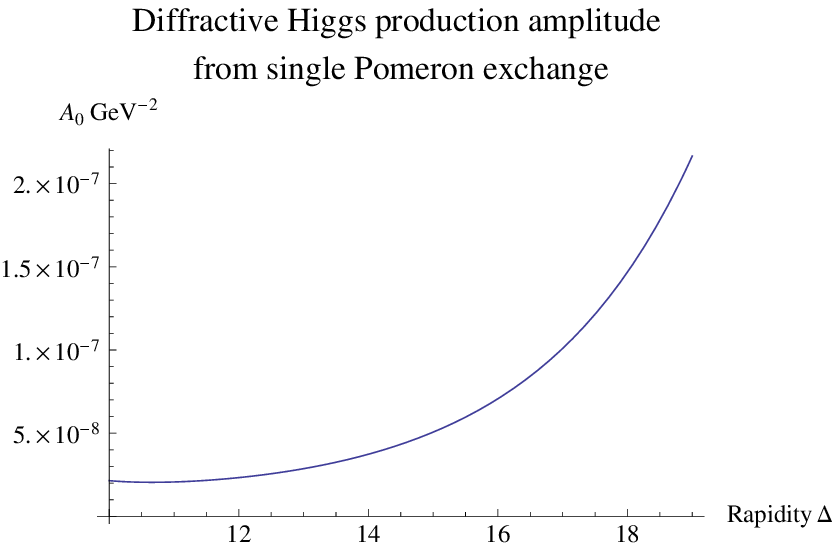,width=90mm}}
\end{minipage}
\caption{Plot of $A_{(0)}\Lb\,\De,\de y_H|\mbox{\fig{f1p}}\Rb$ derived in \eq{3(1)} against the rapidity gap
$\De$ between the scattering dipoles. The values for the rapidity gap $\De$ go from $\de y_H=\ln\Lb M_H^2/4 s_0\Rb$ for the Higgs mass $M_H=100$ GeV and $s_0=1\,\mbox{GeV}^2$, up to the typical LHC rapidity 
$\De=19$.  } \label{fazgraph} } 

The following values were used;

\bea
\omega\Lb 0\Rb\,&&=\,4\bas\,\ln 2;\hspace{0.7cm} \omega^{\,\prime\,\prime}\Lb 0\Rb\,=\,28\,\bas\,\zeta\Lb 3\Rb;\nn\\
\De &&=19;\hspace{2.1cm}\de y_H\,=\,\ln\Lb\f{ M_H^2}{4 s_0}\Rb\nn\\
&&\hspace{4cm}=7.824 \label{Rossvalues}\eea

where $\zeta\Lb n\Rb\,$ 
is the Riemann-zeta function, and where the rapidity gap $\De$ between the scattering dipoles is based on the energy $\sqrt{s}\,=\,14\,$ TeV, which is typical for 
proton collisions at the LHC. The values given in \eq{Rossvalues} will be assumed throughout this paper. The difference in the values of \eq{3(1)}  shows that the amplitude depends
critically on the choice of the strong coupling. The plot of the single Pomeron amplitude derived in \eq{3(1)} against rapidity is shown in \fig{fazgraph}, for $\as = 0.2$. The plot shows that
the amplitude of \fig{f1p} is very sensitive to increases in the rapidity gap $\De$. 

\vspace{2cm}

\section{The simple Pomeron loop}
\label{s2}

\FIGURE[h]{ \centerline {\epsfig{file=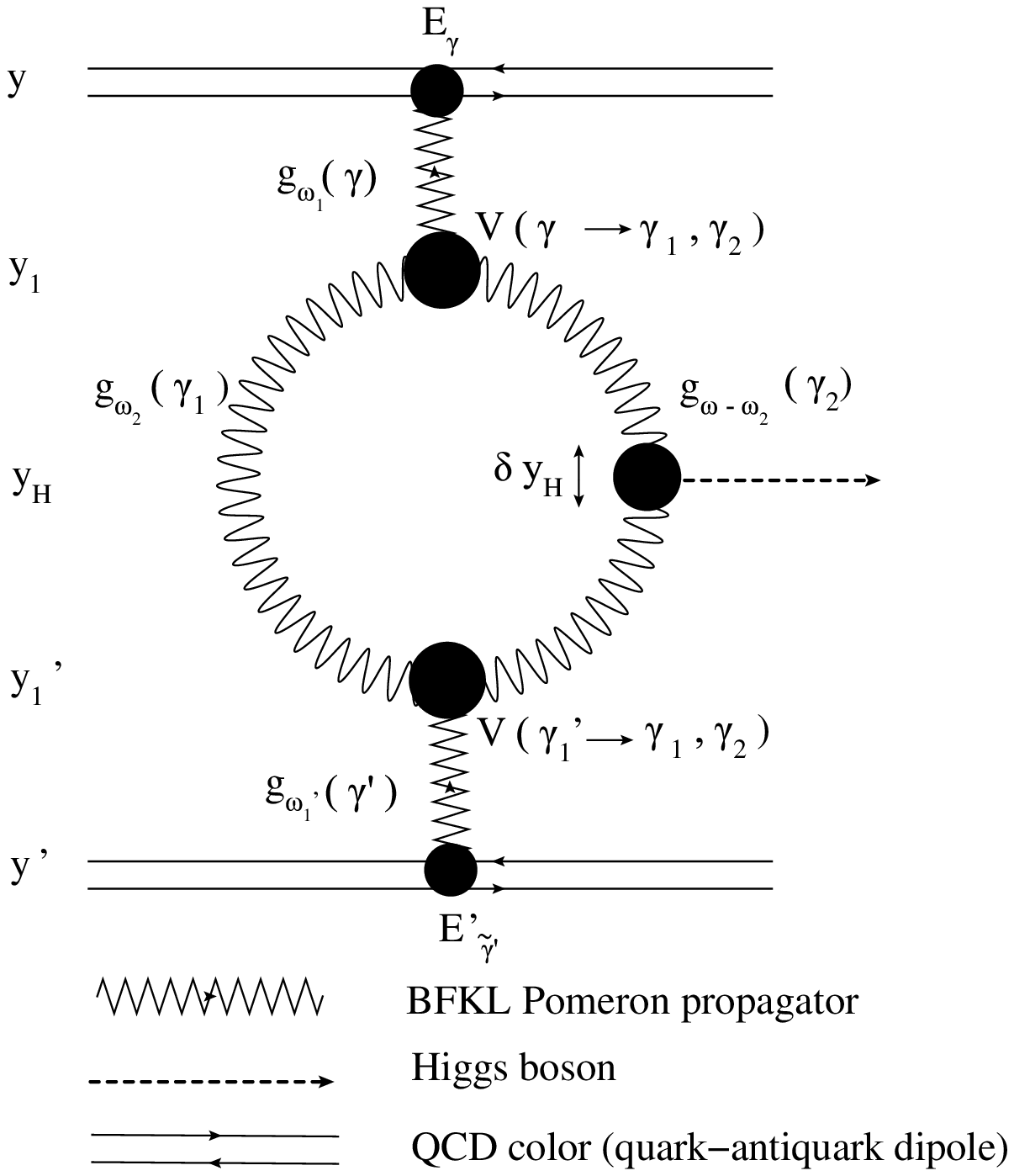,width=100mm}}\caption{ The Pomeron loop diagram.} \label{fPomeronloop2} }

In this section, the Feynman amplitude for the diffractive production of the Higgs boson, from the Pomeron loop diagram is derived.
The diagram shown in \fig{fPomeronloop2} is the scattering of a color dipole at
rapidity $y$ off a color dipole at rapidity $y^{\,\prime}$, with the exchange of a BFKL Pomeron state,
where the correction of one loop is included. Its amplitude is denoted $A_{(1)}\Lb\omega,\omega^{\,\prime},\omega_1\Rb$,
since there is one generation of Pomeron splitting,  and likewise one generation of recombining of two
Pomerons into one, to form the loop.

\bea
A_{(1)}\Lb \omega,\omega^{\,\prime},\omega_1|\mbox{\fig{fPomeronloop2}}\Rb\,&&=\f{\as^2}{4}\!
\int \mathcal{D}\ga \int\mathcal{D}\ga^{\,\prime}E_{\ga}\,E^{\,\prime}_{\,\,-\tilde{\ga}^{\,\prime}} g_{\omega}\Lb\ga\Rb g_{\omega^{\,\prime}}\Lb\ga^{\,\prime}\Rb
D_{(1)}\Lb \ga ,\ga^{\,\prime} | \omega , \omega_1\Rb A_H\hspace{1cm}
\label{1pom}
\eea

The contribution $D_{(1)}\Lb\ga\,,\,\ga^{\,\prime\,}|\omega,\omega_1\Rb$ of
the Pomeron loop to the amplitude of \fig{fPomeronloop2}, is written in $\omega$ representation as a function of the two conformal weights $\ga$ and $\ga^{\,\prime\,}$ of the Pomeron propagators attached to the loop, namely;

\bea
D_{(1)}\Lb\ga\,,\,\ga^{\,\prime\,}|\omega,\omega_1\Rb\,&&=\,\int d^2R \int d^2 R^{\,\prime\,}\,d_{(1)}\Lb\ga\,,\,\ga^{\,\prime\,}|\omega,\omega_1\Rb\label{D(b)}\eea

where $R$ and $R^{\,\prime\,}$ are the center of mass coordinates of the two Pomeron propagators on either side of the loop, and

\bea d_{(1)}\Lb\ga,\ga^{\,\prime\,}|\omega,\omega_1\Rb\!&&=\!\f{1}{S_{(1)}}\int\! d^2R_1\!\! \int\! d^2R_2\!\!\int\! 
\mathcal{D}\ga_1\!\!\int\! \mathcal{D}\ga_2  g_{\omega_1}\Lb \ga_1\Rb\,g_{\omega-\omega_1}\Lb\ga_2\Rb\nn\\
&&\times V\Lb R, R_1, R_2| \ga\to\ga_1,\ga_2\Rb V\Lb R^{\,\prime},R_1,R_2| \ga_1,\ga_2\to\ga^{\,\prime}\Rb\hspace{1.1cm}\label{D}\eea

$S_{(1)}$ in the denominator is a symmetry factor which prevents counting identical diagrams more than once, and its value is determined below. 
  $V\Lb R, R_1, R_2| \ga\to \ga_1,\ga_2 \Rb$
is the vertex for the splitting
of the BFKL Pomeron state labeled $\left\{R,\ga\right\}$,
into the two BFKL Pomeron states labeled by $\left\{ R_1,\ga_1\right\}$ and $\left\{ R_2,\ga_2\right\}$, and it is known as the triple
Pomeron vertex, shown in \fig{ftpv},  ( $\ga\,=\,$ conformal variable and $R\,=\,$ center of mass coordinate).

\FIGURE[h]{\begin{minipage}{100mm}
\centerline{\epsfig{file=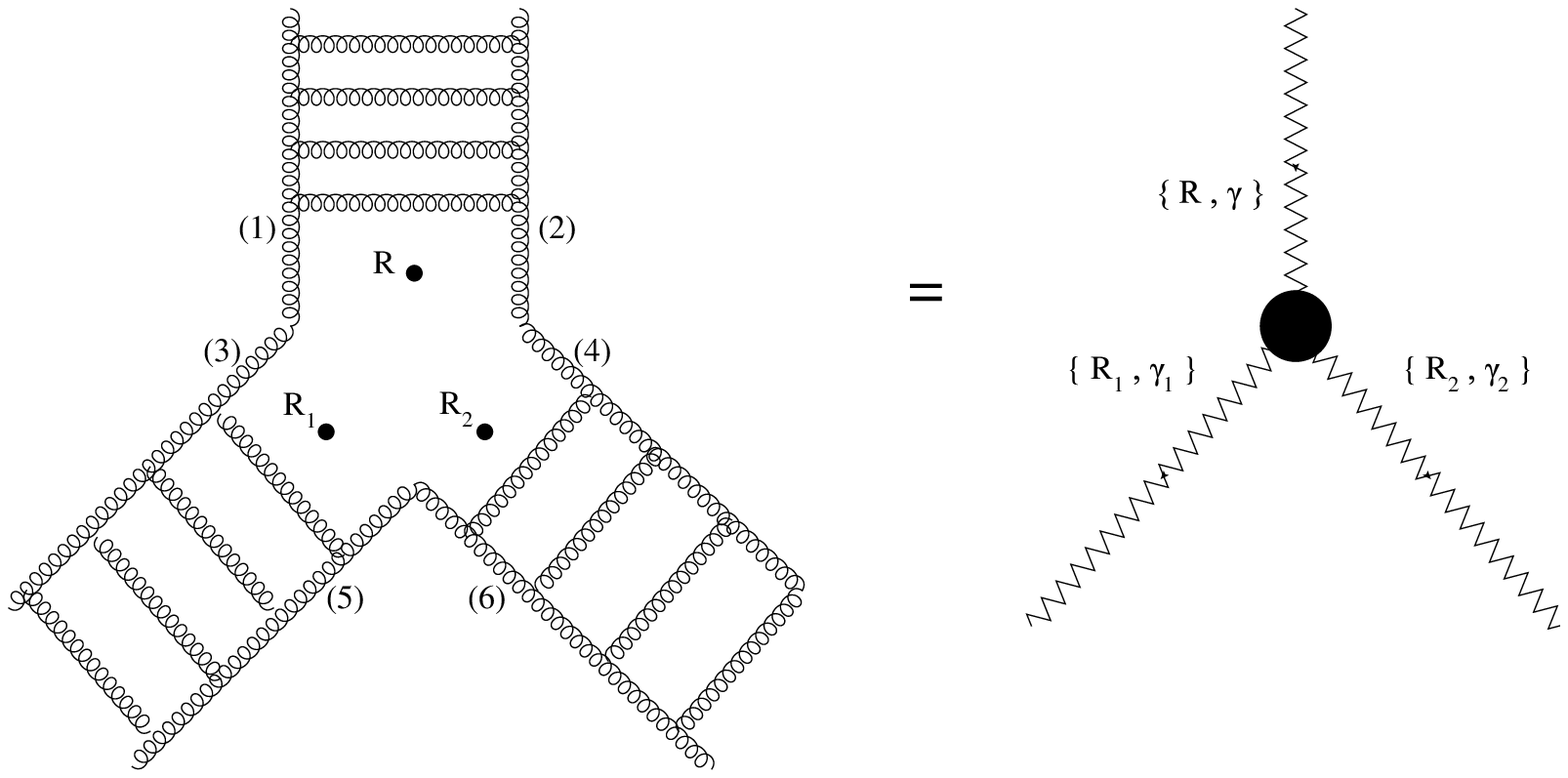,width=100mm}}
\end{minipage}
\caption{The triple Pomeron vertex. } \label{ftpv} } 

The total order of the symmetry group of \fig{fPomeronloop2} is $S_{(1)}=16$. This can be seen by observing that for both vertices of the loop,
there are $2$ permutations of the Pomeron lines which form the $2$ branches of the loop. In addition for each pair of Pomeron lines there are $2$ permutations
formed by swapping the upper with the lower vertex. This makes a total symmetry factor of $4$. The remaining $4$-fold degeneracy stems from 
the triple Pomeron vertex shown in \fig{ftpv}, whereby swapping the reggeized gluon lines $(5)$ and $(6)$ leads to an identical diagram. However swapping
lines $(3)$ and $(4)$ leads to two different diagrams called the planar and  non -planar diagrams shown in \fig{fplanar} and \fig{fnonplanar}. These two diagrams are non identical, and give different
contributions, where the latter is 
suppressed by $1/N_c^2$. \\

Korchemsky pointed out in ref. \cite{Korchemsky:1997fy}, that by the condition of conformal invariance, the triple Pomeron vertex can be factorized
into two pieces which depend only on the center of mass coordinates, and the conformal variables respectively, in the following way;

\vspace{1.5cm}

\bea
V\Lb R,R_1,R_2|\ga\to\ga_1,\ga_2\Rb&&=\,R_{01}^{-\De_{01}}\,R_{12}^{-\De_{12}}\,R_{20}^{-\De_{20}}\,R_{01}^{\ast\,-\,\tilde{\De}_{01}}\,R_{12}^{\ast\,-\,\tilde{\De}_{12}}\,R_{20}^{\ast\,-\tilde{\De}_{20}}\, \Ga\Lb \ga\,|\,\ga_1\,,\,\ga_2\Rb
\label{cinvariance0}\eea

where $R_{0i}=R-R_i\,\,(i=1,2)$ and $R_{12}=R_1-R_2$, and where for example $\De_{01}\,=\,\ga+\ga_1-\ga_2$, and $\De_{12}\,=\,\ga_1+\ga_2-\ga$, with an equivalent definition
 for the $\tilde{\De}$ in terms of the $\tilde{\ga}$. It was found in refs. \cite{Korchemsky:1997fy,Navelet:1997fy}
that;

\bea
\Ga\Lb\,\ga\,|\,\ga_1\,,\,\ga_2\,\Rb\,&&=\,\Lb\f{\as N_c}{\pi}\Rb^2\,16\,\ga\,\Lb 1-\ga\Rb\,\tilde{\ga}\,\Lb 1-\tilde{\ga}\Rb\,\nn\\
&&\times\left\{\Omega\Lb\ga\,|\,\ga_1\,,\,\ga_2\,\Rb\,+\,\f{2\pi}{N_c^2}\,\La\Lb\ga\,|\,\ga_1\,,\,\ga_2\,\Rb\,\Lb\chi\Lb\ga_1\Rb+\chi\Lb\ga_2\Rb
-\chi\Lb\ga\Rb\Rb\right\}\hspace{1cm}\label{v}\\
&&\hspace{0.5cm}\underbrace{\hspace{2.3cm}}\hspace{0.7cm}\underbrace{\hspace{6.8cm}}\nn\\
&&\hspace{0.3cm}\mbox{\footnotesize{planar piece \fig{fplanar}}}\hspace{2.3cm}\mbox{\footnotesize{non-planar piece \fig{fnonplanar}}}\nn\\
\nn\\
\mbox{where}\hspace{0.5cm}\chi\Lb\ga\Rb&&=\Re e \left\{\psi\Lb 1\Rb-\psi\Lb\ga\Rb\right\}\label{definitionoffunctionchi}\eea

where $\psi\Lb\ga\Rb$ is the di-gamma function. The triple Pomeron vertex comes in two types, namely
the planar and the non planar diagram shown in \fig{fplanar} and \fig{fnonplanar} respectively.
The first term in curly brackets is the  contribution of the planar diagram, 
 and the second term is the contribution of the non planar diagram.
 The explicit expressions for $\Omega\Lb\ga\,|\,\ga_1\,,\,\ga_2\,\Rb$ and $\La\Lb\ga\,|\,\ga_1\,,\,\ga_2\,\Rb$ and their derivation, 
can be found in the appendix. 

\DOUBLEFIGURE[h]{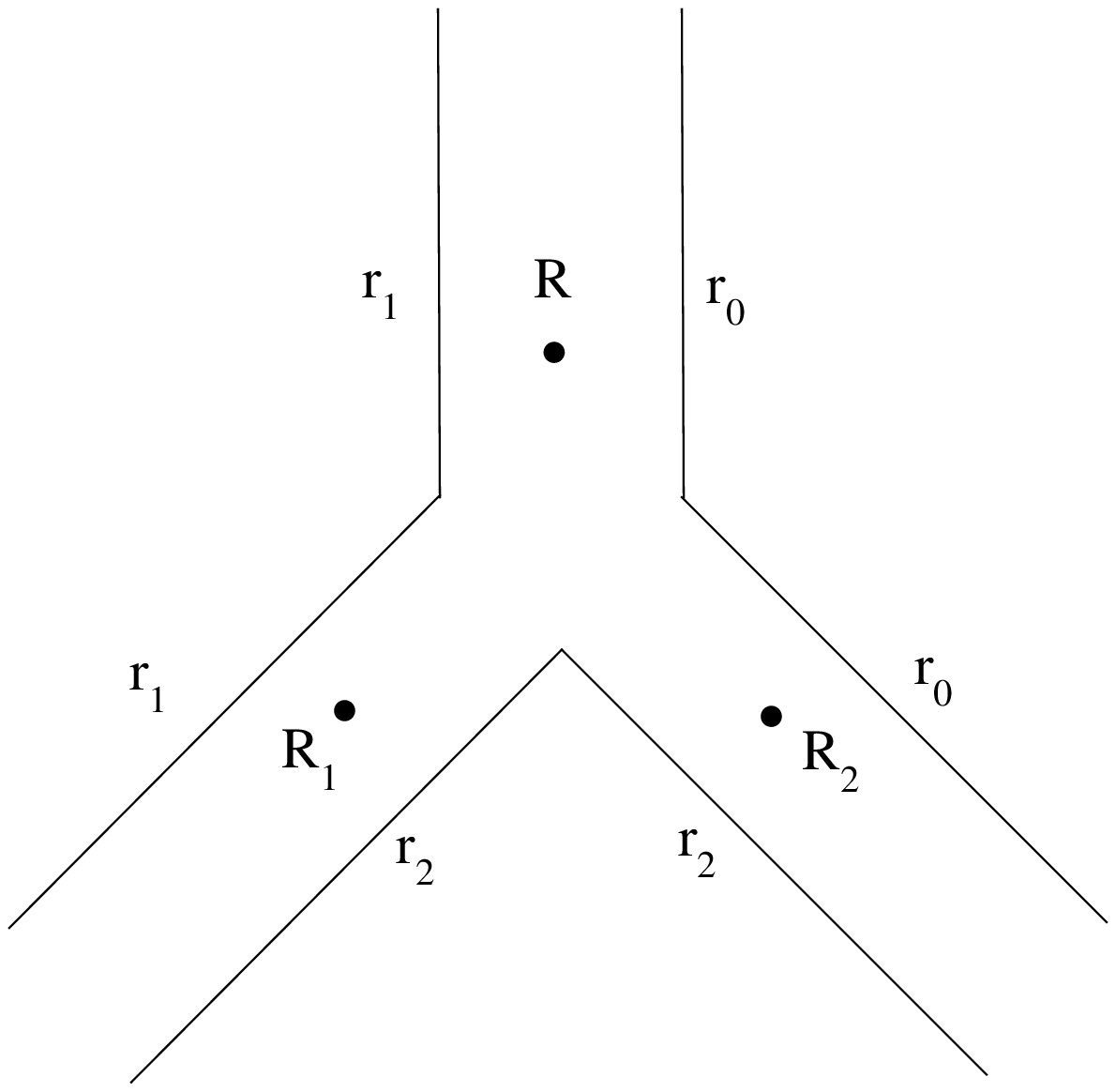,width=55mm,height=40mm}{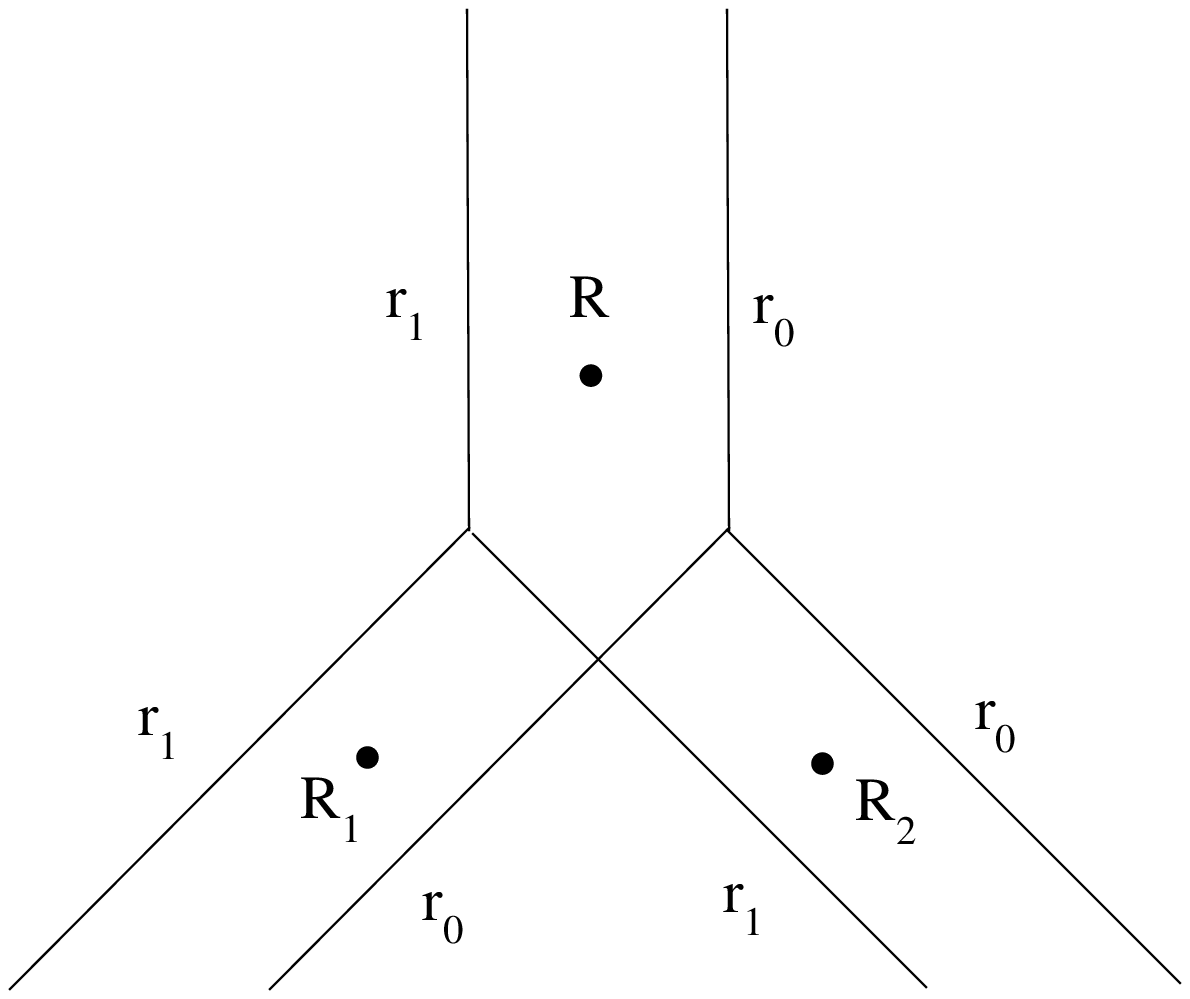,width=55mm,height=40mm}
{Planar diagram \cite{Korchemsky:1997fy}.
\label{fplanar}}{Non planar diagram \cite{Korchemsky:1997fy}.\label{fnonplanar} }

By inserting \eq{cinvariance0} into \eq{D} and evaluating all the integrals over the center of mass coordinates, it was shown in
ref. \cite{Braun:2009fy} that this leads to the following condition;

\bea
d_{(1)}\Lb\ga\,,\,\ga^{\,\prime\,}|\omega,\omega_1\Rb\,&&=\,d_{(1)}\Lb\ga\,|\omega,\omega_1\Rb\,\f{1}{h\Lb \ga\,\Rb}\,\de_{n,n^{\,\prime}}\,\de\Lb\nu-\nu^{\,\prime}\,\Rb\,\de^{(2)}\Lb R-R^{\,\prime\,}\Rb\label{Dclean}\\
\nn\\
\mbox{where}\hspace{1cm}d_{(1)}\Lb\ga\,|\omega,\omega_1\Rb\,&&=\f{1}{16}\int \!\mathcal{D}\ga_1\!\int\! \mathcal{D}\ga_2\,g_{\omega_1}\Lb \ga_1\Rb\,g_{\omega-\omega_1}\Lb\ga_2\Rb\Ga\Lb \ga | \ga_1,\ga_2\Rb \Ga\Lb \bar{\ga} | \bar{\ga}_1 , \bar{\ga}_2\Rb
 \label{D4}\eea

Now inserting the loop amplitude of \eq{D4} into the scattering amplitude of \eq{1pom},
the integration over $\ga^{\,\prime}$ can be eliminated by virtue of the Dirac delta function $\de\Lb\ga-\ga^{\,\prime}\Rb$. Hence after
introducing the explicit expressions for the  integration measure and the conformal propagator given by \eq{dnu} and \eq{conformalpropagator} respectively;

 \bea
A_{(1)}\Lb \omega,\omega^{\,\prime},\omega_1|\mbox{\fig{fPomeronloop2}}\Rb\,&&=\,\f{\as^2}{4}\int \mathcal{D}\ga \,\f{\la^2\Lb\ga\Rb}{\left\{\omega-\omega\Lb\ga\Rb\right\}\,\left\{\omega^{\,\prime}-\omega\Lb\ga\Rb\,\right\}}\,\,E_{\ga}\,E^{\,\prime}_{\,-\,\tilde{\ga}}\,A_H\nn\\
&&\times\,\f{1}{16}\int \!\mathcal{D}\ga_1\!\int \!\mathcal{D}\ga_2\,\f{\la\Lb\ga_1\Rb}{\left\{\omega-\omega\Lb\ga_1\Rb\right\}}\f{\la\Lb\ga_2\Rb}{\left\{\omega-\omega_1-\omega\Lb\ga_2\Rb\,\right\}}
\Ga\Lb \ga|\ga_1,\ga_2\Rb \Ga\Lb \bar{\ga}|\bar{\ga}_1,\bar{\ga}_2\Rb\hspace{1cm}
\label{p2}\eea

The amplitude of the Pomeron loop diagram can be re-expressed as a function of the rapidity instead of the angular momentum $j\,=\,1+\omega$,
using the Mellin transform. The rapidity gap filled up by the loop is $\De_1=y_1-y_1^{\,\prime}$, where $y_1$ and $y_1^{\,\prime}$ are
respectively the rapidity values of the upper and lower vertices of the loop (see \fig{fPomeronloop2}).
In the case of inelastic scattering which results in the production of the heavy Higgs boson from one branch of the loop, there will
be a window of rapidity $\de y_H$ which the
Higgs boson occupies. Therefore $\de y_H$ should be the minimum rapidity gap which the loop occupies, such that the energy of the scattering
is enough to produce the Higgs boson. This affects the upper and lower limits of the rapidity variables $y_1$ and $y_1^{\,\prime}$,
and hence the choice of integration limits are as they appear below in \eq{p3};

 \bea
A_{(1)}\Lb \De,\de y_H|\mbox{\fig{fPomeronloop2}}\Rb&&\!\!=\!\!\int^y_{y^{\,\prime}+\de y_H}\!\!\!\!\!\!\!dy_1\int^{y_1-\de y_H}_{y^{\,\prime}}\!\!\!\!\!\!\!dy_1^{\,\prime}\nn\\
&&\times
\int\!\!\f{d\omega}{2\pi i}e^{\omega\Lb y-y_1\Rb}\!\!\int\!\!\f{d\omega^{\,\prime}}{2\pi i}e^{\omega^{\,\prime}\Lb y_1^{\,\prime}-y^{\,\prime}\Rb}
\!\!\int\!\!\f{d\omega_1}{2\pi i}e^{\omega_1\Lb y_1-y_1^{\,\prime}\Rb}
A_{(1)}\Lb \omega,\omega^{\,\prime}\!\!,\omega_1|\mbox{\fig{fPomeronloop2}}\Rb\nn\\
\nn\\
&&\!\!=\!\!\f{\as^2}{4}\int \mathcal{D}\ga \,\la^2\Lb\ga\Rb\,e^{\,\omega\Lb\ga\Rb\,\De\,}d_{(1)}\Lb\ga\,|\,\De\,,\,\de y_H\Rb\,
E_{\ga}\,E^{\,\prime}_{\,-\,\tilde{\ga}}\,A_H\label{p3}\eea

where $\De\,=\,y\,-\,y^{\,\prime}$ is the rapidity gap between
the scattering dipoles. The loop amplitude in rapidity representation is given by the expression;

\bea
d_{(1)}\Lb \ga |\De , \de y_H\Rb \!\!&&=\f{1}{16}\int \mathcal{D}\ga_1\int \mathcal{D}\ga_2\,\la\Lb\ga_1\Rb\,\la\Lb\ga_2\Rb\,\Ga\Lb \ga\,|\,\ga_1\,,\,\ga_2\Rb\,
\Ga\Lb \bar{\ga}\,|\,\bar{\ga}_1\,,\,\bar{\ga}_2\Rb\nn\\
&&\!\times\int^y_{y^{\,\prime}+\de y_H}\!\!\!\!\!\!\!dy_1\int^{y_1-\de y_H}_{y^{\,\prime}}\!\!\!\!\!\!\!\!\!dy_1^{\,\prime}\,\,\exp\left\{\Lb \omega\Lb\ga_1\Rb+\omega\Lb\ga_2\Rb-\omega\Lb\ga\Rb\Rb\Lb y_1-y_1^{\,\prime}\Rb-\omega\Lb\ga_1\Rb\de y_H\right\}
\hspace{1cm}
\label{d1}\eea

Assuming that the leading contribution at high energy stems from the region where the conformal spins $n\,=\,n_1\,=\,n_2\,=\,0$,
and introducing the explicit expressions given in \eq{dnu}, \eq{h} and \eq{lmbda} for the integration measure and the conformal propagator,
the scattering amplitude of \eq{p3} reduces to

 \bea
A_{(1)}\Lb \De,\de y_H|\mbox{\fig{fPomeronloop2}}\Rb
\!&&\!=\!\f{\as^2}{4}\!\int^\infty_{-\infty}\!\!d\nu  h\Lb\nu\Rb \la^2\Lb\nu\Rb e^{\omega\Lb\nu\Rb \De}d_{(1)}\Lb\nu | \De ,\de y_H\Rb
E_{\nu} E^{\,\prime}_{-\nu} A_H\hspace{1cm}\label{p3a}\eea

where (when the conformal spins $n_1\,=\,n_2\,=\,0$) the loop amplitude simplifies to;

\bea
d_{(1)}\Lb\nu\,|\,\De,\de y_H\Rb\,&&=\,\f{1}{2^{10}\pi^8}\int^\infty_{-\infty}\!\!d\nu_1\int^\infty_{-\infty}\!\! d\nu_2\f{\nu_1^2}{\Lb \nu_1^2+1/4\Rb^2}\f{\nu_2^2}{\Lb \nu_2^2+1/4\Rb^2} 
\begin{vmatrix}\Ga\Lb \nu\,|\,\nu_1\,,\,\nu_2\Rb\,\end{vmatrix}^2\,
\nn\\
 &&\times \,\int^y_{y^{\,\prime}+\de y_H}\!\!\!\!\!\!\!\!dy_1\int^{y_1-\de y_H}_{y^{\,\prime}}\!\!\!\!\!\!\!\!dy_1^{\,\prime}
 \exp\left\{\Lb\omega\Lb \nu_1\Rb+\omega\Lb \nu_2\Rb-\omega\Lb\nu\Rb\Rb\Lb y_1-y_1^{\,\prime}\Rb\,-\omega\Lb\nu_1\Rb\de y_H\right\}\, \label{d2}\eea

A convention used in \eq{d2} and throughout this paper, is that wherever $\nu$ appears
alone without $n$, it has been assumed that $n\,=\,0$ and it has been suppressed.
For example $\Ga\Lb\ga|\ga_1,\ga_2\Rb_{n=n_1=n_2=0}$  is labeled $\Ga\Lb\nu|\nu_1,\nu_2\Rb$, and
$\omega\Lb n=0,\nu\Rb$ is labeled $\omega\Lb \nu\Rb$.
In the region where $n\!=\!n_1\!=\!n_2\!=\!0$, then
$\Ga\Lb\bar{\ga}|\bar{\ga}_1,\bar{\ga}_2\Rb_{n=n_1=n_2=0}\equiv\Ga\Lb -\nu|-\nu_1,-\nu_2\Rb$,
and hence in \eq{d2}, the notation\\
$\begin{vmatrix}\Ga\Lb \nu|\nu_1,\nu_2\Rb\end{vmatrix}^2
=\Ga\Lb \nu|\nu_1,\nu_2\Rb\Ga\Lb \nu|\nu_1,\nu_2\Rb^\ast
\equiv\Ga\Lb \nu|\nu_1,\nu_2\Rb\Ga\Lb -\nu|-\nu_1,-\nu_2\Rb$.\\

Two main regions are considered when evaluating the $\left\{\nu,\nu_1,\nu_2\right\}$ integrals of \eq{p3a} and \eq{d2}. Region I is 
$\left\{\nu,|\nu_1|,|\nu_2|\right\}=\left\{0,1/2,1/2\right\}$ and region II is $\left\{|\nu|,\nu_1,\nu_2\right\}=\left\{1/2,0,0\right\}$.
To find the contribution from region I, after integrating over the rapidity variables $y_1$ and $y_1^{\,\prime}$, the loop amplitude of \eq{d2} takes the form;

\bea
d^{\,\footnotesize{\mbox{I}}}_{(1)}\Lb\nu\,|\,\De,\de y_H\Rb\,&&=\,\f{e^{-\omega\Lb\nu\Rb\de y_H}}{2^{10}\pi^8}\int^\infty_{-\infty}\!\!d\nu_1\int^\infty_{-\infty}\!\!d\nu_2\f{\nu_1^2}{\Lb \nu_1^2+1/4\Rb^2}
\f{\nu_2^2}{\Lb \nu_2^2+1/4\Rb^2}\begin{vmatrix}\Ga\Lb \nu\,|\,\nu_1\,,\,\nu_2\Rb\,\end{vmatrix}^2\,\nn\\
&&\times \,e^{\omega\Lb\nu_2\Rb\,\de y_H}\int^\infty_0\!d\be\,
e^{\,\Lb\omega\Lb\nu\Rb-\omega\Lb \nu_1\Rb\,-\,\omega\Lb \nu_2\Rb\,\Rb\,\be}\,\nn\\
&&\times\,\Lb\De-\de y_H\,+\,\be\,-\,\be\,e^{\Lb\omega\Lb \nu_1\Rb\,+\,\omega\Lb \nu_2\Rb-\omega\Lb\nu\Rb\Rb\,\Lb\,\De -\de y_H \Rb}\Rb
\label{d4}\eea

\FIGURE[h]{\begin{minipage}{70mm}
\centerline{\epsfig{file=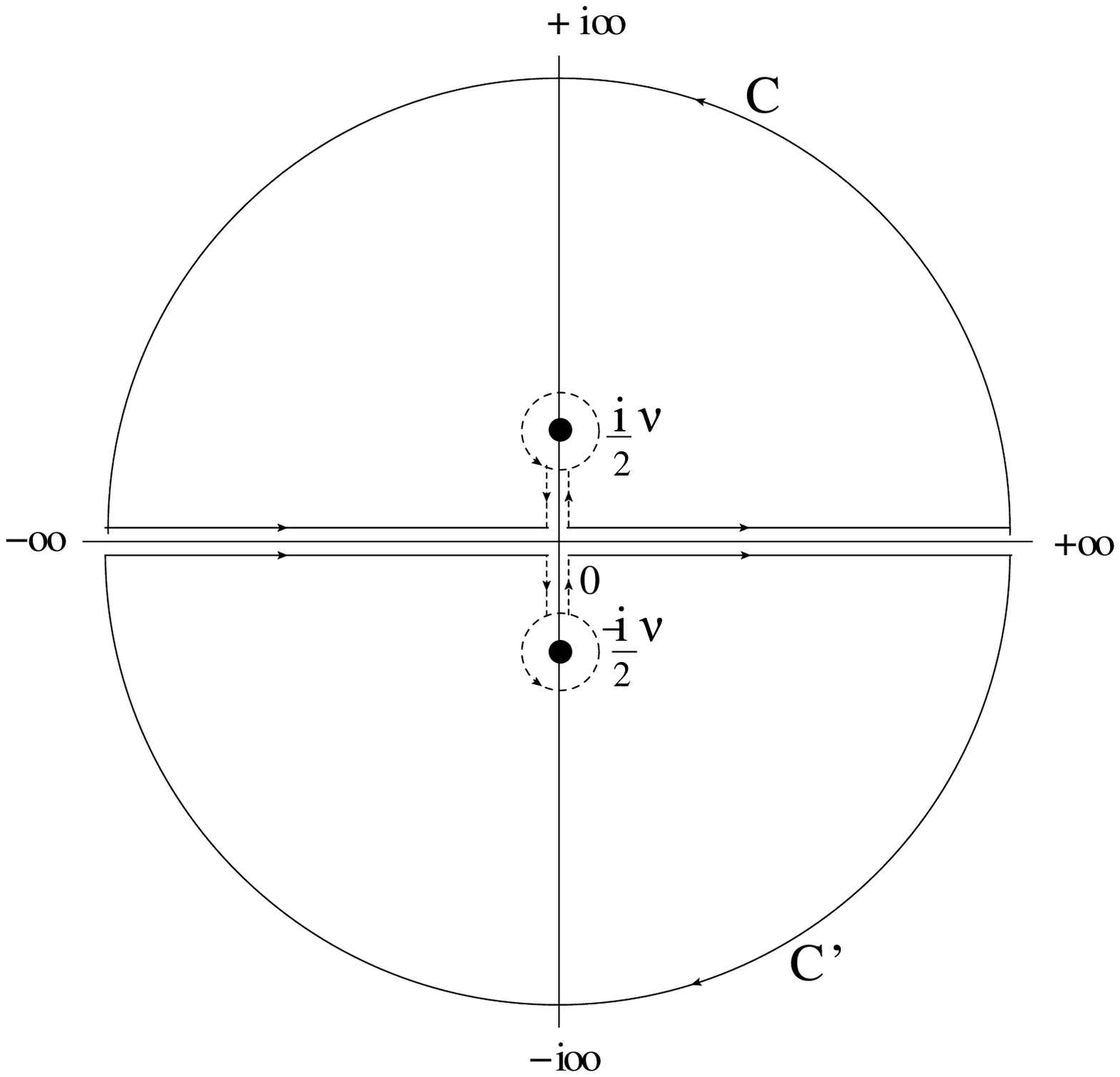,width=70mm}}
\end{minipage}
\caption{The integration contour $C$ ($C^{\,\prime}$) which is closed on the upper (lower) half plane, in order to
enclose the pole $i\nu=1/2\, (-1/2)$. } \label{fintegrationcontour} }

where in \eq{d4} the identity $1/x^n=1/\Ga\Lb n\Rb\,\times\,\int^\infty_0\!d\be\,e^{-\be x}\,\be^{n-1}$ 
 has been used. The singularities in the integrand at $\left\{|\nu_1|,|\nu_2|\right\}=1/2$ suggest closing the $\nu_1$ and $\nu_2$ integration paths 
along the contour $C$ (or $C^{\,\prime}$) shown in \fig{fintegrationcontour} which runs along the real axis from $-\infty$ to $+\infty$ and along the semi circle at infinity in the upper (lower) half plane.  $C$  ($C^{\,\prime}$)  encloses the points 
$\left\{i\nu_1,i\nu_2\right\}=1/2\,\,(-1/2)$, and the two solutions which stem from both contours are identical. Therefore the contour $C$ is chosen,
 so that the solution is $2\,\times\,\Lb 2\pi i\Rb^2\,\sum\Lb\mbox{residues at } \left\{i\nu_1,i\nu_2\right\}=1/2\Rb$, where the extra factor of 2 takes into account the identical
contribution from the residues at $\left\{i\nu_1,i\nu_2=-1/2\right\}$. The non-singular part of the integrand of \eq{d4}, namely\\ $\Lb 1/2+i\nu_1\Rb^{-2}\Lb 1/2+i\nu_1\Rb^{-2}$ tends to zero
as $1/\nu_1^2\nu_2^2$, hence the convergence on the semi-circle at infinity is satisfied, allowing $C$ to be used as the integration contour.
  This choice of contour will be used repeatedly in similar calculations
throughout this paper, therefore wherever a specified integration contour is labeled $C$, it is assumed that the path shown in \fig{fintegrationcontour} is intended, and 
that the non-singular part of the integrand has a good convergence on $C$. \\

The triple Pomeron vertex in region I which should be inserted into \eq{d4} is derived in the appendix
 in \eq{G123combined}. After taking into account all the singularities that stem from \eq{G123combined} at $\left\{i\nu_1,i\nu_2\right\}=1/2$,
and the singularities that stem from $\omega\Lb i\nu_{1,2}\to1/2\Rb=\bas/\Lb 1/2-i\nu_{1,2}\Rb$ and 
$\chi\Lb i\nu_{1,2}\to1/2\Rb=1/\Lb 1/2-i\nu_{1,2}\Rb$ (see \eq{BFKLnu(1)} and \eq{definitionoffunctionchi}), the integration over $\nu_1$ 
in \eq{d4} takes the form;

\bea
&&\oint_C\!d\nu_1\f{\nu_1^2}{\Lb 1/2+i\nu_1\Rb^2\Lb 1/2-i\nu_1\Rb^4}\exp\left\{\f{\bas}{\Lb 1/2-i\nu_1\Rb}\be\right\}\nn\\
&&\hspace{3cm}=\oint_C\!d\nu_1\,\f{\nu_1^2}{\Lb 1/2+i\nu_1\Rb^2\Lb 1/2-i\nu_1\Rb^2}
\Lb\f{1}{\bas}\f{d}{d\be}\Rb^2\exp\left\{\f{\bas}{\Lb 1/2-i\nu_1\Rb}\be\right\};\label{nu1integral0}\eea

where the derivative acts on the exponential to bring down a factor of $\Lb 1/2-i\nu_1\Rb^{-2}$.
The $\Lb 1/2-i\nu_1\Rb^{-2}$ term remaining
is canceled by the Jacobian of the coordinate transformation $u\,=\,\bas\Lb 1/2-i\nu_1\Rb^{-1}+\bas\Lb 1/2-i\nu_2\Rb^{-1}-\omega\Lb\nu_2\Rb\de y_H/\beta$, for which the Jacobian is
$|\D u/\D\nu_1|\,=\,\Lb 1/2-i\nu_1\Rb^2/\bas$. Instead integrating over $u$ yields an integration of the form $\oint_C\! du\,\exp\Lb -iu\,\be\Rb\,\equiv\,\pi i\,\de\Lb\be\Rb$ or
$\oint_C\! du\,\exp\Lb -iu\,\Lb \be - \De +\de y_H\Rb \Rb\,\equiv\,2\pi i\,\de\Lb \be -\De +\de y_H\Rb$.
After taking the residue at $i\nu_1\,=\,1/2$, the integration over $\nu_2$ has now reduced to;

\bea
\oint_C\!d\nu_2\,\f{\nu_2^2\Lb 1-i\nu_1-i\nu_2-2/N_c^2\,\Rb}{\Lb 1/2+i\nu_2\Rb^2\Lb 1/2-i\nu_2\Rb^4}
&&\xrightarrow{i\nu_1\to 1/2}\oint_C\!d\nu_2\,
f\Lb \nu_2\Rb\Lb\f{1}{\Lb 1/2-i\nu_2\Rb^3}-\f{2/N_c^2}{\Lb 1/2-i\nu_2\Rb^4}\Rb\nn\\
&&=2\pi i\Lb\,\f{\Lb -1\Rb^2}{2!}f^{(2)}\Lb i\nu_2=1/2\Rb-\f{2}{N_c^2}\f{\Lb -1\Rb^3}{3!}f^{(3)}\Lb i\nu_2=1/2\Rb\Rb\nn\\
&&=\f{2\pi\,i}{4}\Lb 1-\f{2}{N_c^2}\Rb;\hspace{0.5cm}\mbox{where}\hspace{0.5cm}f\Lb\nu_2\Rb=\f{\nu_2^2}{\Lb 1/2+i\nu_2\Rb^2}
\label{residue}\eea

Overall using the above described method for taking the integrals over $\nu_1$ and $\nu_2$, and evaluating the integration
over $\be$, the loop amplitude of \eq{d4} becomes;

\bea
d^{\,\footnotesize{\mbox{I}}}_{(1)}\Lb\nu |\De,\de y_H\Rb&&=d\Lb\h+i\nu\Rb^3\,\Lb\h-i\nu\Rb^3\chi\Lb\nu\Rb
\left\{\omega\Lb\nu\Rb+\h\omega^2\Lb\nu\Rb\Lb\De -\de y_H\Rb\right\}e^{-\omega\Lb\nu\Rb\de y_H}\hspace{1cm}\label{d7}\\
\nn\\
\mbox{where}\hspace{0.5cm}d&&=\bas\Lb 1-\f{1}{N_c^2}\Rb\Lb 1-\f{2}{N_c^2}\Rb\label{definitionofd}
\eea

Now inserting the result for the loop amplitude of \eq{d7}  into the  scattering amplitude of \eq{p3a},
the steepest descent method described in \sec{s1} can be used for the $\nu$ integration in \eq{p3a},
so that overall the contribution to the scattering amplitude which stems from the region
$\left\{\nu,|\nu_1|,|\nu_2|\right\}=\left\{0,1/2,1/2\right\}$ is;

\bea
A^{\,\footnotesize{\mbox{I}}}_{(1)}\Lb \De,\de y_H|\mbox{\fig{fPomeronloop2}}\Rb\,
&&=\f{\Lb 2\pi\Rb^{1/2}\bas^2\,d\,A_H}{128 \pi^2N_c^2}\f{e^{\omega\Lb 0\Rb\,\Lb \De -\de y_H\Rb}}{\Lb \omega^{\,\prime\,\prime}\Lb 0\Rb\Lb\De -\de y_H\Rb\Rb^{3/2}}
\chi\Lb 0\Rb\!\left\{\!\omega\Lb 0\Rb+\h\omega^2\Lb 0\Rb\Lb\De -\de y_H\Rb\!\right\}\hspace{0.7cm}\label{p5(0)}\\
\nn\\
&&=\,\,2.34
\,\times\,10^{-11}\,\,\mbox{GeV}^{-2}\hspace{1cm}(\as\,=\,0.12)\nn\\
&&\hspace{0.55cm}
1.3\,\,\,\,\times\,10^{-9}\,\,\mbox{GeV}^{-2}\hspace{1.15cm}(\as\,=\,0.2)\label{p5}\eea

The results of \eq{p5} show that the Pomeron loop in region I is small compared with the basic diagram of \fig{f1p} (see \eq{3(1)}).
It turns out that the dominant contribution to the Pomeron loop amplitude stems from region II, namely
$\left\{|\nu|,\nu_1,\nu_2\right\}\,=\,\left\{1/2,0,0\right\}$. In this region the expansion of \eq{BFKLexpansion(1)} for the BFKL eigenfunctions
$\omega\Lb\nu_1\Rb$ and $\omega\Lb\nu_2\Rb$ in \eq{d2} can be used. The triple
Pomeron vertex of \eq{tpvuseful} derived in the appendix for region II should be inserted
into \eq{d2}. The expression of \eq{tpvuseful} is novel for the following reason. From \eq{v} the triple Pomeron vertex in region II takes the following form
\cite{Korchemsky:1997fy} ($n\!=\!n_1\!=\!n_2\!=\!0$ at high energy);

\bea
&&\Ga\Lb|\nu||\nu_1,\nu_2\Rb\Lb\mbox{region II}\Rb=\Ga\Lb\h|0,0\Rb=16\Lb\f{\as N_c}{\pi}\Rb^2\nn\\
&&\times\lim_{i\nu\to 1/2}\Lb\h+i\nu\Rb^2\Lb\h-i\nu\Rb^2\left\{\Omega\Lb\h|0,0\Rb+\f{2\pi}{N_c^2}
\La\Lb\h|0,0\Rb\,\Lb\chi\Lb i\nu\to\h\Rb-2\chi\Lb0\Rb\Rb\right\}\hspace{1cm}\label{vsecond}\\
&&\hspace{5.5cm}\underbrace{\hspace{1.7cm}}\hspace{1cm}\underbrace{\hspace{4cm}}\nn\\
&&\hspace{5cm}\mbox{\footnotesize{planar piece \fig{fplanar}}}\hspace{1cm}\mbox{\footnotesize{non-planar piece \fig{fnonplanar}}}\nn\eea

$\Omega\Lb\h|0,0\Rb$ and $\La\Lb\h|0,0\Rb$ were calculated in the appendix (see \eq{j1ihf} and \eq{Js1halfnu1}) and they both have a second order pole at $i\nu=1/2$, canceled by the $\Lb 1/2-i\nu\Rb^2$ term in front
in \eq{vsecond}.
However an extra 
pole from the non planar piece arises due to the singularity which stems from $\chi\Lb i\nu\to 1/2\Rb\to\Lb 1/2-i\nu\Rb^{-1}$ (see \eq{definitionoffunctionchi}).
Therefore overall for region II the divergent part of the triple Pomeron vertex
stems from the non planar piece, and the contribution from the planar piece is non-singular.
Hence a remarkable feature of the Pomeron loop arises which has not been taken into account before, namely
that the dominant contribution to the Pomeron loop amplitude comes from the non planar piece $\La\Lb\nu|\nu_1,\nu_2\Rb$ of the triple Pomeron vertex shown in
\fig{fnonplanar}.
This property has been so far ignored because previous publications assumed that $N_c\to\infty$, and since in \eq{vsecond} the non planar
piece is suppressed by $2\pi/N_c^2$ compared to the planar piece, it was neglected.
Taking this into account and after evaluating the
$\nu_1$ and $\nu_2$ integrations in \eq{d2} using the method of steepest descents, the contribution of region II ($\left\{|\nu|,\nu_1,\nu_2\right\}=\left\{1/2,0,0\right\}$) to the loop of \fig{fPomeronloop2}
simplifies to;

\bea
d^{\,\footnotesize{\mbox{II}}}_{(1)}\Lb\nu\,|\,\De,\de y_H\Rb&&=\f{a\,e^{-\, \omega\Lb 0\Rb\,\de y_H}}{\Lb 1/2+i\nu\Rb\Lb 1/2-i\nu\Rb}
\int^{y}_{y^{\,\prime}+\de y_H}\!\!\!\!\!\!\!\!dy_1\int^{y_1-\de y_H}_{y^{\,\prime}}\!\!\!\!\!\!\!\!dy_1^{\,\prime}
\f{e^{\Lb2\omega\Lb 0\Rb\,-\omega\Lb \nu\Rb\Rb\De_1}}{\De_1^{3/2}\Lb\De_1-\de y_H\Rb^{3/2}}\hspace{1cm}\label{addition}\\
\nn\\
\mbox{where}\hspace{0.5cm}a&&=\f{2^9\bas^4}{N_c^4\pi[\omega^{\,\prime\,\prime}\Lb 0\Rb]^3};\hspace{2cm}\De_1=y_1-y_1^{\,\prime}.\label{m}\eea

Inserting \eq{addition} into \eq{p3a}, taking into account the singularities that stem from
$\omega\Lb i\nu\to 1/2\Rb\to\Lb 1/2-i\nu\Rb^{-1}$ (see \eq{BFKLnu(1)}) and closing the contour of integration
on the path $C$ shown in \fig{fintegrationcontour}, the $\nu$ integration takes the form;

\bea
&&\oint_C\!d\nu\f{\nu^2}{\Lb1/2+i\nu\Rb^5\Lb 1/2-i\nu\Rb^5}\,\exp\left\{\f{\bas\Lb\De -\De_1\Rb} {1/2-i\nu}\right\}\nn\\
&&\hspace{4cm}=\oint_C\!d\nu\f{\nu^2}{\Lb1/2+i\nu\Rb^5\Lb 1/2-i\nu\Rb^2}\Lb\f{-1}{\bas}\f{d}{d\De}\Rb^{3}
\exp\left\{\f{\bas\Lb\De -\De_1\Rb} {1/2-i\nu}\right\}\label{trick}
\eea

The remaining $\Lb 1/2-i\nu\Rb^{-2}$ term is canceled by the Jacobian of the transformation $u\,=\,\bas/\Lb 1/2-i\nu\Rb$, and integrating over $u$
yields the derivative of the Dirac delta function $2\pi i\de^3\Lb\De-\De_1\Rb/\bas^3$, which is absorbed by the
integration over the rapidity variables $\left\{y_1,y_1^{\,\prime}\right\}$ in \eq{p3a}. Overall the contribution of region II 
($\left\{|\nu|,\nu_1,\nu_2\right\}=\left\{1/2,0,0\right\}$)  to the scattering
amplitude of \fig{fPomeronloop2} is;

\bea
A^{\,\footnotesize{\mbox{II}}}_{(1)}\Lb \De,\de y_H|\mbox{\fig{fPomeronloop2}}\Rb
\!&&\!=\!\f{\bas\,a\,A_H}{2^9 N_c^2\pi} \Lb\De-\de y_H\Rb\Lb\f{-1}{\bas}\f{d}{d\De}\Rb^3
\left\{
\f{e^{2\omega\Lb 0\Rb\,\Lb\De-\de y_H/2\Rb}}{\De^{3/2}\Lb\De-\de y_H\Rb^{3/2}}\right\}\\
\nn\\
&&\!=2.05
\,\times\,10^{-10}\,\,\mbox{GeV}^{-2}\hspace{1cm}(\as\,=\,0.12)\nn\\
&&\hspace{0.35cm}2.14
\,\times\,10^{-7}\,\,\mbox{GeV}^{-2}\hspace{1.2cm}(\as\,=\,0.2)\label{p3a4}\eea

An obvious divergence in the integrand of \eq{d1} originates from the region where the conformal weights \cite{Braun:2009fy} $n_1\,=\,\pm\,1$
and $n_2\,=\,\pm\,1$ (see \eq{lmbda}). However it was proven in ref. \cite{Braun:2009fy} that BFKL Pomeron states with odd $n_1$ or $n_2$
cannot couple to BFKL states where $n$ is even, and therefore this particular divergence does not arise.\\

The complete scattering amplitude for \fig{fPomeronloop2} is given by the sum of the two contributions of \eq{p5} and \eq{p3a4}, namely;

\bea
A_{(1)}\Lb \De,\de y_H|\mbox{\fig{fPomeronloop2}}\Rb&&=
A^{\,\footnotesize{\mbox{I}}}_{(1)}\Lb \De,\de y_H|\mbox{\fig{fPomeronloop2}}\Rb
+A^{\,\footnotesize{\mbox{II}}}_{(1)}\Lb \De,\de y_H|\mbox{\fig{fPomeronloop2}}\Rb\label{thecompleteloopamplitude}\\
\nn\\
&&\!=2.28
\,\times\,10^{-10}\,\,\mbox{GeV}^{-2}\hspace{1cm}(\as\,=\,0.12)\nn\\
&&\hspace{0.35cm}2.15
\,\times\,10^{-7}\,\,\mbox{GeV}^{-2}\hspace{1.2cm}(\as\,=\,0.2)\label{thesompleteloopamplitude}\eea

Comparing the results of \eq{p5} and \eq{p3a4}, it is clear that for the simple Pomeron loop amplitude of \fig{fPomeronloop2}, the dominant contribution stems from region II. From an observation of \eq{p3a4}, 
the simple loop amplitude can be written as

\bea
A^{\,\footnotesize{\mbox{II}}}_{(1)}\Lb \De,\de y_H|\mbox{\fig{fPomeronloop2}}\Rb&&=k_{(1)}e^{2\omega\Lb 0\Rb\De}\label{noninteractingPomerons}\eea

where $k_{(1)}$ contains all the other terms included in the simple loop amplitude. From \eq{noninteractingPomerons} it is clear that the dominant contribution to the loop amplitude is equivalent
to the amplitude of 2 non interacting Pomerons. This can be seen from \fig{fPomeronloop2},  where by taking the 2 branches of the loop outside,  one observes
the exchange of 2 non interacting Pomerons. This was first noted by A.Mueller, who
commented that at high rapidities, loop diagrams reduce to the exchange of 
multiple non interacting Pomerons,
with renormalized Pomeron vertices (see ref. \cite{toy}).\\

Comparing the results of \eq{3(1)} and \eq{thesompleteloopamplitude}, the Pomeron loop of \fig{fPomeronloop2} is potentially very close to the basic diagram of \fig{f1p} 
in amplitude, and suggests that Pomeron loop diagrams potentially suppress the enhanced survival probability $\SPE$ considerably.
However the contribution of higher order multiple Pomeron
loop diagrams needs to be included, before any precise statements can be made about the enhanced survival probability. In the next section, the technique used for deriving the
amplitude for diagrams with  multiple Pomeron loops, will be explained. A good knowledge of how to calculate these type of
diagrams is a necessary tool for estimating the contribution of the full set of enhanced diagrams, to the survival probability.

\section{The main idea}
\label{smi}

\DOUBLEFIGURE[h]{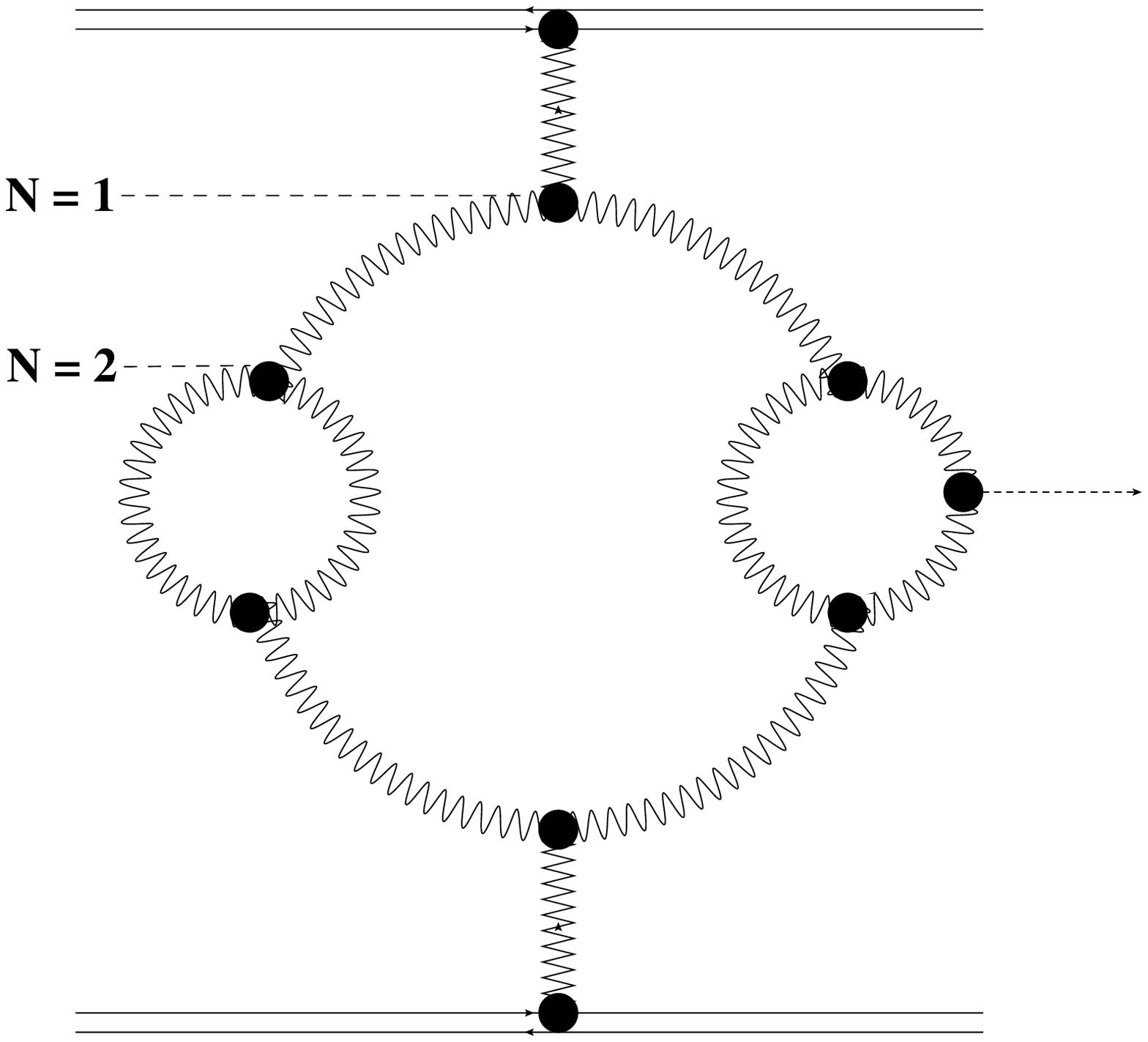,width=65mm,height=60mm}{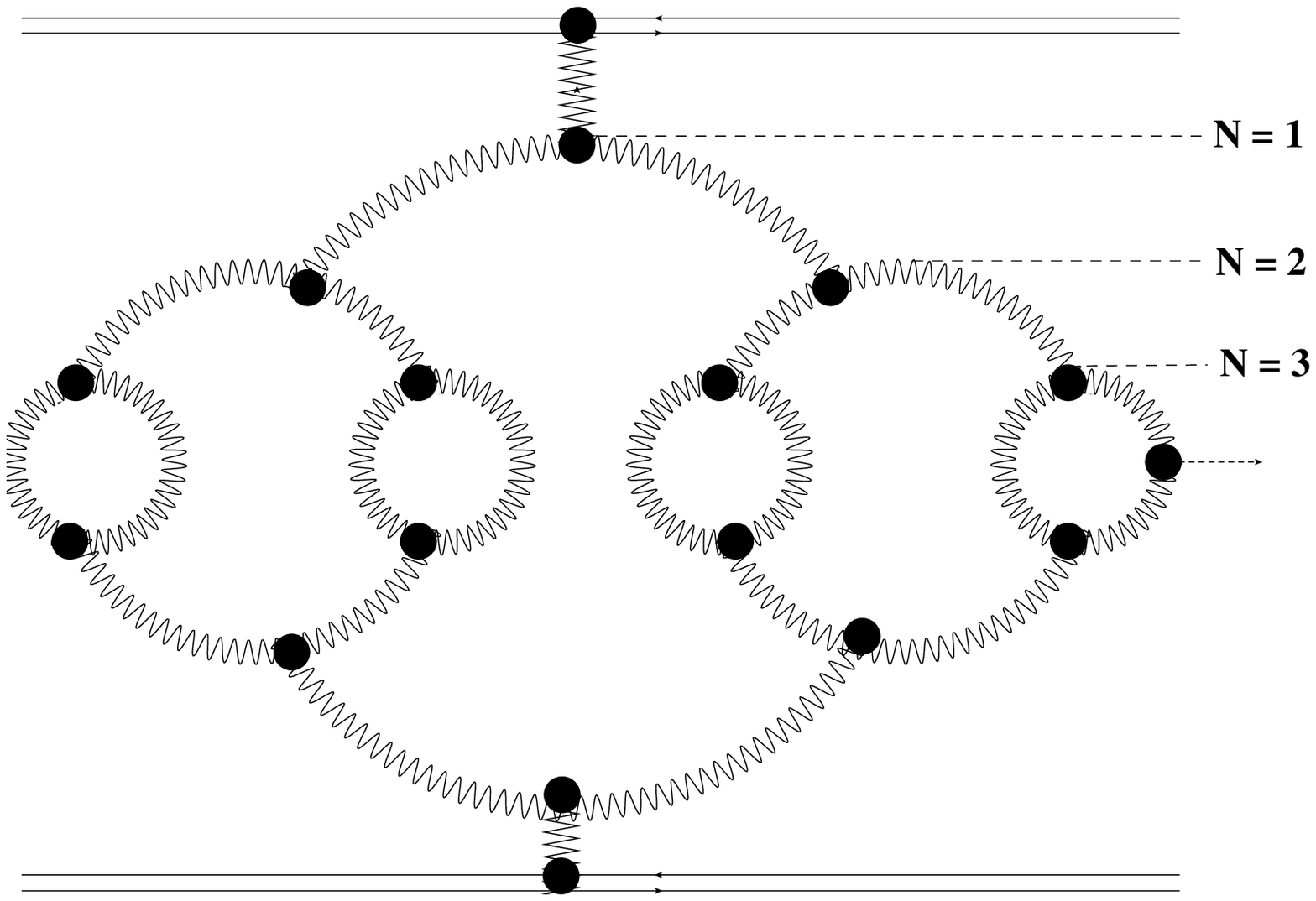,width=90mm,height=60mm}
{The Pomeron enhanced diagram with $N=2$ generations of branching.
\label{fembeddedpict}}{The Pomeron enhanced diagram with $N=3$ generations of branching.\label{fnpict} }

In this section the approach used to calculate the amplitude of more complicated Pomeron diagrams with 
more generations of Pomeron branching, such as the diagrams shown in \fig{fembeddedpict} and \fig{fnpict}, will be
explained. Until now the generally accepted technique for estimating the amplitude for
Pomeron enhanced diagrams of this type, has been the well known mean field approximation of
Mueller-Patel-Salam-Iancu (MPSI). In the previous section, the major contribution to the loop amplitude was shown to be equivalent to 2 non interacting Pomerons.
 In ref. \cite{Levin:2007wc} we showed that the amplitude of any
 general Pomeron enhanced diagram, is equivalent to the amplitude of the diagram of non interacting Pomerons,
with renormalized Pomeron vertices. This was one of the first steps towards deriving the amplitude of more complicated Pomeron
enhanced diagrams, other than the simple loop of \fig{fPomeronloop2} in perturbative QCD. The goal of this section is to show that 
the amplitude of an arbitrary Pomeron enhanced diagram, can be estimated in the precise QCD approach instead of the MPSI approximation.\\

To illustrate some examples of Pomeron enhanced diagrams, \fig{fembeddedpict} shows the enhanced diagram
with $N=2$ generations of branching, and \fig{fnpict} shows the enhanced diagram with $N=3$ generations of branching.
 In \fig{fembeddedpict}, the first generation of splitting forms the larger ``embedded'' loop, and
the second generation of splitting forms the two simple loops at the center of the diagram to form a total of $2^2-1=3$ loops. Likewise in
\fig{fnpict} there are $2^3-1=7$ loops. Diagrams such as these can be calculated in perturbative QCD using the following approach. Consider for
example \fig{fpictN} which shows the diagram with $N$ generations of Pomeron branching, leading to a total of $2^N-1$ loops. In this picture the $N$ generation diagram is equivalent to
2 sets of diagrams with $N-1$ generations of branching, embedded in one large loop as shown in \fig{fpictN}.

\FIGURE[h]{ \centerline {\epsfig{file=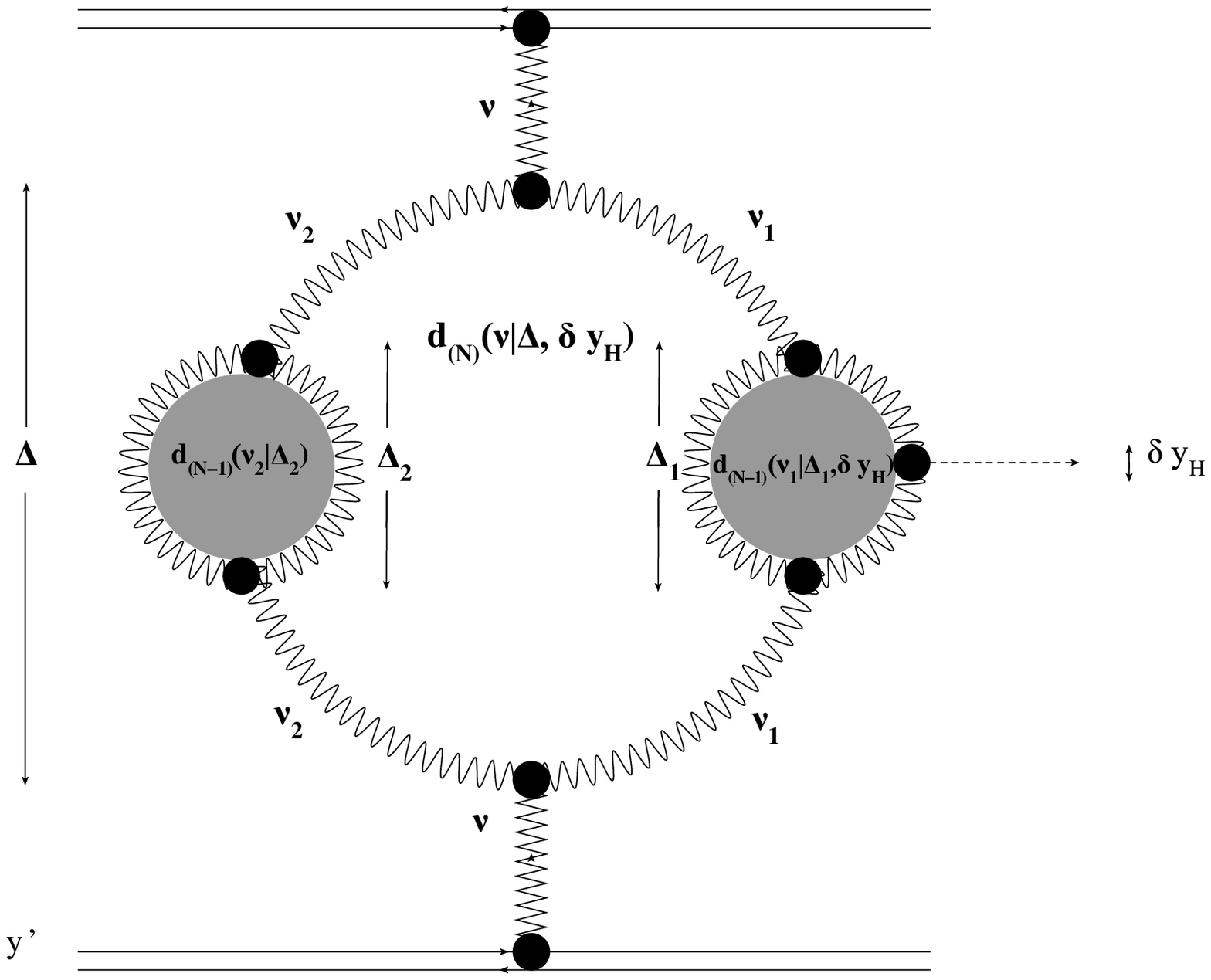,width=120mm}}\caption{ Pictorial representation of the Pomeron enhanced diagram
with $N$ generations of branching, as the product of 2 sets of diagrams each with $N-1$ generations of branching, 
embedded in one large loop.} \label{fpictN} }

From this picture the amplitude $d_{(N)}\Lb\nu|\De,\de y_H\Rb$ of the $N$ generation diagram which occupies a rapidity gap $\De$,
 is a function of the two sets of $N-1$ generation diagrams embedded in the larger loop,
 which occupy a rapidity gap $\De_1$ and $\De_2$ respectively, so that one can write for the amplitude;

\bea
d_{(N)}\Lb\nu|\De,\de y_H\Rb=d_{(N-1)}\Lb\nu_1|\De_1,\de y_H\Rb\,\bigotimes\,d_{(N-1)}\Lb\nu_2|\De_2\Rb\label{itrtv}\eea

where $\bigotimes$ denotes all necessary integrations over the conformal variables $\nu_1$ and $\nu_2$, and also over the rapidity gaps $\De_1$ and $\De_2$,
where $0\leq\De_1\leq \De$, and  $\de y_H\leq\De_2\leq \De$. \eq{itrtv} shows
the iterative expression which will be used to derive the expression for the arbitrary Pomeron enhanced diagram with $N$ generations of Pomeron branching.
The formalism which will be followed is to start from the basic simple loop amplitude $d_{(1)}\Lb \nu|\De,\de y_H\Rb$ derived in \sec{s2},
and plug this into the expansion of \eq{itrtv} to yield the amplitude $d_{(2)}\Lb \nu|\De,\de y_H\Rb$ of the 
$N=2$ generation diagram shown in \fig{fembeddedpict}. Similarly inserting $d_{(2)}\Lb \nu|\De,\de y_H\Rb$ into \eq{itrtv} gives
the amplitude $d_{(3)}\Lb \nu|\De,\de y_H\Rb$ of the $N=3$ generation diagram shown in \fig{fnpict}.
 Continuing with this
technique, the general expression $d_{(N)}\Lb \nu|\De,\de y_H\Rb$ for the amplitude of the $N$ generation diagram
shown in \fig{fpictN1} can be found, using proof by induction.\\

\DOUBLEFIGURE[h]{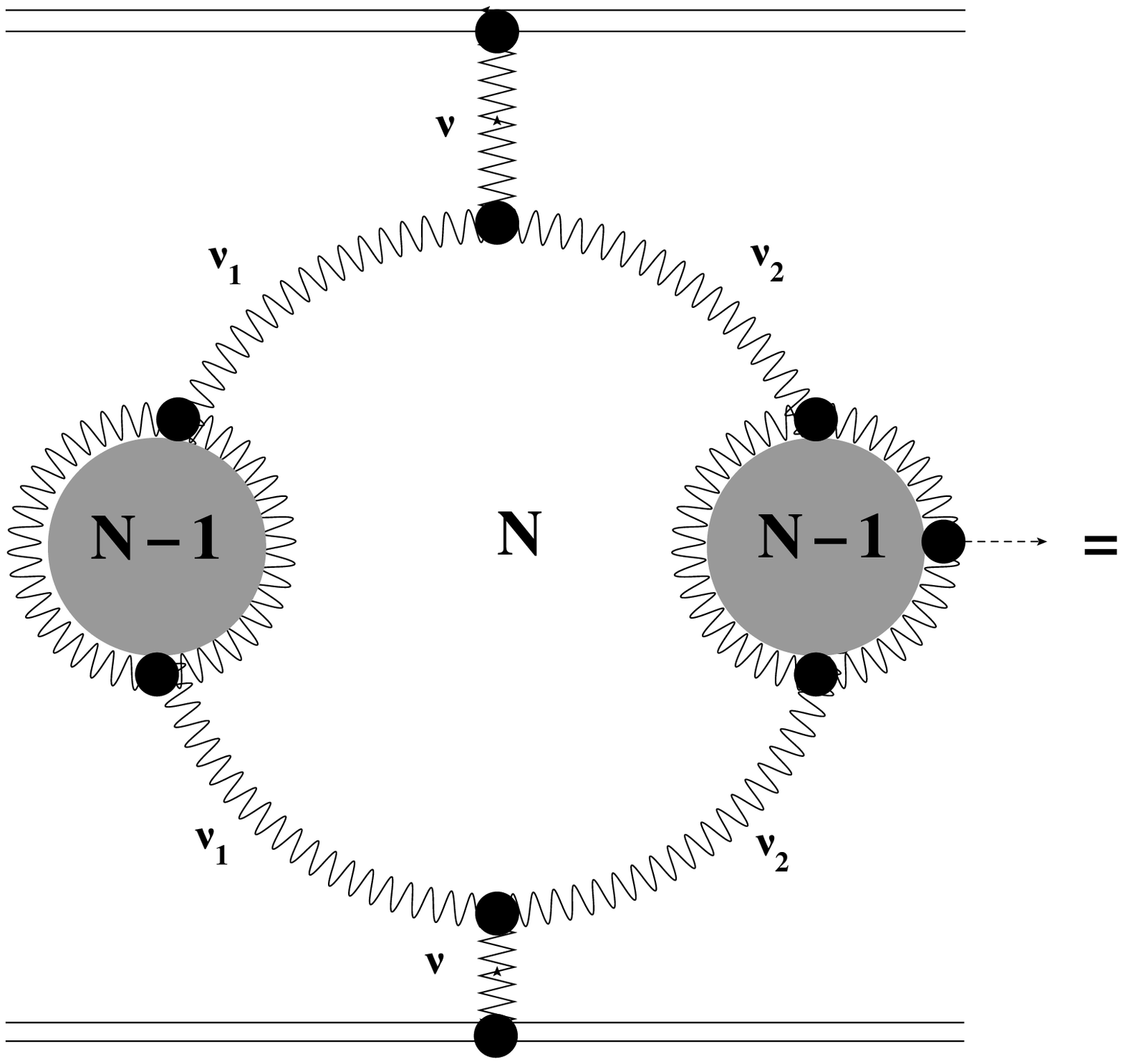,width=55mm,height=55mm}{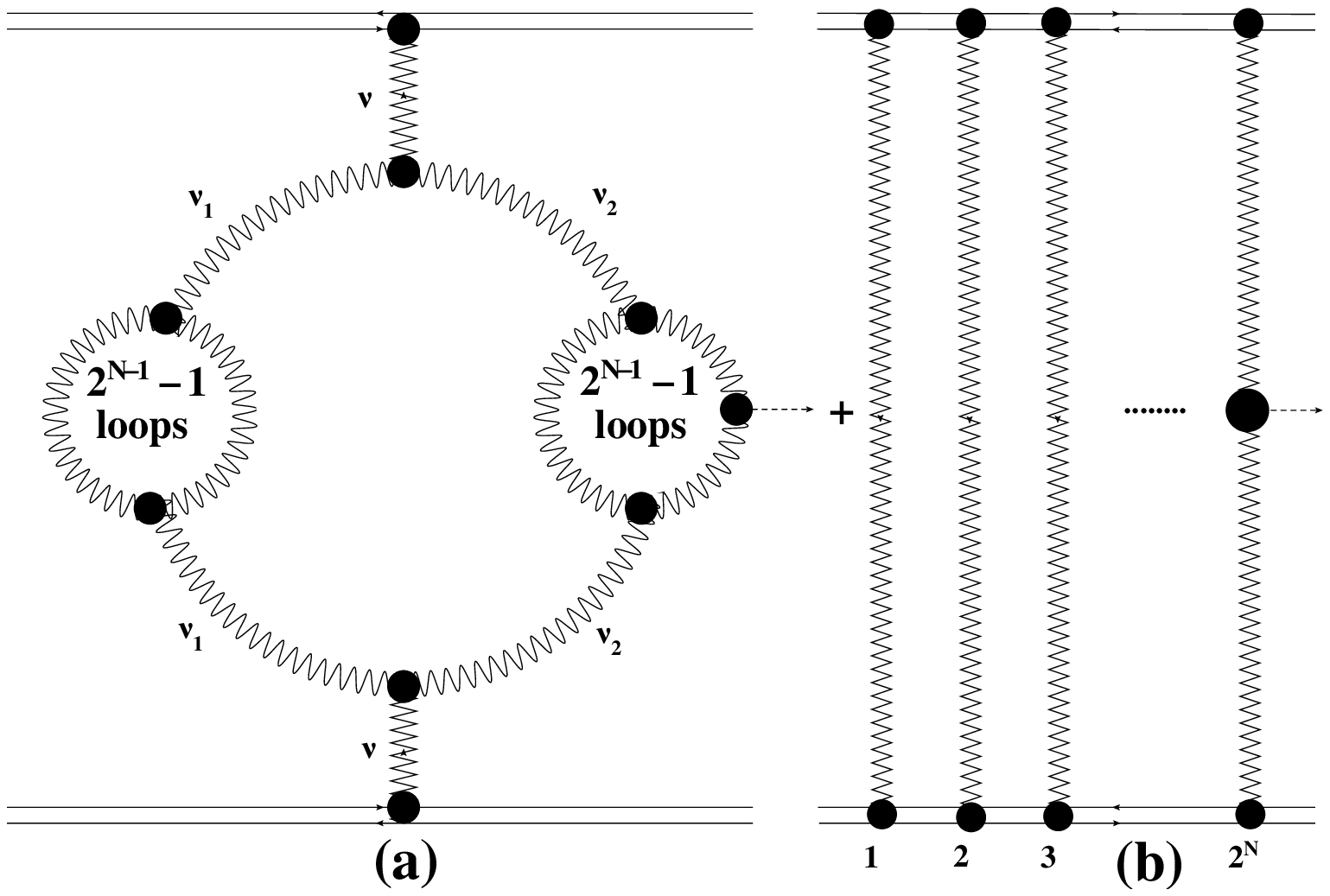,width=90mm,height=60mm}
{The diagram with $N$ generations of branching, shown as two sets of $N-1$ generation diagrams embedded in one larger Pomeron loop.
\label{fpictN1}}{(a) shows the contribution to the $N$ generation diagram amplitude which stems from the region I:  
$\left\{\nu,|\nu_1|,|\nu_2|\right\}=\left\{0,1/2,1/2\right\}$ and region II:  $\left\{|\nu|,\nu_1,\nu_2\right\}=\left\{1/2,0,0\right\}$.  Regions I and II lead to the renormalization of 
the Pomeron intercept, and the loops are preserved. (b) shows the contribution to the amplitude of the $N$ generation diagram, which is equivalent to the amplitude of $2^N$ non
interacting Pomerons, and stems from the region $\left\{|\nu|,|\nu_1|,|\nu_2|\right\}=\left\{1/2,1/2,1/2\right\}$\label{fpictN2} }

In \sec{s2} the simple Pomeron loop amplitude was calculated for the two regions I: $\left\{\nu,|\nu_1|,|\nu_2|\right\}=\left\{0,1/2,1/2\right\}$ and II:  $\left\{|\nu|,\nu_1,\nu_2\right\}=\left\{1/2,0,0\right\}$.
Region I led to the contribution which is the renormalization of the Pomeron intercept (see \eq{p5(0)}). Region II led to the contribution to the
amplitude which is equivalent to the amplitude of 2 non interacting Pomerons, with renormalized Pomeron vertices (see \eq{p3a4}). Likewise, calculating the amplitude of the $N$ generation
diagram shown in \fig{fpictN1}, leads to 2 contributions. The first is the contribution from the renormalization of the Pomeron intercept (see \fig{fpictN2}a). The second contribution
is from the diagram which is equivalent to $2^N$ non interacting Pomerons, with renormalized Pomeron vertices, shown in \fig{fpictN2}b.
These two contributions stem from 3 regions for the conformal variables, where in the notation of \fig{fpictN1},
these are region I: $\left\{\nu,|\nu_1|,|\nu_2|\right\}=\left\{0,1/2,1/2\right\}$; region II:  $\left\{|\nu|,\nu_1,\nu_2\right\}=\left\{1/2,0,0\right\}$;
and region III: $\left\{|\nu|,|\nu_1|,|\nu_2|\right\}=\left\{1/2,1/2,1/2\right\}$. For the diagrams with $N\geq 2$, regions I and II will lead
to the contribution shown in \fig{fpictN2}a which is equivalent to the renormalization of the Pomeron intercept. For $N\geq 2$ Region III will yield the 
contribution to the amplitude which is equivalent to $2^N$ non interacting Pomerons, with renormalized Pomeron vertices.
The contributions of all three regions to the amplitude of the $N$ generation diagram can be summarized as; 

\bea
A_{(N)}\Lb \De,\de y_H|\mbox{\fig{fpictN1}}\Rb
\!&&\!=\!A^{\footnotesize{\mbox{I}}}_{(N)}\Lb\De,\de y_H|\mbox{\fig{fpictN2} a}\Rb
+A^{\footnotesize{\mbox{II}}}_{(N)}\Lb\De,\de y_H|\mbox{\fig{fpictN2}a}\Rb
+A^{\footnotesize{\mbox{III}}}_{(N)}\Lb\De,\de y_H|\mbox{\fig{fpictN2}b}\Rb
\nn\\
\nn\\
&&\!=\!k^{\footnotesize{\mbox{I}}}_{(N)}e^{\omega\Lb 0\Rb\De}+
k^{\footnotesize{\mbox{II}}}_{(N)}e^{2\omega\Lb 0\Rb\De}+
k^{\footnotesize{\mbox{III}}}_{(N)}e^{2^N\omega\Lb 0\Rb\De}\label{mi2}\eea

where $k_{(N)}$ are constants which contain all other terms which are included in the amplitude.

\section{Multiple-loop Pomeron enhanced diagrams}
\label{s3}

In this section, the amplitude for diffractive Higgs production from the Pomeron enhanced diagrams with $N=2$ and $N=3$ generations of branching, shown in \fig{fembedded} and \fig{fn} respectively, are derived
using the iterative technique outlined in \sec{smi}. The resulting expressions for these diagrams, will provide the tools necessary to form a general expression for the diagram with $N$ generations of Pomeron branching, shown in
\fig{figN}. 

\FIGURE[h]{ \centerline {\epsfig{file=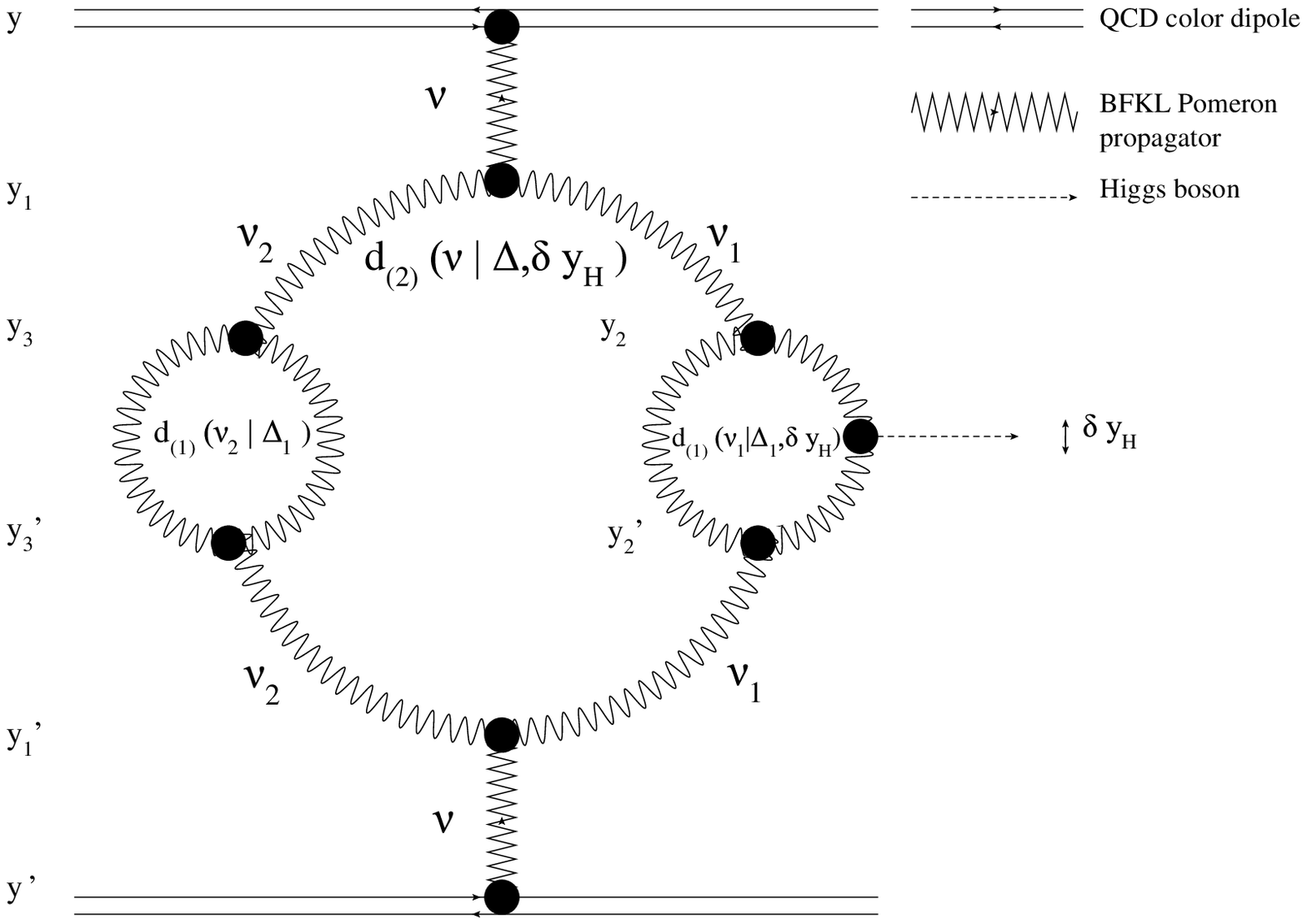,width=150mm}}\caption{ The Pomeron enhanced diagram with $N=2$ generations of branching, leading to $2^2-1=3$ loops.} \label{fembedded} }

 The scattering amplitude of \fig{fembedded} is given by a straightforward extension
of \eq{p3a}, namely;

 \bea
A_{(2)}\Lb \De,\de y_H|\mbox{\fig{fembedded}}\Rb\!
\!&&\!=\f{\as^2}{4}\int^\infty_{-\infty}\!d\nu  h\Lb\nu\Rb \la^2\Lb\nu\Rb e^{\omega\Lb\nu\Rb \De }d_{(2)}\Lb\nu | \De , \de y_H\Rb
E_{\nu} E^{\prime}_{-\nu}A_H\hspace{0.8cm}\label{e}\eea

where $d_{(2)}\Lb\nu\,|\,\De,\de y_H\Rb$ labels the contribution to the scattering amplitude of the 
$2^2-1=3$ loops which arise from the $N=2$ generations of Pomeron branching. 
The order of the symmetry group of the diagram of \fig{fembedded} is $S_{(1)}S_{(2)}$, where
 $S_{(1)}=16$ is associated with the large outer loop from the same considerations discussed in \sec{s2}, and $S_{(2)}=8$ is associated
with permutations of the two internal loops, which lead to identical diagrams.\\

Following the formalism described in \eq{itrtv}, $d_{(2)}\Lb\nu|\De,\de y_{H}\Rb$ is a function of the two simple Pomeron loop amplitudes $d_{(1)}\Lb\nu_1|\De_1,\de y_H\Rb$
and $d_{(1)}\Lb\nu_2|\De_1\Rb$
embedded in the large loop as shown in \fig{fembedded}. Here $\De_1=y_1-y_1^{\,\prime}$ is the rapidity gap occupied by the large outer  loop in \fig{fembedded}. Hence in this approach, including the 
symmetry factor $1/S_{(1)}S_{(2)}=1/2^7$;

 \bea
d_{(2)}\Lb\nu\,|\,\De,\de y_H\Rb\,&&=\f{1}{2^7}\int^\infty_{-\infty}\!\!d\nu_1h\Lb\nu_1\Rb\,\la^2\Lb\nu_1\Rb
\int^\infty_{-\infty}\!\!d\nu_2 h\Lb\nu_2\Rb\la^2\Lb\nu_2\Rb \begin{vmatrix}\Ga\Lb \nu|\nu_1,\nu_2\Rb\end{vmatrix}^2\,\nn\\
 &&\times \int^y_{y^{\,\prime}+\de y_H}\!\!\!\!\!\!dy_1
\int^{y_1-\de y_H}_{y^{\,\prime}}\!\!\!\!\!\!dy_1^{\,\prime}\,
e^{\Lb\,\omega\Lb \nu_1\Rb\,+\,\omega\Lb \nu_2\Rb\,-\omega\Lb\nu\Rb\,\Rb \De_1}\,d_{(1)}\Lb\,\nu_1\,|\,\De_1,\de y_H\Rb\,d_{(1)}\Lb\nu_2\,|\,\De_1\,\Rb \label{e0}
\eea

The simple loop amplitude was derived in \eq{d7} and \eq{addition} for two different regions. $d_{(1)}\Lb\nu_2\,|\,\De_1\,\Rb$ which labels the internal loop
in \fig{fembedded} to the left without the production of the Higgs boson, is derived from \eq{d7} or \eq{addition} by setting $\de y_H$ equal to zero. 
By inserting \eq{d7} into \eq{e0}, one can evaluate the integration over the conformal variables for region I: $\left\{\nu,|\nu_1|,|\nu_2|\right\}=\left\{0,1/2,1/2\right\}$ and region II:
$\left\{|\nu|,\nu_1,\nu_2\right\}=\left\{1/2,0,0\right\}$. For region I the triple Pomeron vertex derived in \eq{G123combined} should be inserted, and then integrate over
over $\nu_1$ and $\nu_2$ using the contour $C$ shown in \fig{fintegrationcontour} and sum over the residues at $\left\{i\nu_1,i\nu_2\right\}=1/2$. An extra factor of $2$ is included to take into account the identical contribution from
 the residues at $\left\{i\nu_1,i\nu_2\right\}=-1/2$. Overall after integrating over
the rapidity variables, the contribution of region I to \eq{e0} is given by the expression;

\bea
d^{\,\footnotesize{\mbox{I}}}_{(2)}\Lb\nu|\De,\de y_H\Rb\!&&\!=\!d^{\,\prime}d^2\Lb\h+i\nu\Rb^3\!\Lb\h-i\nu\Rb^3\!\!\chi\Lb\nu\Rb e^{-\omega\Lb\nu\Rb\de y_H}\nn\\
&&\times\left\{\!\Lb\! 6-\f{3}{4}\bas\de y_H\!\Rb\omega\Lb\nu\Rb\Lb\omega\Lb\nu\Rb+\f{\omega^2\Lb\nu\Rb\,\Lb\De-\de y_H\Rb}{3}\Rb
\!\right\}\hspace{1cm}\label{e8}\\
\nn\\
\mbox{where}\hspace{0.5cm}d^{\,\prime}\,&&=\,\f{\bas^2}{2^{11}} \Lb 1-\f{1}{N_c^2}\Rb\Lb 1-\f{2}{N_c^2}\Rb\label{dembeddedloop}\eea

Inserting \eq{e8} into \eq{e} and integrating over $\nu$ using the method of steepest descents in the same way described in \sec{s1},
yields the following contribution of region I to the scattering amplitude for \fig{fembedded};

\bea
A^{\,\footnotesize{\mbox{I}}}_{(2)}\Lb\De,\de y_H|\mbox{\fig{fembedded}}\Rb\,&&=\,\f{\Lb2\pi\Rb^{\,1/2}\,\bas^2\,A_H}{128\pi^2N_c^2}
d^{\,\prime}\,d^2
\f{e^{\omega\Lb 0\Rb\Lb \De-\de y_H\Rb }}{\Lb \omega^{\,\prime\,\prime}\Lb 0\Rb\Lb\De-\de y_H\Rb\Rb^{3/2}}\nn\\
&&\times\chi\Lb 0\Rb\left\{\Lb 6-\f{3}{4}\bas\de y_H\Rb\omega\Lb 0\Rb\Lb\omega\Lb 0\Rb+\f{\omega^2\Lb 0\Rb\,\Lb\De-\de y_H\Rb}{3}\Rb
\right\}\label{e9(0)}\\
\nn\\
\nn\\
&&=\,
8.91\,\times\,10^{-17}\,\,\mbox{GeV}^{-2}\hspace{1cm}(\as\,=\,0.12)\nn\\
&&\hspace{0.5cm}
3.3\,\,\,\,\times\,10^{-14}\,\,\mbox{GeV}^{-2}\hspace{1cm}(\as\,=\,0.2)\label{e9}\eea

Alternatively, after inserting \eq{d7} into \eq{e0}, the contribution  from region II, namely $\left\{|\nu|,\nu_1,\nu_2\right\}=\left\{1/2,0,0\right\}$ is found by substituting for the triple Pomeron vertex \eq{tpvuseful}
and integrating over the conformal variables $\nu_1$ and $\nu_2$ using the steepest descents method, which yields the expression;

\bea
d^{\,\footnotesize{\mbox{II}}}_{(2)}\Lb\nu\,|\,\De,\de y_H\Rb\,&&=a^{\,\prime}d^2\f{\,e^{-\omega\Lb 0\Rb \de y_H}}{\Lb 1/2+i\nu\Rb\Lb 1/2-i\nu\Rb}\,\int^y_{y^{\,\prime}+\de y_H}\!\!\!\!\!\!dy_1
\int^{y_1-\de y_H}_{y^{\,\prime}}\!\!\!\!\!\!dy_1^{\,\prime}\,
\f{e^{\Lb2\omega\Lb 0\Rb-\omega\Lb\nu\Rb\Rb\De_1}}{\Lb\De_1\Lb\De_1-\de y_H\Rb\Rb^{3/2}}\nn\\
&&\times\chi^2\Lb 0\Rb\!\left\{\!\omega\Lb 0\Rb+\h\omega^2\Lb0\Rb\!\Lb\De_1 -\de y_H\Rb\!\right\}\!
\left\{\!\omega\Lb0\Rb\!+\h\omega^2\Lb0\Rb\De_1 \!\right\}\hspace{1cm} \label{e1(1)}\\
\nn\\
\mbox{where}\hspace{0.5cm}a^{\,\prime}&&=\f{\bas^4}{2^6\pi N_c^4\,[\omega^{\,\prime\prime}\Lb 0\Rb]^3}\eea

Inserting \eq{e1(1)} into \eq{e} and using the contour $C$ shown in \fig{fintegrationcontour} for the $\nu$ integral, the solution is 
the sum over residues at $i\nu=1/2$. In the same way an additional factor of $2$ includes the identical contribution from the sum over residues at $i\nu=-1/2$.
Using this approach, the contribution of region II to the scattering amplitude of \fig{fembedded} is;

 \bea
&&A^{\,\footnotesize{\mbox{II}}}_{(2)}\Lb\,\De,\de y_H|\mbox{\fig{fembedded}}\Rb\,=\f{\bas a^{\,\prime}d^2A_H}{2^9\pi\,N_c^2}\chi^2\Lb 0\Rb\Lb\De-\de y_H\Rb\nn\\
&&\hspace{1cm}\times
\Lb\f{-1}{\bas}\f{d}{d\De}\Rb^3\!\left\{\f{e^{2\omega\Lb 0\Rb\Lb\De-\de y_H/2\Rb}}{\Lb\De\Lb\De-\de y_H\Rb\Rb^{3/2}}\!
\Lb\!\omega\Lb 0\Rb+\h\omega^2\Lb0\Rb\!\Lb\De\! -\!\de y_H\Rb\!\Rb\!
\Lb\!\omega\Lb0\Rb\!+\h\omega^2\Lb0\Rb\De\! \Rb\!\right\}\hspace{0.7cm}\label{renpomint}\\
\nn\\
&&\hspace{1cm}=
4.94\,\times\,10^{-17}\,\,\mbox{GeV}^{-2}\hspace{0.5cm}(\as=0.12)\nn\\
\hspace{1.4cm} 
&&\hspace{1.4cm}3.23\,\times\,10^{-13}\,\,\mbox{GeV}^{-2}\hspace{0.6cm}(\as=0.2)\label{e04}\eea

\eq{e9(0)} and \eq{renpomint} are the contributions to the scattering amplitude of \fig{fembedded}, which lead to the renormalization of the Pomeron intercept. Alternatively, the contribution to \fig{fembedded} which is equivalent to
$4$ non interacting Pomerons, is found by substituting for $d_{(1)}\Lb\nu_1|\De_1,\de y_H\Rb$ and $d_{(1)}\Lb\nu_2|\De_1\Rb$ the simple loop amplitude of \eq{addition} in \eq{e0}, and evaluating the integrals
over the conformal variables, taking into account the singularities which arise from  region III: $\left\{|\nu|,|\nu_1|,|\nu_2|\right\}=\left\{1/2,1/2,1/2\right\}$. Inserting the triple Pomeron vertex of \eq{G123combined} into \eq{e0},
the $\nu_1$ and $\nu_2$ integrations are evaluated by summing over the identical residues at $\left\{i\nu_1,i\nu_2\right\}=1/2$ and $-1/2$, which arise from integrating along 
 the contours $C$ and $C^{\,\prime}$ shown in \fig{fintegrationcontour}, yielding the result; 

 \bea
d^{\mbox{\footnotesize{III}}}_{(2)}\Lb\nu|\De,\de y_H\Rb&&=a^2be^{-\omega\Lb 0\Rb\de y_H}\Lb 1/2+i\nu\Rb^3\Lb 1/2-i\nu\Rb^3\chi\Lb\nu\Rb
\int^y_{y^{\,\prime}+\de y_H}\!\!\!\!\!\! dy_1\int^{y_1-\de y_H}_{y^{\,\prime}}\!\!\!\!\!\! dy_1^{\,\prime}e^{-\omega\Lb\nu\Rb\De_1}\nn\\
&&\times\Lb\De_1-\de y_H\Rb\De_1\left\{\Lb\f{-1}{\bas}\f{d}{d\De_1}\Rb^3\f{e^{2\omega\Lb 0\Rb\De_1}}{\De_1^{3/2}\Lb\De_1-\de y_H\Rb^{3/2}}\right\}
\left\{\Lb\f{-1}{\bas}\f{d}{d\De_1}\Rb^3\f{e^{2\omega\Lb 0\Rb\De_1}}{\De_1^3}\right\}\hspace{0.7cm}\label{nonint2}\\
\nn\\
\mbox{where}\hspace{0.5cm}b&&=\f{\bas^2}{2^{10}}\Lb1-\f{1}{N_c^2}\Rb^2\label{definitionofb}
\eea

Inserting \eq{nonint2} into \eq{e}, the $\nu$ integration is solved using the same method of residues described above, so that
 the contribution of region III to the scattering amplitude of \fig{fembedded},  which is equivalent to the amplitude of non interacting Pomerons is;

\bea
&&A^{\mbox{\footnotesize{III}}}_{(2)}\Lb\De,\de y_H|\mbox{\fig{fembedded}}\Rb=\f{\bas\,A_H}{2^9 N_c^2\pi}a^2be^{-\omega\Lb 0\Rb\de y_H}
\nn\\
&&\hspace{1cm}\times\Lb\De-\de y_H\Rb^2\De\left\{\Lb\f{-1}{\bas}\f{d}{d\De}\Rb^3\f{e^{2\omega\Lb 0\Rb\De}}{\De^{3/2}\Lb\De-\de y_H\Rb^{3/2}}\right\}
\left\{\Lb\f{-1}{\bas}\f{d}{d\De}\Rb^3\f{e^{2\omega\Lb 0\Rb\De}}{\De^3}\right\}\label{nonintPomsolution}\\
\nn\\
&&\hspace{1cm}=
2.85\,\times\,10^{-15}\,\,\mbox{GeV}^{-2}\hspace{1cm}(\as\,=\,0.12)\nn\\
&&\hspace{1.4cm}
1.39\,\times\,10^{-7}\,\,\mbox{GeV}^{-2}\hspace{1.2cm}(\as\,=\,0.2)\label{nonint3}
\eea

From an observation of \eq{nonintPomsolution};

\bea
A^{\mbox{\footnotesize{III}}}_{(2)}\Lb\De,\de y_H|\mbox{\fig{fembedded}}\Rb&&=k_{(2)}e^{4\omega\Lb 0\Rb\De}\label{nonint4}\eea

where $k_{(2)}$ contains all the other terms contained in the amplitude. The above results show that the main contribution to the
  the scattering amplitude of \fig{fembedded} is just the equivalent of 4 non interacting Pomerons, with renormalized Pomeron vertices. The compete scattering amplitude of the $N=2$ generation
  diagram of \fig{fembedded}, is the sum over the contributions of \eq{e9(0)}, \eq{renpomint} and \eq{nonintPomsolution}, namely;
 
 \bea
 A\Lb\De,\de y_H|\mbox{\fig{fembedded}}\Rb&&=
A^{\mbox{\footnotesize{I}}}_{(2)}\Lb\De,\de y_H|\mbox{\fig{fembedded}}\Rb+
A^{\mbox{\footnotesize{II}}}_{(2)}\Lb\De,\de y_H|\mbox{\fig{fembedded}}\Rb+ A^{\mbox{\footnotesize{III}}}_{(2)}\Lb\De,\de y_H|\mbox{\fig{fembedded}}\Rb\label{completeembeddedsolution}\\
\nn\\
&&=
2.85\,\times\,10^{-15}\,\,\mbox{GeV}^{-2}\hspace{1cm}(\as\,=\,0.12)\nn\\
&&\hspace{0.4cm}
1.39\,\times\,10^{-7}\,\,\mbox{GeV}^{-2}\hspace{1.2cm}(\as\,=\,0.2)\label{completeembeddedvalues}\eea
  
Comparing the values of \eq{thesompleteloopamplitude} and \eq{completeembeddedvalues},  \fig{fembedded} is of the same order of magnitude as the simple loop amplitude of \fig{fPomeronloop2}. 
This implies that in order to estimate the enhanced survival probability, it is not enough to just take into account the simple Pomeron loop, but rather also enhanced diagrams with
multiple Pomeron loops need to be included in the estimate.

\FIGURE[h]{ \centerline {\epsfig{file=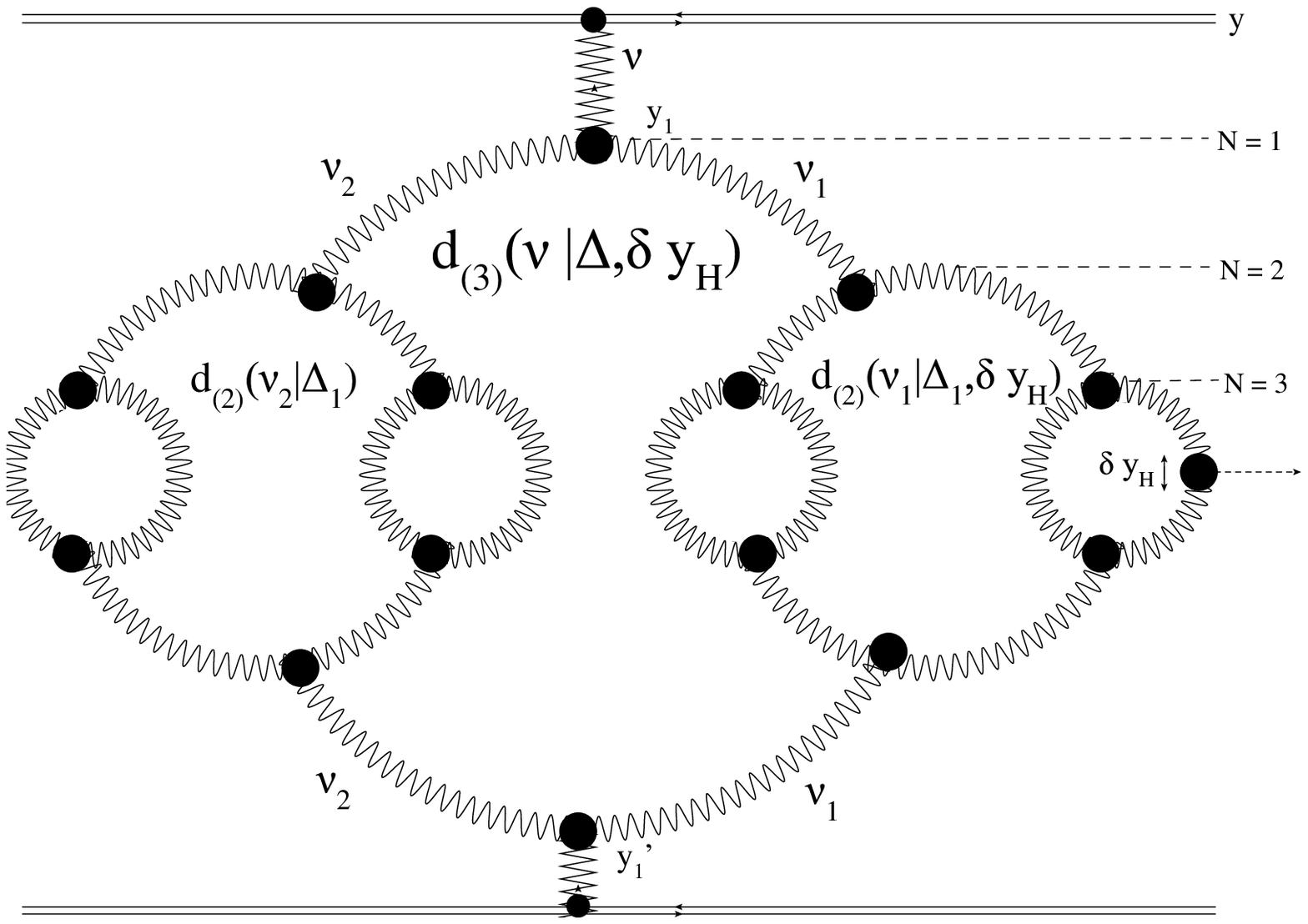,width=120mm}}\caption{ The $N$=3 generation Pomeron loop diagram, leading to $2^3-1=7$ loops. } \label{fn} }

\fig{fn} shows the diagram for diffractive Higgs production in t-channel Pomeron exchange, where there are 3 generations of Pomeron branching,
which recombine to form 1 large loop at the $N=1$ level, 2 smaller internal loops at the $N=2$ level and 4 simple loops at the $N=3$ level at the center of
the diagram. The scattering amplitude of \fig{fn} is given by the expression;

 \bea
A_{(3)}\Lb \De,\de y_H|\mbox{\fig{fn}}\Rb\!
&&\!=\!\f{\as^2}{4}\!\!\int^\infty_{-\infty}\!\!\!d\nu h\Lb\nu\Rb  \la^2\Lb\nu\Rb e^{ \omega\Lb\nu\Rb\De }d_{(3)}\Lb\nu | \De , \de y_H\Rb
E_{\nu} E^{\,\prime}_{\,-\,\nu} A_H\hspace{1cm}\label{N3}\eea

where $d_{(3)}\Lb\De,\de y_H\Rb$ is the contribution of the $2^3-1\,=\,7$ loops in \fig{fn} to the scattering amplitude. 
Following the approach of \eq{itrtv}, $d_{(3)}\Lb\De,\de y_H\Rb$ is a function of $d_{(2)}\Lb\De_1,\de y_H\Rb$ and  $d_{(2)}\Lb\De_1,\de y_H\Rb$ embedded in the larger outer loop,
as shown in \fig{fn}. The $N=3$ generation amplitude  $d_{(3)}\Lb\De,\de y_H\Rb$ should also have in the denominator the same symmetry factor
$1/2^7$ as for the case of the $N=2$ generation amplitude, for the same above explained reasons. As such  $d_{(3)}\Lb\De,\de y_H\Rb$ is given by the expression;

 \bea
d_{(3)}\Lb\nu\,|\,\De,\de y_H\Rb\,&&=\f{1}{2^7}\int^\infty_{-\infty}\!\! d\nu_1h\Lb\nu_1\Rb\la^2\Lb\nu_1\Rb
\int^\infty_{-\infty}\!\! d\nu_2 h\Lb\nu_2\Rb\la^2\Lb\nu_2\Rb \begin{vmatrix}\Ga\Lb \nu | \nu_1 , \nu_2\Rb\end{vmatrix}^2 \nn\\
 &&\times \int^y_{y^{\,\prime}+\de y_H}\!\!\!\!\!\!dy_1
\int^{y_1-\de y_H}_{y^{\,\prime}}\!\!\!\!\!\!dy_1^{\,\prime}
e^{\Lb\omega\Lb \nu_1\Rb+\omega\Lb \nu_2\Rb-\omega\Lb\nu\Rb\Rb\De_1 }\,
d_{(2)}\Lb\,\nu_1\,|\,\De_1,\de y_H\Rb\,d_{(2)}\Lb\nu_2\,|\,\De_1\,\Rb \label{N31}\eea

Following the same steps as \eq{e0} - \eq{nonintPomsolution}, the contributions of region I and II to \fig{fn} which leads to the renormalization of the Pomeron intercept, and  the contribution of region III
which is equivalent to the amplitude of non interacting Pomerons  are given by the expressions;

  \bea
A^{\,\footnotesize{\mbox{I}}}_{(3)}\Lb \De,\de y_H|\mbox{\fig{fn}}\Rb\,
&&=\,\f{\Lb2\pi\Rb^{\,1/2} \bas^3\,d^4d^{\,\prime\,3}A_H}{128\pi^2N_c^2}\,\f{e^{\omega\Lb0\Rb\Lb\De-\de y_H\Rb}\,}{\Lb\omega^{\,\prime\,\prime}\Lb 0\Rb\Lb\De-\de y_H\Rb\Rb^{3/2}}\,\chi\Lb 0\Rb\nn\\
&&\times6\Lb 6-\f{3}{2}\bas\de y_H\Rb\!\left\{\!\Lb\f{16}{3}-\f{4}{9}\Lb 2\bas\Rb\de y_H\Rb\omega^2\Lb 0\Rb\!\Lb\!\omega\Lb 0\Rb+\f{\omega^2\Lb 0\Rb\!\Lb\De-\de y_H\Rb}{4}\Rb\!\right\}
\hspace{1cm}\label{N35}\\
\nn\\
\nn\\
&&=
3.93\,\times\,10^{-22}\,\,\mbox{GeV}^{-2}\hspace{1cm}(\as\,=\,0.12)\nn\\
&&\hspace{0.4cm}
6.27\,\times\,10^{-18}\,\,\mbox{GeV}^{-2}\hspace{1cm}(\as\,=\,0.2)\label{3values}\eea

\bea
&&A^{\,\footnotesize{\mbox{II}}}_{(3)}\Lb\De,\de y_H|\mbox{\fig{fn}}\Rb\!\!=\!\f{\bas a^{\,\prime}d^4d^{\,\prime\,2}A_H}{2^9\pi N_c^2}
6\!\Lb6-\f{3}{4}\bas\de y_H\Rb\!\chi^2\Lb 0\Rb\omega^2\Lb0\Rb\,\,\Lb\De-\de y_H\Rb\nn\\
&&\hspace{1cm}\times\!\!\Lb\f{-1}{\bas}\f{d}{d\De}\Rb^3\left\{\!\f{e^{2\omega\Lb 0\Rb\Lb\De-\de y_H/2\Rb}}{\Lb\De\Lb \De - \de y_H\Rb\Rb^{3/2}}\!\Lb\!\omega\Lb 0\Rb+\f{\omega^2\Lb 0\Rb\!\Lb\De_1-\de y_H\Rb}{3}\!\Rb
\!\!\Lb\!\omega\Lb 0\Rb+\f{\omega^2\Lb 0\Rb\!\Lb\De_1-\de y_H\Rb}{3}\!\Rb\!\right\}\hspace{0.6cm}\label{AII(3)}\\
\nn\\
&&\hspace{3.3cm}=
2.52\,\times\,10^{-26}\,\,\mbox{GeV}^{-2}\hspace{1cm}(\as\,=\,0.12)\nn\\
&&\hspace{3.7cm}
8.53\,\times\,10^{-21}\,\,\mbox{GeV}^{-2}\hspace{1cm}(\as\,=\,0.2)
\hspace{1cm}\label{d3second2}\eea

\bea
A^{\mbox{\footnotesize{III}}}_{(3)}\Lb\De,\de y_H|\mbox{\fig{fn}}\Rb&&=\f{\bas\,A_H}{2^9 N_c^2\pi}a^4b^3e^{-\omega\Lb 0\Rb\de y_H}
\nn\\
&&\times\De^4\Lb\De-\de y_H\Rb^3
\left\{\Lb\f{-1}{\bas}\f{d}{d\De}\Rb^3\f{e^{2\omega\Lb 0\Rb\De}}{\De^{3/2}\Lb\De-\de y_H\Rb^{3/2}}\right\}
\left\{\Lb\f{-1}{\bas}\f{d}{d\De}\Rb^3\f{e^{2\omega\Lb 0\Rb\De}}{\De^3}\right\}^3\hspace{0.4cm}\label{AIII(3)}\\
\nn\\
&&=
3.34\,\times\,10^{-24}\,\,\mbox{GeV}^{-2}\hspace{1cm}(\as\,=\,0.12)\nn\\
&&\hspace{0.4cm}
1.28\,\times\,10^{-7}\,\,\mbox{GeV}^{-2}\hspace{1.2cm}(\as\,=\,0.2)\label{nonint5}
\eea

\eq{AIII(3)} is the contribution which is equivalent to the
amplitude of 8 non interacting Pomerons, as can be seen clearly by recasting \eq{AIII(3)} in the following form;

\bea
A^{\mbox{\footnotesize{III}}}_{(3)}\Lb\De,\de y_H|\mbox{\fig{fn}}\Rb
&&=k_{(3)}e^{8\omega\Lb 0\Rb\De}\label{nonint6}\eea

where $k_{(3)}$ contains all other terms included in the amplitude. The major contribution comes from the non interacting Pomeron contribution of \eq{nonint5}.
The full scattering amplitude of \fig{fn} is the sum over the contributions of \eq{N35}, \eq{AII(3)} and \eq{AIII(3)}, namely;

\bea
A\Lb\De,\de y_H|\mbox{\fig{fn}}\Rb&&=
A^{\,\footnotesize{\mbox{I}}}_{(3)}\Lb\De,\de y_H|\mbox{\fig{fn}}\Rb+
A^{\,\footnotesize{\mbox{II}}}_{(3)}\Lb\De,\de y_H|\mbox{\fig{fn}}\Rb+A^{\,\footnotesize{\mbox{III}}}_{(3)}\Lb\De,\de y_H|\mbox{\fig{fn}}\Rb\label{A3FULL}\\
\nn\\
&&=
3.34\,\times\,10^{-24}\,\,\mbox{GeV}^{-2}\hspace{1cm}(\as\,=\,0.12)\nn\\
&&\hspace{0.4cm}
1.28\,\times\,10^{-7}\,\,\mbox{GeV}^{-2}\hspace{1.2cm}(\as\,=\,0.2)\label{A3fullvalues}
\eea
 The value in \eq{A3fullvalues} is  very close to the amplitude of \fig{fembedded} for $\as =0.2$ (see \eq{completeembeddedvalues}) and the same order of magnitude
as the amplitude of the simple Pomeron loop of \fig{fPomeronloop2} (see \eq{thesompleteloopamplitude}). This shows that in order to obtain an accurate estimate of the enhanced survival probability $\SPE$,
the sum over the complete set of Pomeron loops needs to be found. This is the subject matter of the next section. \\

\section{The summation over Pomeron loop diagrams}
\label{s4}

In this section the diffractive Higgs production amplitude from the general multiple Pomeron loop diagram, with $N$ generations of branching is derived. This general expression provides the basis for summing over 
the complete set of Pomeron loop diagrams, in perturbative QCD, instead of using the MPSI approach. 
In  a diagram with $N$ generations of Pomeron branching shown in \fig{figN}, the $k^{th}$ generation of branching gives rise to $2^{k-1}$ loops, so that  in \fig{figN}, there
are a total of $
\sum^N_{k=1}2^{k-1}\,=\,2^N-1$ loops.  So far 3 contributing regions to the amplitude of Pomeron loops  have been discussed. An analytical formula for the diagram with $N$ generations of branching can be derived,
 for all 3 of these regions using  the iterative technique outlined in \sec{smi}, without any
a priori assumptions. For the diagram of diffractive Higgs production in t-channel Pomeron exchange, where there are $N$ generations of Pomeron branching as shown in
\fig{figN}, the high energy scattering amplitude is labeled;

\FIGURE[h]{ \centerline {\epsfig{file=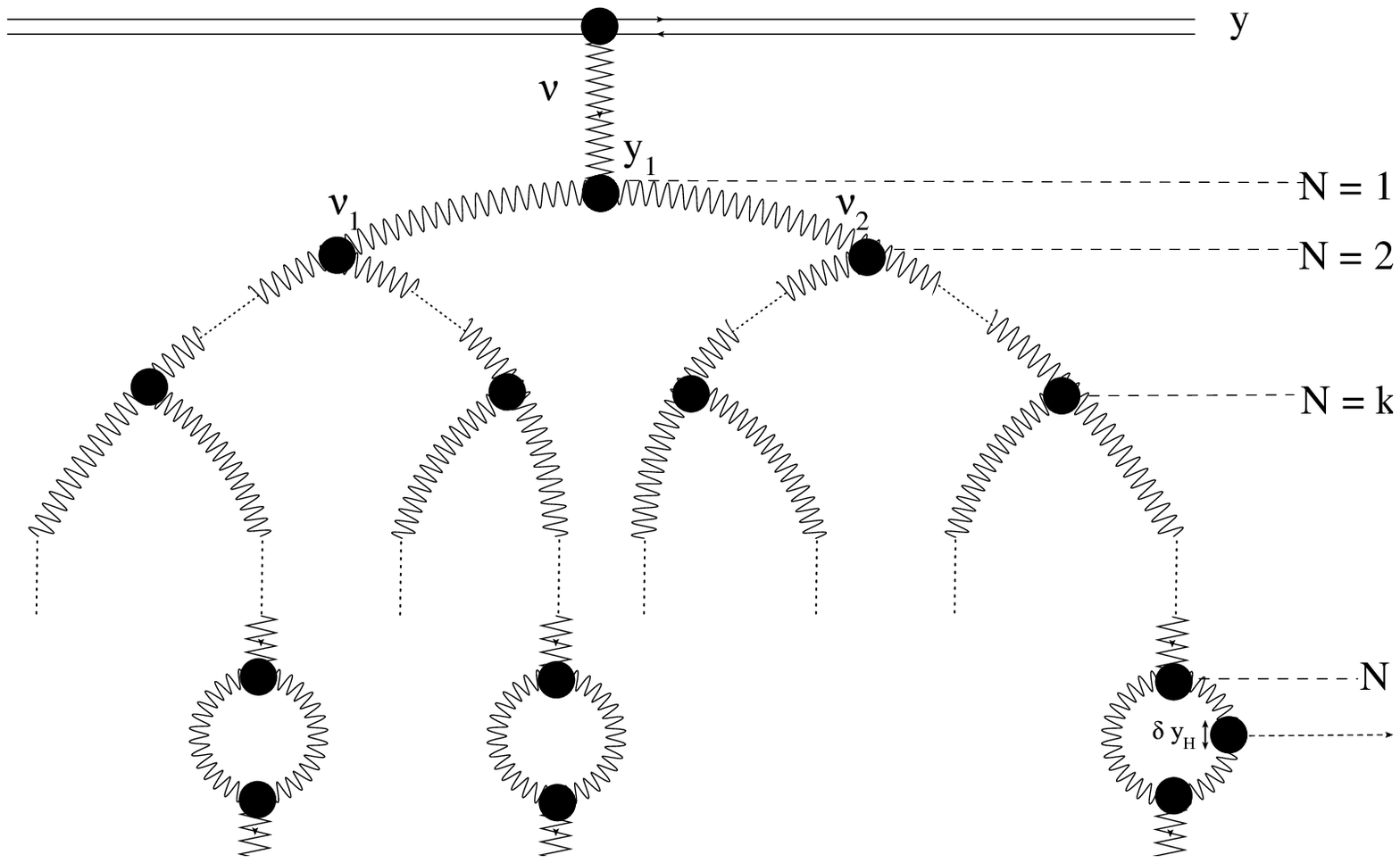,width=120mm}}\caption{ The diagram with $N$ generations of Pomeron branching, leading to $2^N-1$ loops.}  \label{figN}}

\bea
A_{(N)}\Lb \De,\de y_H|\mbox{\fig{figN}}\Rb
\!&&\!=\!\f{\as^2}{4}\int^\infty_{-\infty}\!\!  d\nu h\Lb\nu\Rb \la^2\Lb\nu\Rb e^{ \omega\Lb\nu\Rb \De}d_{(N)}\Lb\nu | \De,\de y_H\Rb
E_{\nu}E^{\,\prime}_{\,-\,\nu}A_H\hspace{1cm}\label{N1}\eea

where the amplitude for the $N$ generations of loops $d_{(N)}\Lb\nu | \De,\de y_H\Rb$ is found from the iterative equation;

 \bea
d_{(N)}\Lb\nu | \De,\de y_H\Rb&&=\f{1}{2^7}\int^\infty_{-\infty}\!\! d\nu_1\,h\Lb\nu_1\Rb\,\la^2\Lb\nu_1\Rb
\int^\infty_{-\infty}\!\! d\nu_2 h\Lb\nu_2\Rb \la^2\Lb\nu_2\Rb \begin{vmatrix}\Ga\Lb \nu | \nu_1 , \nu_2\Rb\end{vmatrix}^2 e^{-\omega\Lb\nu\Rb\de y_H} \nn\\
 &&\times e^{ \omega\Lb\nu_2\Rb \de y_H}\int^y_{y^{\,\prime}+\de y_H}\!\!\!\!\!\!dy_1
\int^{y_1-\de y_H}_{y^{\,\prime}}\!\!\!\!\!\!dy_1^{\,\prime}
e^{\Lb \omega\Lb \nu_1\Rb + \omega\Lb \nu_2\Rb -\omega\Lb\nu\Rb \Rb\Lb \De_1 - \de y_H \Rb }\,\nn\\
&&\times
d_{(N-1)}\Lb\,\nu_1\,|\,\De_1,\de y_H\Rb\,d_{(N-1)}\Lb\nu_2\,|\,\De_1\,\Rb
\hspace{3.5cm}\forall\,\,\,N\,\geq\,1\,. \label{II}\eea

The expression for $d_{(1)}\Lb\nu | \De,\de y_H\Rb$, $d_{(2)}\Lb\nu | \De,\de y_H\Rb$ and $d_{(3)}\Lb\nu | \De,\de y_H\Rb$ for the diagrams of \fig{fPomeronloop2}, \fig{fembedded} and \fig{fn},
were derived using the iterative expression of \eq{II}. Continuing this process using identical techniques, from observing the outcome expression for $d_{(N)}\Lb\nu | \De,\de y_H\Rb$ for higher values of $N$,
a predictable sequence emerges. From an observation of this sequence, a general formula for $d_{(N)}\Lb\nu | \De,\de y_H\Rb$ can be written for the contribution of region I: $\Lb\,\left\{\nu,|\nu_1|,|\nu_2|\right\}=\left\{0,1/2,1/2\right\}\,\Rb$, region II:\\ $\Lb\,\left\{|\nu|,\nu_1,\nu_2\right\}=\left\{1/2,0,0\right\}\,\Rb$  and region III: 
$\Lb\,\left\{|\nu|,|\nu_1|,|\nu_2|\right\}=\left\{1/2,1/2,1/2\right\}\,\Rb$. This expression can be proved by induction, whereby provided the formula for $d_{(N)}\Lb\nu | \De,\de y_H\Rb$  is true for $N=2$, and 
yields the predicted formula for $N+1$ by plugging  $d_{(N)}\Lb\nu | \De,\de y_H\Rb$   into \eq{II}, then the proof is complete.  In this formalism the following expressions for the contribution of the above  three regions are derived
for  $d_{(N)}\Lb\nu | \De,\de y_H\Rb$ ;

\bea
d^{\,\footnotesize{\mbox{I}}}_{(N)}\Lb \nu|\De,\de y_H\Rb\,
&&=\f{\Lb d\,d^{\,\prime}\Rb^{2^{[N-1]}}}{d^{\,\prime}}\Lb\h+i\nu\Rb^3\Lb\h-i\nu\Rb^3\chi\Lb\nu\Rb\,e^{-\omega\Lb\nu\Rb \de y_H}\nn\\
&&\times
\prod^{N}_{k=2}\bas^{\Lb k-2\Rb2^{[N-k]}}
\Lb\f{k+1}{k}-\f{\Lb k+3\Rb\Lb k+2\Rb\Lb k+1\Rb}{2 k^2}\Rb^{2^{[N-k]}-1}\,\nn\\
&& \times\prod^{N}_{k=2}\Lb\f{k+1}{k}-\f{\Lb k+3\Rb\Lb k+2\Rb\Lb k+1\Rb}{2 k^2}+\f{k+1}{k^2}\bas\de y_H\Rb\,\nn\\
&&\times\,\,\omega^{N-1}\Lb \nu\Rb\Lb\omega\Lb \nu\Rb+\f{\omega^2\Lb \nu\Rb\Lb\De_1-\de y_H\Rb}{N+1}\Rb\,\hspace{4.5cm}\forall N\geq 1;\label{dIgeneral}\\
\nn\\
\nn\\
d^{\,\footnotesize{\mbox{II}}}_{(N)}\Lb\nu| \De,\de y_H\Rb&&=
 \f{a^{\,\prime}d^{2^{[N-1]}} \Lb d^{\,\prime}\Rb^{2^{[N-2]}}}{\Lb 1/2+i\nu\Rb\Lb 1/2-i\nu\Rb}\!
f\Lb N-1,\de y_H\Rb\! f\Lb N-1\Rb\,e^{-\omega\Lb 0\Rb\de y_H}\!\nn\\
&&\times\!\!\int^y_{y^{\,\prime}+\de y_H}\!\!\!\!\! dy_1\int^{y_1-\de y_H}_{y^{\,\prime}}\!\!\!\!\! dy_1^{\,\prime}
\f{e^{\Lb2 \omega\Lb 0\Rb-\omega\Lb\nu\Rb\Rb\De_1 } }{\Lb\De_1\Lb\De_1-y_H\Rb\Rb^{3/2}}
\nn\\
&&\times \chi^2\Lb0\Rb\omega^{2N-4}\Lb0\Rb\!\Lb\!\omega\Lb0\Rb+\f{\omega^2\Lb0\Rb\Lb\De_1-\de y_H\Rb}{N}\!\Rb\!
\Lb\!\omega\Lb0\Rb+\f{\omega^2\Lb0\Rb\De_1}{N}\!\Rb\,\hspace{0.4cm}\forall N\geq 2;\hspace{0.9cm}\label{dIIgeneral}\eea

where the function $f\Lb N,\de y_H \Rb$ is defined by;

\bea
f\Lb N,\de y_H\Rb&&=\prod^{N}_{k=2}\,\bas^{\Lb k-2\Rb2^{[N-k]}}
\Lb\f{k+1}{k}-\f{\Lb k+3\Rb\Lb k+2\Rb\Lb k+1\Rb}{2 k^2}\Rb^{2^{[N-k]}-1}\,\nn\\
&&\times \prod^{N}_{k=2}\,\Lb\f{k+1}{k}-\f{\Lb k+3\Rb\Lb k+2\Rb\Lb k+1\Rb}{2 k^2}+\f{k+1}{k^2}\,\bas\,\de y_H\Rb\,;\label{definitionoff}\eea

and  $f\Lb N-1\Rb=f\Lb N-1,\de y_H=0\Rb$ can be read off \eq{definitionoff} by setting $\de y_H=0$, and

\bea
d^{\mbox{\footnotesize{III}}}_{(N)}\Lb\nu|\De,\de y_H\Rb&&=
\f{\Lb ab\Rb^{2^{[N-1]}}}{b}e^{-\omega\Lb 0\Rb\de y_H}\int^y_{y^{\,\prime}+\de y_H}\!\!\!\!\! dy_1\int^{y_1-\de y_H}_{y^{\,\prime}}\!\!\!\!\! dy_1^{\,\prime} \De_1^{2^{N}-N-1}\Lb\De_1-\de y_H\Rb^{N-1}e^{-\omega\Lb\nu\Rb\De_1}\nn\\
&&\times 
\left\{\!\Lb\f{-1}{\bas}\f{d}{d\De_1}\Rb^3\!\f{e^{2\omega\Lb 0\Rb\De_1}}{\De_1^{3/2}\Lb\De_1-\de y_H\Rb^{3/2}}\!\right\}\!
\left\{\!\Lb\f{-1}{\bas}\f{d}{d\De_1}\Rb^3\!\f{e^{2\omega\Lb 0\Rb\De_1}}{\De_1^3}\!\right\}^{2^{[N-1]}-1}\hspace{0.5cm}\forall N\geq 2\,.\hspace{0.5cm}\label{dIIIgeneral}\eea

where the full set of constants are;

\bea
d&&=\bas\Lb 1-\f{1}{N_c^2}\Rb\Lb 1-\f{2}{N_c^2}\Rb;\hspace{1cm}d^{\,\prime}=\f{\bas^2}{2^{11}}\Lb 1-\f{1}{N_c^2}\Rb\Lb 1-\f{2}{N_c^2}\Rb;\nn\\
a&&=\f{2^9\bas^4}{N_c^4\pi[\omega^{\,\prime\prime}\Lb 0\Rb]^3};\hspace{3cm}a^{\,\prime}=\f{\bas^4}{2^6N_c^4\pi[\omega^{\,\prime\prime}\Lb 0\Rb]^3};\hspace{1cm}
b=\f{\bas^2}{2^{10}}\Lb 1-\f{1}{N_c^2}\Rb^2\,.\label{fullsetofconstants}
\eea

Inserting the formulae of \eq{dIgeneral}, \eq{dIIgeneral} and \eq{dIIIgeneral} into \eq{N1}, one finds the following expressions for the contributions of regions I, II and III to the
 scattering amplitude for diffractive Higgs production, from the $N$ generation diagram of
\fig{figN};

\bea
A^{\,\footnotesize{\mbox{I}}}_{(N)}\Lb \De,\de y_H|\mbox{\fig{figN}}\Rb\,
&&=\f{\Lb 2\pi\Rb^{1/2}\bas^2A_H}{128\pi^2N_c^2}\f{\Lb d\, d^{\,\prime}\Rb^{2^{[N-1]}}}{d^{\,\prime}}\,\f{e^{\,\omega\Lb 0\Rb\,\Lb\De-\de y_H\Rb}}{\Lb\omega^{\,\prime\,\prime}\Lb 0\Rb\Lb\De-\de y_H\Rb \Rb^{3/2}}
\,\chi\Lb 0\Rb \nn\\
&&\times
\prod^{N}_{k=2}\bas^{\Lb k-2\Rb2^{[N-k]}}
\Lb\f{k+1}{k}-\f{\Lb k+3\Rb\Lb k+2\Rb\Lb k+1\Rb}{2 k^2}\Rb^{2^{[N-k]}-1}\,\nn\\
&& \times\prod^{N}_{k=2}\Lb\f{k+1}{k}-\f{\Lb k+3\Rb\Lb k+2\Rb\Lb k+1\Rb}{2 k^2}+\f{k+1}{k^2}\bas\de y_H\Rb\,\nn\\
&&\times\,\,\omega^{N-1}\Lb 0\Rb\Lb\omega\Lb 0\Rb+\f{\omega^2\Lb 0\Rb\Lb\De-\de y_H\Rb}{N+1}\Rb\hspace{4cm}\forall N\geq 1;\hspace{0.5cm}\label{dn10}\\
\nn\\
\nn\\
A^{\,\footnotesize{\mbox{II}}}_{(N)}\Lb \De,\de y_H|\mbox{\fig{figN}}\Rb&&=\!\f{\bas A_H}{2^9\pi N_c^2}
 a^{\,\prime}d^{2^{[N-1]}} \Lb d^{\,\prime}\Rb^{2^{[N-2]}}\!\nn\\&&\times
f\Lb N-1,\de y_H\Rb\! f\Lb N-1\Rb\!\chi^2\Lb0\Rb\omega^{2N-4}\Lb0\Rb\,\Lb\De-\de y_H\Rb\nn\\
&&\times\!\Lb\f{-1}{\bas}\f{d}{d\De}\Rb^3\!\!\left\{\!
\f{e^{2 \omega\Lb 0\Rb\Lb \De-\de y_H/2\Rb } }{\Lb\De\Lb\De-y_H\Rb\Rb^{3/2}}
\Lb\omega\Lb0\Rb+\f{\omega^2\Lb0\Rb\Lb\De-\de y_H\Rb}{N}\Rb\right.\nn\\
&&\hspace{2.2cm}\left.
\Lb\omega\Lb0\Rb+\f{\omega^2\Lb0\Rb\De}{N}\Rb\! \right\}\hspace{5cm}\forall N\geq 2;\hspace{0.5cm}\label{neededformula!}\\
\nn\\
\nn\\
A^{\mbox{\footnotesize{III}}}_{(N)}\Lb\De,\de y_H|\mbox{\fig{figN}}\Rb&&=\f{\bas\,A_H}{2^9 N_c^2\pi}
\f{\Lb ab\Rb^{2^{[N-1]}}}{b}e^{-\omega\Lb 0\Rb\de y_H}\De^{2^{N}-N-1}\Lb\De-\de y_H\Rb^N\nn\\
&&\times 
\left\{\Lb\f{-1}{\bas}\f{d}{d\De}\Rb^3\f{e^{2\omega\Lb 0\Rb\De}}{\De^{3/2}\Lb\De-\de y_H\Rb^{3/2}}\!\right\}\!
\left\{\!\!\Lb\f{-1}{\bas}\f{d}{d\De}\Rb^3\!\!\f{e^{2\omega\Lb 0\Rb\De}}{\De^3}\!\!\right\}^{2^{[N-1]}-1}\forall N\geq 2;\hspace{0.9cm}\label{nonint7}\eea

\eq{dn10} and \eq{neededformula!} are the contributions to the amplitude of the $N$ generation diagram of \fig{figN} which leads to the
renormalization of the Pomeron intercept. \eq{nonint7} is the contribution to \fig{figN} which is equivalent to the
amplitude of $2^N$ non interacting Pomerons, with renormalized Pomeron vertices.  This can be seen by observing that \eq{nonint7} can be written in the form; 

\bea
A^{\mbox{\footnotesize{III}}}_{(N)}\Lb\De,\de y_H|\mbox{\fig{figN}}\Rb&&=k_{(N)}e^{2^N\omega\Lb 0\Rb\De}\label{nonint8}
\eea

where $k_{(N)}$ includes all other terms contained in the amplitude.  The complete expression for the scattering amplitude of \fig{figN} is the sum of the contributions of \eq{dn10}, \eq{neededformula!} and \eq{nonint7}, namely;

\bea
&&A_{(N)}\Lb\De,\de y_H|\mbox{\fig{figN}}\Rb\nn\\
&&\hspace{1cm}=A^{\mbox{\footnotesize{I}}}_{(N)}\Lb\De,\de y_H|\mbox{\eq{dn10}}\Rb+
A^{\mbox{\footnotesize{II}}}_{(N)}\Lb\De,\de y_H|\mbox{\eq{neededformula!}}\Rb+
A^{\mbox{\footnotesize{III}}}_{(N)}\Lb\De,\de y_H|\mbox{\eq{nonint7}}\Rb\label{thecompletegeneralamplitude}\eea

\TABLE[ht]{
\begin{tabular}{||c|c||}
\hline \hline
&\\ 
   & $A_{(N)}\Lb \De=19,\de y_H=\ln\Lb M_H^2/4 s_0\Rb \Rb$ \\
 & $\as=0.12$\hspace{1.8cm}$\as=0.2$\\
 \hline
$N=0$ & $1.57\times10^{-8}\,\,\,\,$$\hspace{1cm}\,\,2.17\times10^{-7}$ \\ \hline
$N=1$&   $2.05\times10^{-10}\,\,$$\hspace{1cm}\,\,2.15\times10^{-7}$\\ \hline
$N=2$   
& $2.85\times10^{-15}\,\,$$\hspace{1cm}\,\,1.39\times10^{-7}$ \\
\hline
$N=3$  
 & $3.40\times10^{-24}\,\,$$\hspace{1cm}\,\,1.28\times10^{-7}$\\
 \hline
$N=4$  
& $8.20\times10^{-42}\,\,$$\hspace{1cm}\,\,2.31\times10^{-8}$\\ \hline
$N=5$  
 & $8.13\times10^{-77}\,\,$$\hspace{1cm}\,\,5.11\times10^{-9}$ \\
\hline
$N=6$ 
 & $1.36\times10^{-146}\,\,$$\hspace{1cm}4.26\times10^{-10}$   \\ 
\hline \hline
\end{tabular}
  \caption{
 Results for the scattering amplitude for diffractive Higgs
production from the multi Pomeron loop diagram with $N$ generations of Pomeron branching, for the contribution of regions I , II and III. The mass of the Higgs boson
is assumed to be $M_H=100\,\mbox{GeV}$, and the rapidity gap $\De$ between the scattering protons is taken to be $\De=19$, based on
proton proton collisions at the typical LHC energy $\sqrt{s}=14$ TeV.
 }\label{t3}}

 \Tab{t3} lists the results for the multiple Pomeron loop amplitudes for $\as=0.12$ and $\as=0.2$. The greatest contribution
to Pomeron enhanced diagrams stems from region III, which leads to the contribution to the amplitude which is equivalent to the amplitude of non interacting Pomerons. 
The values in the table indicate that the amplitude of the $N$ generation diagram becomes smaller as $N$ grows, for energies within the LHC range. The difference in values for different values of
$\as$ show that the amplitude is very sensitive to the choice of the Pomeron intercept.
The sum over Pomeron loop diagrams from $N=0$ (so that the basic diagram of \fig{f1p} is included), takes the following form;

\vspace{2.4cm}

\bea
\Sigma\Lb\De,\de y_H\Rb&&=A_{(0)}\Lb\De,\de y_H|\mbox{\fig{f1p}}\,;\,\mbox{\eq{3(1)}}\Rb+\sum_{N=1}\Lb -1\Rb^NA_{(N)}\Lb\De,\de y_H|\mbox{\eq{thecompletegeneralamplitude}}\Rb\hspace{1cm}
\label{sopl}\eea

\FIGURE[h]{ \centerline {\epsfig{file=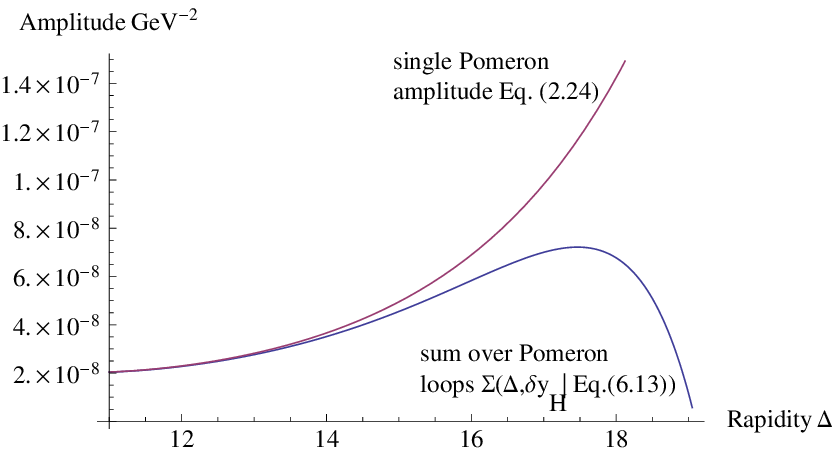,width=120mm}}\caption{ Plot for the single Pomeron amplitude $A_{(0)}\Lb\De,\de y_H|\mbox{\fig{f1p}}\Rb$ derived in \eq{3(1)} (upper line), next
to the plot for the sum over Pomeron loops  $\Sigma\Lb\De,\de y_H|\,\mbox{\eq{sopl}}\Rb$ up to $N=20$ (lower line), against the rapidity separation 
$\De = y-y^{\,\prime}$ between the scattering
protons. The values of $\De$ go from $\de y_H\,=\,\ln\Lb M_H^2/4s_0 \Rb\,$ (assuming
the Higgs mass $M_H\,=\, 100$ GeV and $s_0=1\,\mbox{GeV}^2$), up to the typical LHC rapidity $\De=19$. $\as = 0.2$. }  \label{fsIIgraph}}

\fig{fsIIgraph} shows the plot for the amplitude 
of the single Pomeron diagram derived in \eq{3(1)}, for the basic diagram shown in \fig{f1p} (upper line), next to the plot for the sum over Pomeron loops $\Sigma\Lb\De,\de y_H|\mbox{\eq{sopl}}\Rb$ up to $N=20$
(lower line),
 against the rapidity separation $\De$ between the incoming projectiles. From an observation of \fig{fsIIgraph} it is clear that
for rapidity values approaching the typical LHC range $\De=19$,  $\Sigma\Lb\De,\de y_H|\mbox{\eq{sopl}}\Rb$
starts to be substantially less than the basic single Pomeron amplitude, as the two graphs grow further apart. This
proves that the shadowing correction of Pomeron loops to the basic diagram of \fig{f1p}, is large within the LHC range of energies. 
As the rapidity separation between the scattering protons increases, the effect of this shadowing correction increases.\\

\newpage

\section{The survival probability for exclusive Higgs production}
\label{ssurvivalprobability}

This section is limited to an estimate of the contribution of enhanced diagrams to the survival probability $\SPE$
 of large rapidity gaps,
 in exclusive diffractive Higgs production shown in \fig{f1p}.
In \sec{s4} the sum over Pomeron enhanced diagrams in \eq{sopl} was derived in the exact QCD approach, thanks to the expressions of \eq{dn10} , \eq{neededformula!}  and \eq{nonint7}.
This means that the complete set of
hard re-scattering contributions to the enhanced survival probability can be estimated in QCD to any order, without relying on the mean field approximation approach.\\

The enhanced survival probability $\SPE$
  is estimated by
subtracting from the basic diagram of \fig{f1p}, the first enhanced diagram of  \fig{fPomeronloop2},
 and subtract from this the second enhanced diagram of \fig{fembedded}\, and so on. Finally divide by the amplitude of the basic diagram of \fig{f1p} to give the correctly normalized survival probability as the expression;

\bea
&&\SPE\nn\\
&&\hspace{0.4cm}=\f{A_{(0)}\Lb\De,\de y_H|\mbox{\fig{f1p}}\Rb-A_{(1)}\Lb\De,\de y_H|\mbox{\fig{fPomeronloop2}}\Rb+A_{(2)}\Lb\De,\de y_H|\mbox{\fig{fembedded}}\Rb-\dots-A_{(N)}\Lb\De,\de y_H|\mbox{\fig{fn}}\Rb}{A_{(0)}\Lb\De,\de y_H|\mbox{\fig{f1p}}\Rb}
\nn\\
&&\hspace{0.4cm}=\sum^\infty_{N=0}\Lb -1\Rb^N\f{\,A_{(N)}\Lb \De,\de y_H\Rb}{A_{(0)}\Lb\De,\de y_H\Rb}\label{SPE}\eea

where $A_{(0)}\Lb\De,\de y_H|\mbox{\fig{f1p}}\Rb$ was calculated in \eq{3(1)} and all the subsequent terms $A_{(N)}\Lb\De,\de y_H\Rb\,\,\,\forall\,\,N\geq\,1$ 
are the sum over Pomeron loops given in \eq{dn10} - \eq{thecompletegeneralamplitude}

\FIGURE[h]{ \centerline {\epsfig{file=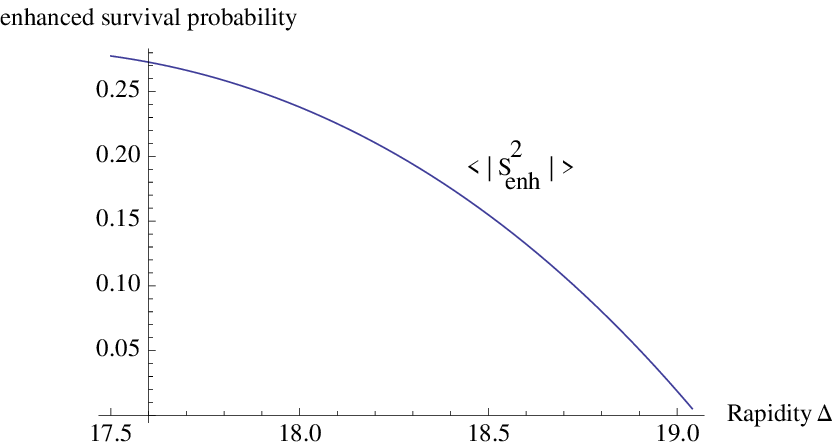,width=100mm}}\caption{ The enhanced survival probability $\SPE$ plotted against the rapidity separation $\De$ of the scattering protons. $\as = 0.2$. }  \label{fspgraph}}

The results for the enhanced survival probability $\SPE$ are shown in the graph of \fig{fspgraph} against the rapidity separation $\De$ between the incoming projectiles, where all the terms in \eq{SPE} up to $N=20$
have been taken into account.
The results of \fig{fspgraph} show that $\SPE$ is small, and could be even less than $1\%$ for the LHC range of rapidity values, in agreement with refs. \cite{Miller:2006bi,Gotsman:2008tr}. The observation from \fig{fspgraph} that $\SPE$ decreases
as the rapidity gap $\De$ between the scattering protons increases, also matches with the findings of refs.  \cite{Miller:2006bi,Gotsman:2008tr}.
However \fig{fspgraph} shows a steeper rise in $\SPE$ as the rapidity separation $\De$ 
decreases than the results found in refs.  \cite{Miller:2006bi,Gotsman:2008tr}, where the improved Mueller-Patel-Salam-Iancu (MPSI) approach was used.
We believe this difference in the slope of the graph for $\SPE$,  shows that there are more complicated diagrams to take into account other than those diagrams considered here, which would lead to a result for $\SPE$ which shows
a more gradual increase as the rapidity gap $\De$ decreases. This will be discussed in more detail in the conclusion section.

\section{Conclusion}
\label{sc}

The main achievements of this paper are the following;

\begin{enumerate}

\item
The observation that the major contribution to the Pomeron loop amplitude stems from the non planar piece of the Korchesky triple Pomeron vertex of \fig{fnonplanar}

\item
The derivation of the general analytical expression for the diagram with an arbitrary number of Pomeron loops, found in  \eq{dn10} - \eq{thecompletegeneralamplitude}. 

\item
The finding that the main contribution to the amplitude of multiple Pomeron loop diagrams, is equivalent to the
amplitude of non interacting Pomerons.

\item
The solution to the summation
over Pomeron loop diagrams, for the first time in the exact QCD approach.

\item
Estimation of the contribution of Pomeron enhanced diagrams to the survival probability $\SPE$ for diffractive Higgs production in the QCD formalism, instead of the MFA approach.
\end{enumerate}

The main achievements of this paper are the formulae of \eq{dn10} - \eq{thecompletegeneralamplitude}
 for the amplitude of the multiple Pomeron loop diagram, with $N$ generations of Pomeron branching as shown \fig{figN}. It was found that
 there are 3 regions
which contribute to Pomeron loop diagrams. Regions I and II lead to the renormalization of the Pomeron intercept. Region III leads to the contribution
to Pomeron loop diagrams, which is equivalent to the amplitude of non interacting Pomerons, with renormalized Pomeron vertices. The contribution from non interacting Pomerons,
 gives the greatest contribution to the amplitude
of Pomeron loop diagrams. The property of the equivalence of Pomeron loop diagrams to non interacting Pomerons, was first noticed by A. Mueller (see ref. \cite{toy}).  We showed in ref. \cite{Levin:2007wc}, 
that if the sum over Pomeron loops was performed in the QCD approach, then the summation should reduce to the sum over non interacting Pomerons diagrams. Therefore the
findings in this paper are in agreement with the results in the above mentioned papers.\\

The sum over Pomeron loop diagrams yields a significant shadowing correction to the basic single Pomeron diagram of \fig{f1p}. The shadowing correction however is only
significant for a rapidity separation between the incoming protons  $\De\geq14$, and the shadowing correction becomes larger as the rapidity separation grows (see \fig{fsIIgraph}).
We did not take into account the diagrams which contribute to the vertex in the framework of the
Schwinger-Dyson equation. However, since the intercept of the Pomeron $\Delta >0$ as was shown in refs. \cite{toy,Levin:2007wc}, these diagrams give a small contribution.\\

The application of the sum over Pomeron enhanced diagrams to an estimate of the enhanced survival probability $\SPE$, produced a result which shows that $\SPE$ is potentially
less than $1\%$  for a typical LHC rapidity separation $\De = 19$. From an observation of \fig{fspgraph}, $\SPE$ decreases as the rapidity separation $\De$ increases (see \fig{fspgraph}). Both 
of these findings are in agreement with refs. \cite{Miller:2006bi,Gotsman:2008tr}. This comparison shows that the summation over Pomeron enhanced diagrams in the formalism of this paper,  
reproduces the expected behavior of $\SPE$ with rapidity separation $\De$.  The estimates
for the LHC range of rapidity values are close to the previous estimates of refs. \cite{Miller:2006bi,Gotsman:2008tr} (i.e. less than 1\%). \\

The behavior of the enhanced survival probability as a function of the rapidity separation, is a property which stems from the definition of the survival probability.
That is the survival probability is 
a quantitative measure of the effect of shadowing corrections, that stem from inelastic scattering emerging from extra parton showers in the reaction proton + proton
 $\to$ proton + [LRG] + Higgs + [LRG] + proton, (where [LRG] denotes a large rapidity gap).
Increasing
the energy of this reaction leads to a rise in the number of extra parton showers, which spoil the large rapidity gaps as shown in \fig{flrg}.
This leads to the natural conclusion that as the rapidity separation between the scattering protons increases, the survival probability decreases as more unwanted parton showers arise.\\

\DOUBLEFIGURE[ht]{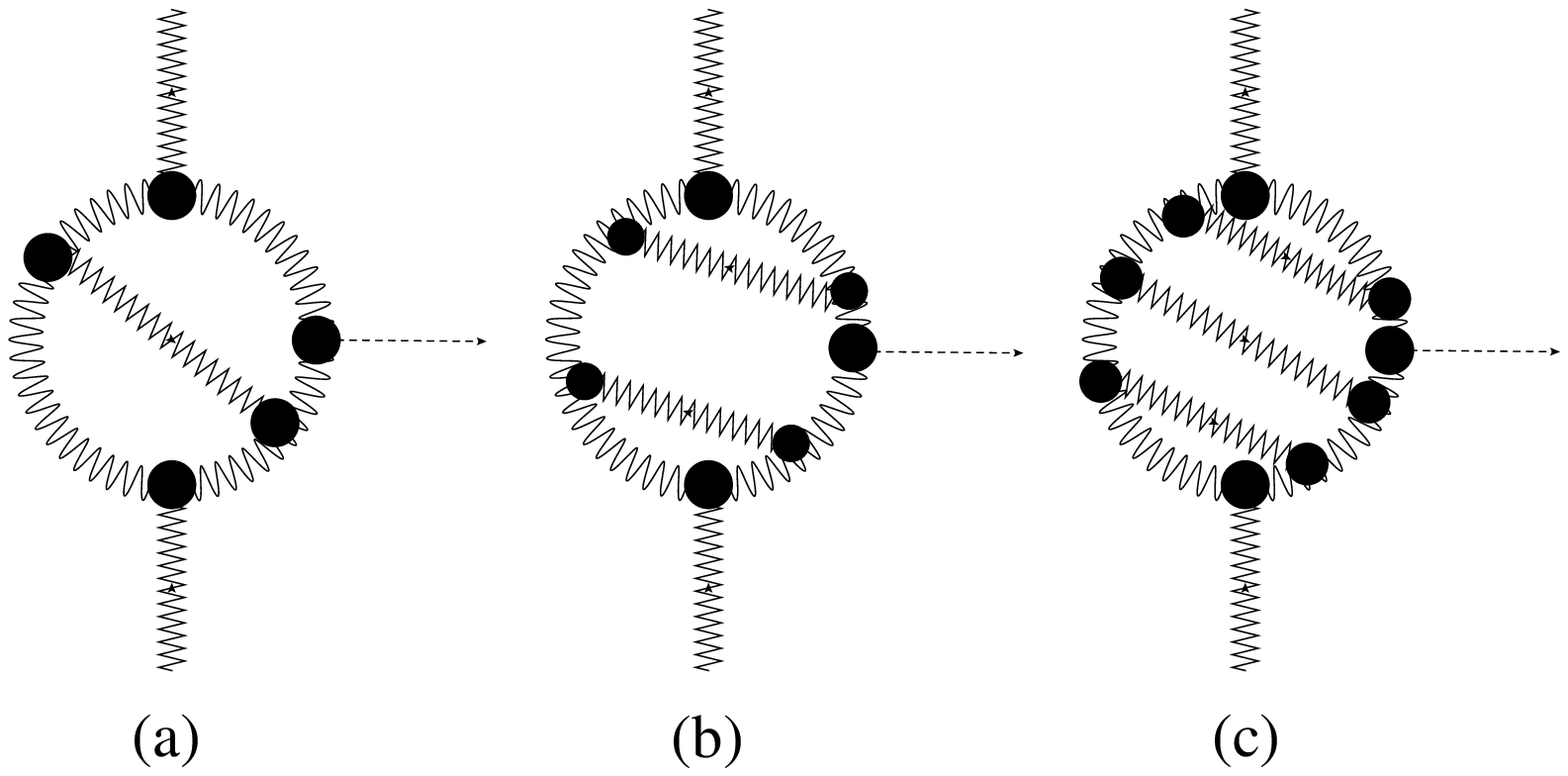,width=85mm,height=50mm}{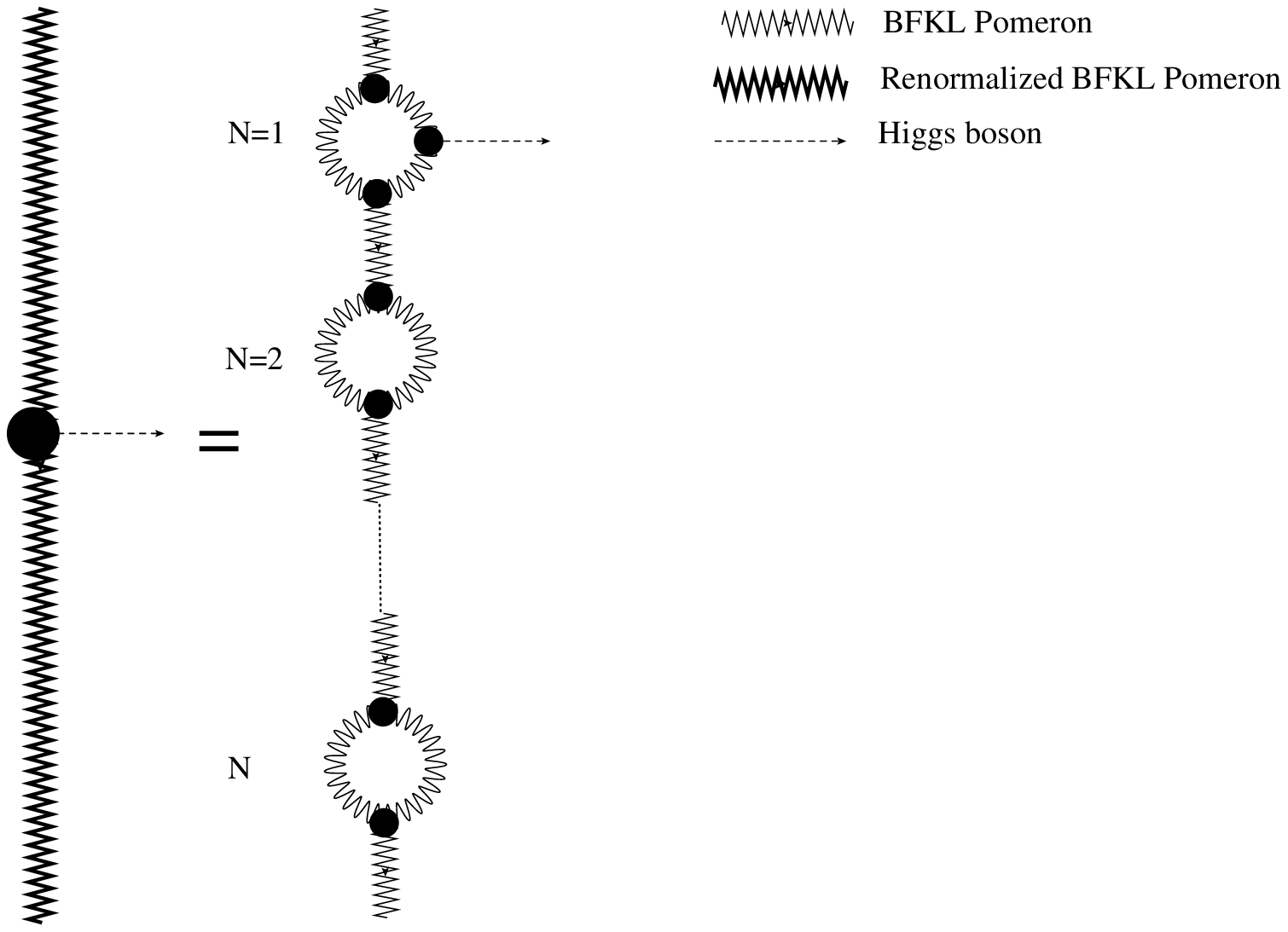,width=75mm,height=60mm}
{More complicated Pomeron enhanced diagrams that have not yet been taken into account. (a) shows a diagram with $N=2$ generations of Pomeron branching, (b) shows $N=3$ and (c) shows $N=4$ generations
of branching.  
 \label{forbital}}
{Schematic representation of loop corrections that leads to the renormalized BFKL Pomeron propagator.\label{fren}}

The slope of $\SPE$ of \fig{fspgraph} with the rapidity separation is steeper than the findings of refs. \cite{Miller:2006bi,Gotsman:2008tr},
which relied on an improved (MPSI) model. This observation we believe shows that the
summation over Pomeron loops in this paper, while  this is an important step forward, is possibly not yet complete.\\

For example, more complicated enhanced diagrams such as those shown in \fig{forbital}  could provide an important contribution to the enhanced survival probability, which would lead
to a behavior with a less steep increase with rapidity separation. The diagrams in \fig{forbital} are not topologically equivalent to any of the other
enhanced diagrams discussed in this paper. Since multiple Pomeron loop diagrams decrease with the number of generations of branching $N$, diagrams (a), (b) and (c) in 
\fig{forbital} which belong to the $N=2$, $N=3$ and $N=4$ class of loop diagrams are not expected to be larger than the simple Pomeron loop diagram of \fig{fPomeronloop2}, however their contribution
to the summation over enhanced diagrams could affect the slope of \fig{fspgraph} to make it less steep. 
There is also the issue of the renormalization of the BFKL Pomeron which needs to be taken into account. For example diagrams such as \fig{fren}, which although are not expected to be large for any value of $N$  
could still provide an important contribution to $\SPE$, which would make the slope with rapidity separation not as steep.

We hope that the findings of this paper will be useful for experiments aimed at discovering the Higgs boson at the LHC. We also hope that the
methods developed in this paper will provide an important foundation for future calculations of BFKL Pomeron loop diagrams.

\section{Acknowledgements}
\label{sack}

This paper is dedicated to my Grandmother. I would like to
thank E. Levin for helpful advice in writing this paper. I would
also like to thank L. Apolin$\acute{a}$rio, M. Braun, J. Dias De Deus, and A. Vale  for
fruitful discussions on the subject. I would also like to thank my wife, Sharon.
This research was supported by the Funda\c{c}$\tilde{a}$o para ci$\acute{e}$ncia e a tecnologia (FCT), and CENTRA - Instituto Superior T$\acute{e}$cnico (IST), Lisbon.

\appendix

\section{The triple Pomeron vertex}
\label{sa}

In this section of the appendix the derivation of the triple Pomeron vertex, which was first done by Korchemsky in ref. \cite{Korchemsky:1997fy} will be outlined. The triple Pomeron vertex is the vertex of three BFKL Pomeron states, which couple either as shown in the planar diagram of \fig{fplanar},
or they couple in the way shown in the non planar diagram of \fig{fnonplanar}.
The coupling of three BFKL Pomeron states with the center of mass coordinates $R$, $R_1$ and $R_2$ and the conformal weights $\ga$, $\ga_1$
and $\ga_2$ is given by for the planar and non-planar diagrams respectively;

\bea
\Omega\Lb\nu\,|\,\nu_1,\nu_2|\fig{fplanar}\Rb\,&&=\,\int\!d^2R\int\!d^2R_1\int\!d^2R_2\f{1}{|R_{01}|^2|R_{12}|^2|R_{20}|^2}\,\, E_{\ga}\,E_{\ga_1}\,E_{\ga_2}
\label{V0}\\
\La\Lb\nu\,|\,\nu_1,\nu_2|\fig{fnonplanar}\Rb\,&&=\,\int\!d^2R_0\int\!d^2R_1\f{1}{|R_{01}|^4}\,\,E_{\ga}\,E_{\ga_1}\,E_{\ga_2}
\label{V1}
\eea

The leading contribution to the high energy scattering amplitude stems from the region where all the conformal spins vanish, namely $n=n_1=n_2=0$.
Adding together  \eq{V0} and \eq{V1} and integrating over the
complex coordinates, the triple Pomeron vertex takes the form;

\bea
&&\Ga\Lb \nu\,|\,\nu_1,\nu_2\Rb=\Lb\f{\as N_c}{\pi}\Rb^2\,\left\{ 4\nu^2+1\right\}^2\,\Lb\Omega\Lb \nu|\nu_1,\nu_2\Rb+\f{2\pi}{N_c^2}\La\Lb \nu|\nu_1,\nu_2\Rb
\left\{\chi\Lb\nu\Rb-\chi\Lb\nu_1\Rb-\chi\Lb\nu_2\Rb\right\}\Rb\hspace{0.8cm}\label{V5lo}\eea

where $\chi\Lb\nu\Rb$ is the function;

\bea
\chi\Lb\nu\Rb\,=\,\Re e\left\{\psi\Lb 1\Rb-\psi\Lb\h+i\nu\Rb\right\}\label{chiappendix}\eea

and where
\bea\Omega\Lb \nu\,|\,\nu_1,\nu_2\Rb\,&&=\,\pi^3\,\left\{\Ga^{2}\Lb\,\h+i\nu\,\Rb\,\Ga^{2}\Lb\,\h+i\nu_1\,\Rb\,\Ga\Lb\,\h-i\nu\,\Rb\,\Ga\Lb\,\h-i\nu_1\,\Rb\,\Ga\Lb\,\h-i\nu_2\,\Rb\right\}^{-1}\nn\\
&&\times\,J^a\Lb\,\nu\,,\,\nu_1\,,\,\nu_2\,\Rb\,\tilde{J}_a\Lb\,\nu\,,\,\nu_1\,,\,\nu_2\,\Rb\label{a}\eea

where the contracted $a$ index implies summation over $a\,=1,2,3$.
The integrals $J_a\Lb\,\nu\,,\,\nu_1\,,\,\nu_2\,\Rb$ and $\tilde{J}_a\Lb\nu\,,\,\nu_1\,,\,\nu_2\,\Rb$
were evaluated in ref. \cite{Korchemsky:1997fy}, where they were found to be;

\bea J_1\Lb \nu,\nu_1,\nu_2\Rb\,
&&=\,\Ga\Lb\,\h+i\nu+i\nu_1-i\nu_2\,\Rb\,\Ga\Lb\,\h+i\nu\,\Rb\Ga\Lb\,\h-i\nu\,\Rb\Ga\Lb\,\h+i\nu_1\,\Rb\Ga\Lb\,\h-i\nu_1\,\Rb\nn\\
&&\times\,\int^1_0dx\,\Lb 1-x\Rb^{-1/2-i\nu_2}\,_2F_1\Lb\h+i\nu,\h-i\nu;1|x\Rb\,_2F_1\Lb\h+i\nu_1,\h-i\nu_1;1|x\Rb \nn\\
J_2\Lb \nu,\nu_1,\nu_2\Rb\,
&&=\,\Ga\Lb\,\h+i\nu+i\nu_1-i\nu_2\,\Rb\,\Ga\Lb\h+i \nu\Rb\,\Ga\Lb \h-i\nu\Rb\,\Ga\Lb \h+i\nu_1\Rb\,\Ga\Lb \h-i\nu_1\Rb\nn\\
&&\times\,\f{\Ga^2\Lb \h-i\nu_2\Rb}{\Ga\Lb 1+i\nu-i\nu_2\Rb\,\Ga\Lb 1-i\nu-i\nu_2\Rb}\nn\\
&&\times\,_4F_3\Lb\h+i \nu_1\,,\,\h-i\nu_1\,,\,\h-i\nu_2\,,\,\h-i\nu_2\,|\,1+i\nu-i\nu_2\,,\,1-i\nu-i\nu_2\,,\,1\,|\,1\,\Rb\hspace{1cm}\nn\\
J_3\Lb \nu,\nu_1,\nu_2\Rb\,
&&=\,\Ga\Lb\,\h+i\nu+i\nu_1-i\nu_2\,\Rb\,\Ga\Lb\h+i \nu\Rb\,\Ga\Lb \h-i\nu\Rb\,\Ga\Lb\h+i \nu_1\Rb\,\Ga\Lb \h-i\nu_1\Rb\nn\\
&&\times\,\f{\Ga^2\Lb \h-i\nu_2\Rb}{\Ga\Lb 1+i\nu_1-i\nu_2\Rb\,\Ga\Lb 1-i\nu_1-i\nu_2\Rb}\nn\\
&&\times\,_4F_3\Lb \h+i\nu\,,\,\h-i\nu\,,\,\h-i\nu_2\,,\,\h-i\nu_2\,|\,1+i\nu_1-i\nu_2\,,\,1-i\nu_1-i\nu_2\,,\,1\,|\,1\,\Rb\hspace{1cm}\nn\\
\tilde{J}_1\Lb\,\nu,\nu_1,\nu_2\Rb\,
&&=\,\f{\Ga\Lb\h+i\nu\Rb\,\Ga^2\Lb\h-i\nu\Rb\,\Ga\Lb\h+i\nu_1\Rb\,\Ga^2\Lb\h-i\nu_1\Rb}{\Ga\Lb\h-i\nu_2\Rb\,\Ga\Lb1/2-i\nu-i\nu_1+i\nu_2\Rb}\nn\\
&&\times\,\int^1_0dx\,x^{\,\,-1/2-i\nu_2}\,\Lb 1-x\Rb^{-1/2+i\nu_2-i\nu-i\nu_1}\,_2F_1\Lb \h-i\nu,\h-i\nu;1|x\Rb\,_2F_1\Lb \h-i\nu_1,\h-i\nu_1;1|x\Rb\nn\\
\tilde{J}_2\Lb\,\nu,\nu_1,\nu_2\Rb\,
&&=\f{\Ga\Lb\h+i\nu_1\Rb\Ga\Lb\h+i\nu_2\Rb\Ga\Lb\h-i\nu\Rb\Ga^2\Lb\h-i\nu_1\Rb}{\Ga\Lb 1/2-i\nu-i\nu_1+i\nu_2\Rb}\nn\\
&&\times\,\int^1_0\Lb 1-x\Rb^{-1/2-i\nu-i\nu_1 +i\nu_2}\,_2F_1\Lb \h-i\nu,\h-i\nu;1|x\Rb\,_2F_1\Lb \h-i\nu_1,\h-i\nu_1;1|x\Rb\hspace{1cm}\nn\\
\tilde{J}_3\Lb\,\nu,\nu_1,\nu_2\Rb\,
&&=\f{\Ga\Lb\h+i\nu\Rb\Ga\Lb\h+i\nu_2\Rb\Ga\Lb\h-i\nu_1\Rb\Ga^2\Lb\h-i\nu\Rb}{\Ga\Lb 1/2-i\nu-i\nu_1+i\nu_2\Rb}\nn\\
&&\times\,\int^1_0\Lb 1-x\Rb^{-1/2-i\nu-i\nu_1 +i\nu_2}\,_2F_1\Lb \h-i\nu,\,\h-i\nu;1|x\Rb\,_2F_1\Lb \h-i\nu_1,\h-i\nu_1;1|x\Rb\label{1J2(a)}\eea


For the non planar piece;

\bea
\La\Lb\,\nu\,|\,\nu_1\,,\,\nu_2\,\Rb\,
&&=\,2\pi^2\,\f{\,\Ga\Lb \h-i\nu_2\Rb\,\Ga\Lb\h+ i\nu+i\nu_1-i\nu_2\Rb\,}{\Ga\Lb \h+i\nu\Rb\,\Ga\Lb \h+i\nu_1\Rb\Ga\Lb 1+i\nu-i\nu_2\Rb\,\Ga\Lb 1-i\nu-i\nu_2\Rb}\nn\\
&&\times\,_3F_2\Lb\,\h+i\nu_1\,,\,\h-i\nu_1\,,\,\h-i\nu_2\,|\,1+i\nu-i\nu_2\,,\,1-i\nu-i\nu_2\,|\,1\Rb\hspace{1cm}\nn\\
&&\times\,\f{\Ga\Lb \h-i\nu_1\Rb\,\Ga\Lb \h-i\nu\Rb\,\Ga\Lb\h+i\nu+i\nu_1+i\nu_2\Rb\,\Ga\Lb\h-i\nu+i\nu_1+i\nu_2\Rb}{\Ga\Lb1-i\nu+i\nu_2\Rb\,\Ga\Lb1+i\nu_1+i\nu_2\Rb}\nn\\
&&\times\,_3F_2\Lb \h+i\nu_1\,,\,\h-i\nu\,,\,\h-i\nu+i\nu_1+i\nu_2\,|\,1-i\nu+i\nu_2\,,\,1+i\nu_1+i\nu_2\,|\,1\Rb
\hspace{1cm}\label{Js1}\eea


In ref. \cite{Korchemsky:1997fy}  it was shown that $\Omega\Lb\nu\,|\,\nu_1\,,\,\nu_2\,\Rb$ and $\La\Lb\nu\,|\,\nu_1\,,\,\nu_2\,\Rb$ are symmetric under cyclic permutations of $\left\{\nu\,,\,\nu_1\,,\,\nu_2\right\}$. That is;

\bea
\Omega\Lb\nu\,|\,\nu_1\,,\,\nu_2\,\Rb \,&&=\,\Omega\Lb\nu_2\,|\,\nu\,,\,\nu_1\,\Rb\,=\,\Omega\Lb\nu_1\,|\,\nu_2\,,\,\nu\,\Rb\nn\\
\nn\\
\La\Lb\nu\,|\,\nu_1\,,\,\nu_2\,\Rb \,&&=\,\La\Lb\nu_2\,|\,\nu\,,\,\nu_1\,\Rb\,=\,\La\Lb\nu_1\,|\,\nu_2\,,\,\nu\,\Rb\label{cp}\eea

with an identical result for the complex conjugates. 
For specific values of the conformal variables $\left\{\nu,\nu_1,\nu_2,\right\}$, inserting the appropriate values into \eq{1J2(a)};

\bea
\Omega\Lb i \nu|i\nu_1,i\nu_2\Rb=\Omega\Lb i\nu|\h,\h\Rb\hspace{0.75cm}&&=\,2\pi^3
\f{\Lb 1-i\nu_1-i\nu_2\Rb}{\Lb 1/2-i\nu_1\Rb^2\Lb1/2-i\nu_2\Rb^2}\label{omega123}\\
\nn\\
\Omega^\ast\Lb i\nu|i\nu_1,i\nu_2\Rb=\Omega\Lb i\nu|-\h,-\h\Rb&&=\,\f{4\pi^3}{\Lb 1/2+i\nu\Rb\Lb1/2-i\nu\Rb}\Re e\Lb\psi\Lb 1\Rb-\psi\Lb\h+i\nu\Rb\Rb\hspace{0.6cm}
\label{omega123cc}\\
\nn\\
\Omega\Lb i\nu|i\nu_1,i\nu_2\Rb=\Omega\Lb \h|0,0\Rb\hspace{1cm}&&=\f{\pi^5}{1/2-i\nu}+\f{2\pi^3}{\Lb 1/2-i\nu\Rb^2}\,_3F_2\left\{\h\,,\,\h\,,\,\h\,\mid\,1\,,\,\f{3}{2}\,\mid\,1\,\right\}\label{j1ihf}\\
\nn\\
\Omega^\ast\Lb i\nu|i\nu_1,i\nu_2\Rb=\Omega\Lb -\h|0,0,\Rb\hspace{0.5cm}&&=\f{2\pi^5}{\Lb 1/2+i\nu\Rb}\label{omegahh0}
\eea

Similarly from \eq{Js1};

\bea
\La\Lb i\nu|i\nu_1,i\nu_2\Rb=\La\Lb i\nu|\h,\h\Rb\hspace{0.7cm}&&=\f{2\pi^2}{\Lb1/2-i\nu_1\Rb\Lb1/2-i\nu_2\Rb}\label{la123}\\
\nn\\
\La^\ast\Lb i\nu|i\nu_1,i\nu_2\Rb=
\La\Lb i\nu|-\h,-\h\Rb&&=\f{2\pi^2}{\Lb 1/2+i\nu\Rb\Lb1/2-i\nu\Rb}\label{la123cc}\\
\nn\\
\La\Lb i\nu|i\nu_1,i\nu_2\Rb=\La\Lb\h|\,0,0\,\Rb\hspace{0.9cm}&&=\f{\pi^2}{\Lb1/2-i\nu\Rb^2}\label{Js1halfnu1}\\
\nn\\
\La^\ast\Lb i\nu|i\nu_1,i\nu_2\Rb=\La\Lb-\h|0,0\Rb\hspace{0.65cm}&&=\f{\pi^2}{\Lb1/2+i\nu\Rb^2}\label{Js1halfnucc2}\eea

Collecting the above results and inserting them into \eq{V5lo}, one finds that the planar piece gives a first order pole at $\left\{\nu,|\nu_1|,|\nu_2|\right\}=\left\{0,1/2,1/2\right\}$ and the non planar piece is
finite in this region, so that the planar piece is the dominant term for this region. For the region $\left\{|\nu|,\nu_1,\nu_2\right\}=\left\{1/2,0,0\right\}$ one finds the reverse, namely the planar piece is finite whereas the non planar piece offers a first order pole at 
$\left\{|\nu|,\nu_1,\nu_2\right\}=\left\{1/2,0,0\right\}$, so that the non planar piece is the relevant part of the triple Pomeron vertex for this region;

\bea
\Ga\Lb i\nu|i\nu_1,i\nu_2\Rb=
\Ga\Lb i\nu|\h,\h\Rb\hspace{0.8cm}
&&=\,32\,\pi^3\bas^2\f{\Lb\h-i\nu\Rb^2\,\Lb\h+i\nu\Rb^2}{\Lb 1/2-i\nu_1\Rb\,\Lb 1/2-i\nu_2\Rb}\nn\\
&&\times\,
\left\{\f{1-i\nu_1-i\nu_2-2/N_c^2}{\Lb 1/2-i\nu_1\Rb\,\Lb 1/2-i\nu_2\,\Rb}
+\f{2}{N_c^2}\chi\Lb\nu\Rb\!\right\}\hspace{1.1cm}\label{G123}\\
\nn\\
\Ga^\ast\Lb i\nu|i\nu_1,i\nu_2\Rb=
\Ga\Lb i\nu|-\h,-\h\Rb&&=\,\Lb 4\pi\Rb^3\,\bas^2\,\Lb \h+i\nu\,\Rb\,\Lb \h-i\nu\Rb\,
\Lb \,1-\f{1}{N_c^2}\,\Rb
\chi\Lb\nu\Rb\label{G123cc}\\
\Ga\Lb i\nu|i\nu_1,i\nu_2\Rb=\Ga\Lb \h|0,0\Rb\hspace{1cm}&&=\,\f{32\pi^3\bas^2}{N_c^2}\f{1}{1/2-i\nu}\,\hspace{0.5cm}\label{V5lohal}\\
\nn\\
\Ga^\ast\Lb i\nu|i\nu_1,i\nu_2\Rb=\Ga\Lb-\h|0,0\Rb\hspace{0.7cm}&&=\,\f{32\pi^3\bas^2}{N_c^2}\f{1}{1/2+i\nu}\,\hspace{0.5cm}\label{V5lohalcomplexconjugate}
\eea

The following notation is used for the product of two expressions of the triple Pomeron vertex with specific values, where one is the
complex conjugate of the other;

\bea
\Ga\Lb i\nu|i\nu_1,i\nu_2\Rb\,\times\,\Ga^\ast\Lb i\nu|i\nu_1,i\nu_2\Rb\,\equiv\,\begin{vmatrix}\Ga\Lb\nu|\nu_1,\nu_2\Rb\end{vmatrix}^2
\label{tpvcomplexconjugate}\eea

In this notation, combining  \eq{G123} and \eq{G123cc} yields;

\bea
\begin{vmatrix}\Ga\Lb\nu|\nu_1,\nu_2\Rb\end{vmatrix}^2=\begin{vmatrix}\Ga\Lb\nu|\h,\h\Rb\end{vmatrix}^2
&&=\,32\,\Lb\,2\pi\Rb^6\bas^4\,\Lb \,1-\f{1}{N_c^2}\,\Rb
\chi\Lb\nu\Rb\f{\Lb\h-i\nu\Rb^3\,\Lb\h+i\nu\Rb^3}{\Lb 1/2-i\nu_1\Rb\,\Lb 1/2-i\nu_2\Rb}\nn\\
&&\times\,
\left\{\f{1-i\nu_1-i\nu_2-2/N_c^2}{\Lb 1/2-i\nu_1\Rb\,\Lb 1/2-i\nu_2\,\Rb}
+\f{2}{N_c^2}\chi\Lb\nu\Rb\,\right\}\label{G123combined}
\eea

and collecting the results of \eq{V5lohal} and \eq{V5lohalcomplexconjugate};

\bea
\begin{vmatrix}\Ga\Lb\nu|\nu_1,\nu_2\Rb\end{vmatrix}^2=\begin{vmatrix} \Ga\Lb \h|0,0\Rb\end{vmatrix}^2=\Lb \f{32\pi^3\bas^2}{N_c^2}\Rb^2\f{1}{\Lb 1/2+i\nu\Rb\Lb 1/2-i\nu\Rb}
+\,\mbox{finite terms}\label{tpvuseful}\eea

\section{The fan diagram}
\label{sf}


\FIGURE[h]{\begin{minipage}{80mm}
\centerline{\epsfig{file=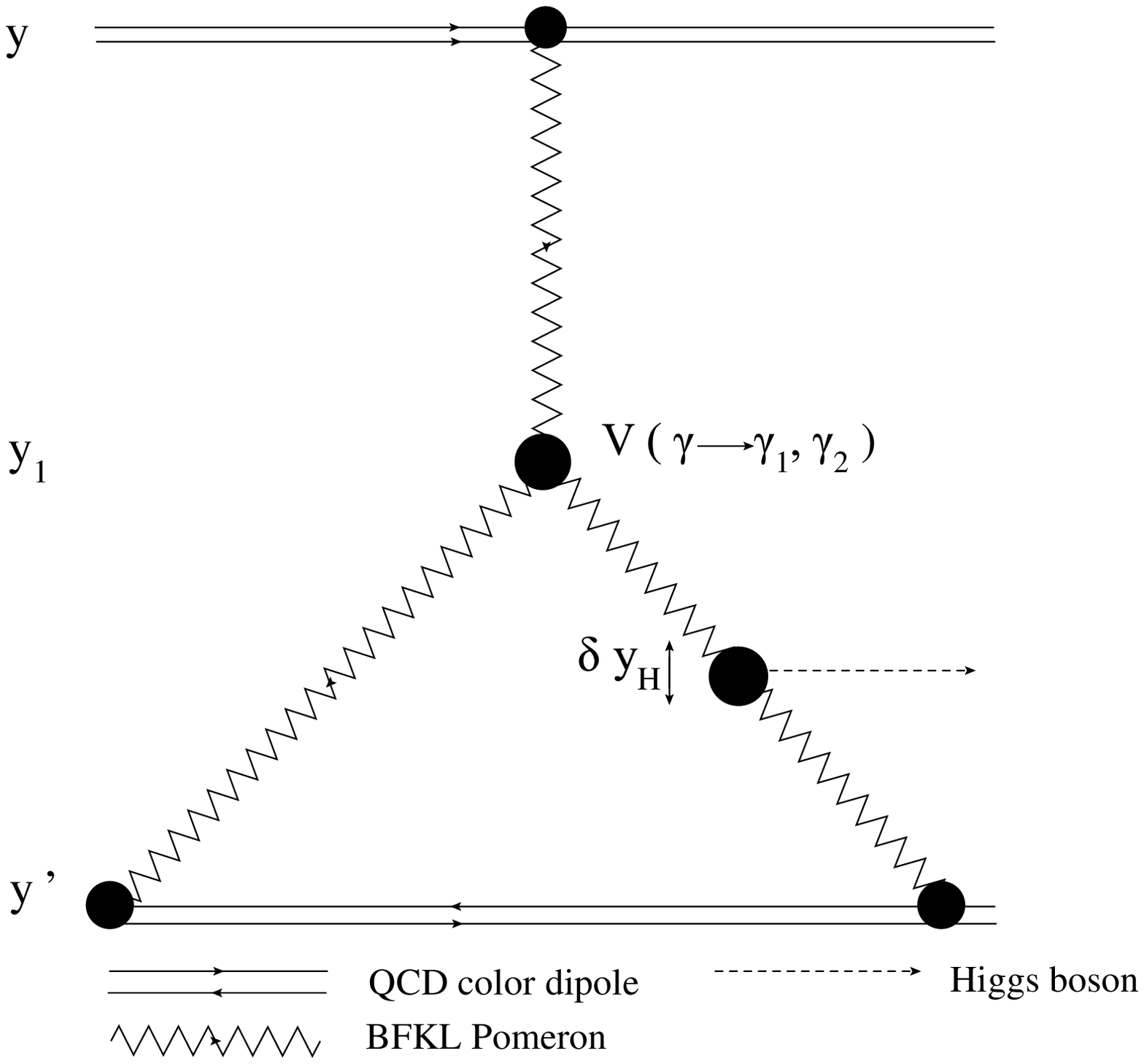,width=80mm}}
\end{minipage}
\caption{The first fan diagram with $N=1$ generations of branching. } \label{ffan} } 

In this section of the appendix the amplitude of the diagram of \fig{ffan} known as the  Pomeron ``fan'' diagram is calculated. The first fan diagram is shown in \fig{ffan},
where the t-channel Pomeron splits 
into two daughter Pomerons, and the Higgs boson is produced from one branch of the fan diagram. The scattering is between two QCD dipoles
separated by a rapidity gap $\De=y-y^{\,\prime}$ in rapidity space, and the Pomeron vertex where the splitting occurs is at rapidity $y_1$ where
$y^{\,\prime}+\de y_H\leq y_1\leq y$. The motivation for this inequality condition is to ensure that the scattering is energetic enough to produce the Higgs boson, and therefore
the separation in rapidity space between the vertex at the top of the fan and the target dipole should be at least $\de y_H$. 
In the representation of complex angular momentum $\omega = 1+j$ the fan diagram amplitude of \fig{ffan}
has the expression;

\bea
F\Lb \omega,\omega_1|\fig{ffan}\Rb&&=\f{1}{S}\Lb\f{\as}{4}\Rb^3\!A_H\!\int\!\mathcal{D}\ga\!\!\int\!\mathcal{D}\ga_1\!\!\int\!\mathcal{D}\ga_2\, g_{\omega}\Lb\ga\Rb g_{\omega_1}\Lb\ga_1\Rb g_{\omega-\omega_1}\Lb\ga_2\Rb E_\ga E_{\ga_1} E_{\ga_2} \nn\\
&&\times \int d^2R\int d^2R_1\int d^2R_2
V\Lb R,R_1,R_2|\ga\to\ga_1,\ga_2\Rb\label{ff1}
\eea

where following all the conventions of \sec{s1}, the $g_{\omega}\Lb\ga\Rb$ are the propagators of the three BFKL Pomeron states in \fig{ffan} and the $E_\ga$ represent the couplings of
the BFKL Pomeron propagators to the QCD color dipoles.  $S$ is the order of the symmetry group of \fig{ffan}. Including the $2$ ways of attaching the Pomeron branch lines at the vertex, and the $2$  permutations of the reggeized gluon lines in the
triple Pomeron vertex shown in \fig{ftpv}, the total symmetry factor  in \eq{ff1} is $S=4$. The reason behind the coefficient $\Lb\as/4\Rb^3$ in \eq{ff1} follows from
similar arguments given in \sec{s1}. In \fig{ffan} there are three vertices where the BFKL Pomeron couples to the QCD color dipole, namely at the top
of the diagram and twice at the lower ends of the two fan branches. Each vertex brings a factor of $\as$ on account of the 2 couplings of each of the
2 reggeized gluons in the BFKL Pomeron structure (see \fig{fPomeronpropagator1}). Each Pomeron - dipole vertex contains a 4 - fold degeneracy, 
since both of the 2 reggeized gluons can couple to either of the 2 quark lines in the color dipole, making a total of 4 
identical possibilities for each vertex. Thus one arrives at a factor of $\as/4$ for each of the three vertices in \fig{ffan}, which ensures
that identical diagrams are not counted more than once.\\

 $V\Lb R,R_1,R_2|\ga\to\ga_1,\ga_2\Rb$
denotes the triple Pomeron vertex for the splitting
of the BFKL Pomeron state labeled by the curly brackets ($\ga$ is the conformal variable and $R$ is the center of mass coordinate)
$\left\{R,\ga\right\}$, into the two BFKL Pomeron states labeled by $\left\{ R_1,\ga_1\right\}$ and $\left\{ R_2,\ga_2\right\}$ (see \fig{ftpv}).
Recall from \sec{s2} , that from the condition of conformal invariance comes from the property for
the triple Pomeron vertex \cite{Korchemsky:1997fy}

\bea
V\Lb R,R_1,R_2|\ga\,\to\,\ga_1\,,\,\ga_2\,\Rb
&&=\,R_{01}^{-\De_{01}}\,R_{12}^{-\De_{12}}\,R_{20}^{-\De_{20}}\,R_{01}^{\ast\,-\,\tilde{\De}_{01}}\,R_{12}^{\ast\,-\,\tilde{\De}_{12}}\,R_{20}^{\ast\,-\tilde{\De}_{20}}
\Ga\Lb \ga\,|\,\ga_1\,,\,\ga_2\Rb
\label{cinvariance1}\eea

where $R_{0i}=R-R_i\,\,(i=1,2)$ and $R_{12}=R_1-R_2$, and where for example $\De_{01}\,=\,\ga+\ga_1-\ga_2$, and $\De_{12}\,=\,\ga_1+\ga_2-\ga$, with an equivalent definition
 for the $\tilde{\De}$ in terms of the $\tilde{\ga}$. Therefore using this property, \eq{ff1} can be recast as;

\bea
F\Lb \omega,\omega_1|\mbox{\fig{ffan}}\Rb&&=\f{1}{4}\Lb\f{\as}{4}\Rb^3\!A_H\!\int\!\mathcal{D}\ga\!\!\int\!\mathcal{D}\ga_1\!\!\int\!\mathcal{D}\ga_2\, g_{\omega}\Lb\ga\Rb g_{\omega_1}\Lb\ga_1\Rb g_{\omega-\omega_1}\Lb\ga_2\Rb E_\ga E_{\ga_1} E_{\ga_2} \nn\\
&&\times g\Lb\ga|\ga_1,\ga_2\Rb\Ga\Lb\ga|\ga_1,\ga_2\Rb\label{ff2}
\eea 

where $g\Lb\ga|\ga_1,\ga_2\Rb$ represents the integral over the center of mass coordinates of the three BFKL states;

\bea
g\Lb\ga|\ga_1,\ga_2\Rb&&=\int\!d^2R\!\int\!d^2R_1\!\int\!d^2R_2\!\Lb R_{01}R_{01}^\ast\Rb^{\be_{01}}\Lb R_{12}R_{12}^\ast\Rb^{\be_{12}}
\Lb R_{20}R_{20}^\ast\Rb^{\be_{20}}\label{comintegral}\\
\mbox{where \cite{Braun:2009fy}}\hspace{0.5cm}\be_{01}&&=-\h-\h\Lb n-n_1+n_2\Rb +\Lb i\nu-i\nu_1+i\nu_2\Rb\nn\\
\be_{12}&&=-\h+\h\Lb n+n_1+n_2\Rb -\Lb i\nu+i\nu_1+i\nu_2\Rb\nn\\
\be_{20}&&=-\h+\h\Lb -n-n_1+n_2\Rb +\Lb i\nu+i\nu_1-i\nu_2\Rb\label{fullsetofpowers}\eea

Substituting for $g_{\omega}\Lb\ga\Rb$ the explicit expression of \eq{conformalpropagator} and switching to rapidity representation (using the
above explained inequality $y^{\,\prime}+\de y_H\leq y_1\leq y$ for the integration limits of the vertex rapidity $y_1$);

\bea
F\Lb \De,\de y_H|\mbox{\fig{ffan}}\Rb&&=\int^y_{y^{\,\prime}+\de y_H}\!\!\!\!\! dy_1 \int^{a+i\infty}_{a-i\infty}\!\!\!\f{d\omega}{2\pi i}\,e^{\omega\De}
\int^{a+i\infty}_{a-i\infty}\!\!\f{d\omega_1}{2\pi i}\,e^{\omega\Lb y_1-y^{\,\prime}+\de y_H\Rb}F_{(1)}\Lb\omega,\omega_1\Rb\nn\\
&&=
\Lb\f{\as}{4}\Rb^3\!A_H\int^y_{y^{\,\prime}+\de y_H}\!\!\!\!\!dy_1\int\!\mathcal{D}\ga\la\Lb\ga\Rb\,e^{\omega\Lb\ga\Rb\,\De}\nn\\
&&\times \f{1}{4}\int\!\mathcal{D}\ga_1\!\!\int\!\mathcal{D}\ga_2\la\Lb\ga_1\Rb\la\Lb\ga_2\Rb\,\exp\left\{\Lb\omega\Lb\ga_1\Rb+\omega\Lb\ga_2\Rb-\omega\Lb\ga\Rb\Rb\Lb y_1-y^{\,\prime}\Rb
-\omega\Lb\ga_1\Rb\de y_H\right\}\nn\\
&&\times E_\ga E_{\ga_1} E_{\ga_2}   g\Lb\ga|\ga_1,\ga_2\Rb\Ga\Lb\ga|\ga_1,\ga_2\Rb\label{ff3}
\eea 

where $\De=y-y^{\,\prime}$ is the rapidity gap between the scattering dipoles in \fig{ffan}. The leading contribution stems from the region where
all conformal spins $\left\{n,n_1,n_2\right\}=0$ since the BFKL eigenfunction $\omega\Lb n,\nu\Rb$ decreases sharply as $n$ increases, and
 is positive only for $n=0$ at high
energies. Therefore ignoring all solutions where $\left\{n,n_1,n_2\right\}\neq 0$, and using the explicit expression of \eq{dnu} for the
integration measure \eq{ff3} reduces to;

\bea
F\Lb \De,\de y_H|\mbox{\fig{ffan}}\Rb&&=
\Lb\f{\as}{4}\Rb^3\!A_H\int^y_{y^{\,\prime}+\de y_H}\!\!\!\!\!dy_1\int^\infty_{-\infty}d\nu h\Lb\nu\Rb\la\Lb\nu\Rb\,e^{\omega\Lb\nu\Rb\,\De}\nn\\
&&\times\f{1}{4}\int^\infty_{-\infty}d\nu_1 h\Lb\nu_1\Rb\la\Lb\nu_1\Rb\int\!d\nu_2 h\Lb\nu_2\Rb\la\Lb\nu_2\Rb
e^{\Lb\omega\Lb\nu_1\Rb+\omega\Lb\nu_2\Rb-\omega\Lb\nu\Rb\Rb\Lb y_1-y^{\,\prime}\Rb
-\omega\Lb\nu_1\Rb\de y_H}\nn\\
&&\times E_\nu E_{\nu_1} E_{\nu_2}  g\Lb\nu|\nu_1,\nu_2\Rb\Ga\Lb\nu|\nu_1,\nu_2\Rb\nn\\
\nn\\
\nn\\
&&=\f{\bas^3\,A_H}{2^8\pi N_c^3}\!\int^\infty_{-\infty}d\nu \f{\nu^2}{\Lb 1/2+i\nu\Rb^2\Lb 1/2-i\nu\Rb^2}\,
e^{\omega\Lb\nu\Rb\,\De}\,E_\nu\,f\Lb \nu|\De,\de y_H\Rb\label{ff4}\\
\nn\\
\nn\\
\mbox{where}\hspace{0.5cm}f\Lb\nu|\De,\de y_H\Rb&&=
\f{1}{2^7\pi^8}\int^\infty_{-\infty}\!\!\f{d\nu_1\nu_1^2}{\Lb 1/2+i\nu_1\Rb^2\Lb 1/2+i\nu_1\Rb^2}\int^\infty_{-\infty}\!\! \f{d\nu_2\nu_2^2}{\Lb 1/2+i\nu_2\Rb^2\Lb 1/2-i\nu_2\Rb^2}
\nn\\
&&\times \int^y_{y^{\,\prime}+\de y_H}\!\!\!\!\!\!\!\!\!\!dy_1\!\exp\left\{\Lb\omega\Lb\nu_1\Rb+\omega\Lb\nu_2\Rb-\omega\Lb\nu\Rb\Rb\Lb y_1-y^{\,\prime}\Rb
-\omega\Lb\nu_1\Rb\de y_H\right\}\nn\\
&&\times E_{\nu_1} E_{\nu_2}   g\Lb\nu|\nu_1,\nu_2\Rb\Ga\Lb\nu|\nu_1,\nu_2\Rb\label{ff5}
\eea 

\eq{ff5} represents the contribution of the fan part of the diagram of \fig{ffan}. The leading contribution to \eq{ff5} stems from the region
  $\left\{|\nu|,\nu_1,\nu_2\right\}=\left\{1/2,0,0\right\}$ which was discussed in the previous sections for
solving the Pomeron loop integrals.  For this region, $g\Lb\nu|0,0\Rb$ can be calculated from \eq{comintegral}, by choosing the frame of reference where one of the center of mass coordinates $R=0$. 
With this choice of frame after evaluating the remaining integrations over $R_1$ and $R_2$ it turns out that;

\bea
g\Lb\nu|0,0\Rb=\f{16\pi^3}{1/2-i\nu}\label{comintegralsolved}\eea

After substituting \eq{comintegralsolved} and the triple Pomeron vertex of \eq{V5lohal} into \eq{ff5}, the integrals over $\nu_1$ and $\nu_2$ can be solved using
the method of steepest descents. Plugging the result back into \eq{ff4} yields the expression;

\bea
F\Lb \De,\de y_H|\mbox{\fig{ffan}}\Rb
&&=\f{2\bas^5\,A_H}{ N_c^5\pi^2 [\omega^{\,\prime\prime}\Lb 0\Rb]^3}\!\oint _C\!d\nu \f{\nu^2}{\Lb 1/2+i\nu\Rb^2\Lb 1/2-i\nu\Rb^4}\,\,E_\nu\nn\\
&&\times\int^y_{y^{\,\prime}+\de y_H}\!\!\!\!\!\!\!\!\!\!dy_1e^{\omega\Lb\nu\Rb\,\Lb y-y_1\Rb}\,\f{e^{2\omega\Lb0\Rb\Lb y_1-y^{\,\prime}-\de y_H/2\Rb}}{\Lb y_1-y^{\,\prime}\Rb^{3/2}\Lb y_1-y^{\,\prime}-\de y_H\Rb^{3/2}} \,\label{ff12}\eea

 where $C$ is the contour shown in \fig{fintegrationcontour} which encloses the pole at $i\nu=1/2$, and a factor of $2$ has been included which takes into account the identical contribution from the contour $C^{\,\prime}$,
 which  encloses the pole at $i\nu=-1/2$.
 One can solve the $\nu$ integral using the same technique shown in \eq{trick}, which generates the derivative of the Dirac delta function  $\de^{(2)}\Lb y-y_1\Rb/\bas^2$, and this is absorbed by the integration over
 the rapidity variable $y_1$ to finally yield;
 
 \bea
F\Lb \De,\de y_H|\mbox{\fig{ffan}}\Rb
&&=\f{2\bas^2\,A_H}{ N_c^5\pi [\omega^{\,\prime\prime}\Lb 0\Rb]^3}\f{d^2}{d\De^2} \Lb\f{e^{2\omega\Lb0\Rb\Lb \De-\de y_H/2\Rb}}{\De^{3/2}\Lb \De-\de y_H\Rb^{3/2}}\Rb\nn\\
\nn\\
&&=\,2.93\times\,10^{-11}\,\,\mbox{GeV}^{-2}\hspace{1cm}(\as\,=\,0.12)\nn\\
&&\hspace{0.4cm}
3.95\,\times\,10^{-8}\,\,\mbox{GeV}^{-2}\hspace{1.2cm}(\as\,=\,0.2) \label{ff12}\eea

\end{document}